\pdfoutput=1
\documentclass[12pt]{article}
\usepackage{caption}
\usepackage{amsmath,amsfonts,amsthm,amssymb}
\usepackage{subfigure}
\usepackage{setspace}
\usepackage{Tabbing}
\usepackage{lastpage}
\usepackage{extramarks}
\usepackage{chngpage}
\usepackage[usenames,dvipsnames]{color}
\usepackage{graphicx,float,wrapfig}

\def\rf#1{(\ref{eq:#1})}
\def\lab#1{\label{eq:#1}}

\def\br{\begin{eqnarray}}
\def\er{\end{eqnarray}}
\def\be{\begin{equation}}
\def\ee{\end{equation}}
\def\({\left(}
\def\){\right)}
\def\pa{\partial}
\def\rlx{\relax\leavevmode}
\def\IR{\rlx\hbox{\rm I\kern-.18em R}}
\def\vp{\varphi}
\def\ve{\varepsilon}

\newcommand{\sbr}[2]{\left\lbrack\,{#1}\, ,\,{#2}\,\right\rbrack}

\def\IZ{\rlx\hbox{\sf Z\kern-.4em Z}}
\def\IR{\rlx\hbox{\rm I\kern-.18em R}}
\def\IC{\rlx\hbox{\,$\inbar\kern-.3em{\rm C}$}}
\def\one{\hbox{{1}\kern-.25em\hbox{l}}}

\def\FAaIA#1#2#3{{\sl Functional Analysis and Its Application} {\bf #1} (#2)
#3}

\def\PHSD#1#2#3{{\sl Physica} {\bf D#1} (#2) #3}
\def\PJA#1#2#3{{\sl Proc. Japan. Acad} {\bf #1A} (#2) #3}

\begin{document}

\begin{titlepage}
\vspace*{-1cm}

\vskip 2cm

\vspace{.2in}
\begin{center}
{\large\bf Quasi-integrable deformations of the $SU(3)$ Affine Toda Theory}
\end{center}

\vspace{.5cm}

\begin{center}
L. A. Ferreira~$^{\star}$
P. Klimas~$^{\nabla}$ and Wojtek J. Zakrzewski~$^{\dagger}$

\small

\par \vskip .2in \noindent
$^{(\star)}$Instituto de F\'\i sica de S\~ao Carlos; IFSC/USP;\\
Universidade de S\~ao Paulo  \\ 
Caixa Postal 369, CEP 13560-970, S\~ao Carlos-SP, Brazil\\
email: laf@ifsc.usp.br
\small

%\par \vskip \vspace{.3 in}.2in \noindent
$^{(\nabla)}$Universidade Federal de Santa Catarina,\\
 Trindade, CEP 88040-900, Florian\'opolis-SC, Brazil\\
email: pawel.klimas@ufsc.br

\small 
\par \vskip .2in \noindent
$^{(\dagger)}$~Department of Mathematical Sciences,\\
 University of Durham, Durham DH1 3LE, U.K.\\
email: W.J.Zakrzewski@durham.ac.uk

\normalsize
\end{center}

%%%%%%%%%%%%%%%%%%%%%%%%%%%%%%%%%%%%%%%%%%%%%

%\vspace{.5in}

\begin{abstract}

We consider deformations of the $SU(3)$ Affine Toda theory (AT) and investigate the integrability properties of the deformed theories. We find that for some special deformations all conserved quantities change to being   conserved only asymptotically, {\it  i.e.}   in the process of the scattering of two solitons these charges do vary in time, but they return, after the scattering, to the values they had prior to the scattering. This phenomenon, which we have called   quasi-integrability,  is related to special properties of the two-soliton solutions under space-time parity transformations. Some  properties of the  AT solitons are discussed, especially those involving interesting static multi-soliton solutions. We support our analytical studies with detailed numerical ones in which the time evolution has been simulated by the 4th order Runge-Kutta method. We find that for some perturbations the solitons repel and for the others they form a quasi-bound state. When we send solitons towards each other they can repel when they come close together with or without  `flipping' the fields of the model. The solitons radiate very little and appear to be  stable. 
These results support the ideas of quasi-integrability, {\it i.e.} that many effects of integrability also approximately hold for the deformed models. 

\end{abstract} 
\end{titlepage}

%%%%%%%%%%%%%%%%%%%%%%%%%%%%%%%%%%%%%%%%%%%%%

\section{Introduction}
\label{sec:intro}
\setcounter{equation}{0}

Solitons play a very important role in the study of non-linear phenomena because often they arise in the mathematical description of the behaviour of some physical systems. 

Many properties of solitons are associated with the integrability of the mathematical models in which they arise. In such cases solitons are described as localised classical field configurations of the model that propagate without dissipation and dispersion. Moreover, when two such solitons are scattered they do not destroy each other but come out of their interaction region essentially unscathed. The only lasting effect of the scattering is a shift in their positions relative to the values they would have had, had they not encountered each other. The usual explanation of this behaviour involves the integrability of the model and associated with it existence of an infinite number of conserved quantities. These conservation laws dramatically constrain the soliton dynamics. The integrable theories are, however, very special as they possess highly non-trivial hidden symmetries. So, even small perturbations of these theories can destroy these symmetries and it is important to check whether any of these properties still hold when the underlying mathematical models are nonintegrable. Afterall, one would expect some `continuity' of the properties as one introduces small (or not so small) perturbations.

We have looked at this problem and recently we have found that some non-integrable field theories in $(1+1)$ dimensions, present properties similar to those of exactly integrable theories \cite{us,us2,recent,vrecent,LuizandVinicius}. They have soliton-like field configurations that behave in a scattering process in a way which is very similar to true solitons. We have also shown that such theories possess an infinite number of quantities which are not exactly time-independent  but are, however, asymptotically conserved. By that we mean that the values of these quantities change during their scattering process, and at times change a lot, but after the scattering, they  return, to the values they have had before it. This is an interesting property since from the point of view of the scattering what matters are the asymptotic states, and so a theory in which solitons behave like this looks a bit as an effectively integrable theory. For these reasons we have named this phenomenon {\em quasi-integrability}. The mechanisms  responsible for this behaviour are not properly understood yet, but we believe that this behaviour will play an important role in the study of many non-linear phenomena. Since integrable theories are rare and, in general, do not describe realistic physical phenomena, the  {\em quasi-integrable} theories may play a significant role in the description of more realistic physical processes. 

Most of the models we studied so far \cite{us,us2,recent,vrecent,LuizandVinicius,blas,baron1,baron2} involved $(1+1)$-dimensional theories of either one real scalar field $\phi$ subjected to a potential which is a deformation of the Sine-Gordon potential or a complex field which satisfied a modified non-linear Schr\"odinger equation or equation of the modified Bullough-Dodd model. 
The original models were integrable and the deformation of their potentials made them non-integrable.

Here we decided to extend our investigations to systems with more fields and so we have had a look at the $SU(N)$ Toda models and their deformations. All such undeformed models are integrable and the lowest of them ($N=2$) 
is, in fact, the Sine-Gordon model in disguise. So, in this paper we report results of our study of the next model in this family of models, namely, of the $SU(3)$ one.

The paper is organized as follows. In section \ref{sec:model},  we present this model and discuss some of its properties and in particular its symmetries.
We also suggest a possible deformation of the model which possesses most of these symmetries.  The following section discusses various properties of both the undeformed
and deformed models such as their quasi-zero curvature conditions and the resulting quasi-conserved quantities. 
Section 4 discusses how the fields of these models change when one Lorentz transforms them and when they lead to charge conservation. We also present the explicit expression for the anomaly terms - which control the situation when the charges are only asymptotically conserved (which corresponds to our ideas of quasi-integrability). 
In section 5 we discuss the well known soliton solutions of the undeformed model paying particular attention to the solutions which describe static solitons.

The following two sections describe the numerical procedure used by us for checking some of these claims and present the results of our numerical investigations.
In fact all our results were obtained using the 4th order Runge-Kutta method to simulate the time dependence of field configurations. First we performed such numerical evolutions of field configurations for which we had analytical expressions. 
This not only checked our numerical schemes but also demonstrated that the soliton solutions of the un-deformed $SU(3)$ model were really stable, with respect to small numerically induced, perturbations. Then we looked at the deformed models for various values of the deformation and for solitons at rest. We followed these studies by looking at solitons moving towards each other at various speeds. In section 8  we present some of our conclusions. 

%%%%%%%%%%%%%%%%%%%%%%%%%%%%%%%%%%%%%%%%%%%%%%

\section{The model}
\label{sec:model}
\setcounter{equation}{0}

In this paper we consider field theories in $(1+1)$-dimensional Minkowski space-time for two complex scalar fields $\phi_a$, $a=1,2$, defined by the Lagrangian
\be
{\cal L}= \frac{1}{12}\left[ \(\partial_{\mu}\phi_1\)^2 + \(\partial_{\mu}\phi_2\)^2 - \partial_{\mu}\phi_1\,\partial^{\mu}\phi_2\right] -V\(\phi_1\, ,\, \phi_2\)= \frac{1}{24}\, \(\partial_{\mu} {\vec \phi}\)^2 - V,
\lab{lagrangian}
\ee
where we have introduced the vector 
\be
{\vec \phi} = {\vec \alpha}_1\, \phi_1+ {\vec \alpha}_2\, \phi_2
\lab{vecphidef}
\ee
and where ${\vec \alpha}_1$ and ${\vec \alpha}_2$ are the simple roots of $SU(3)$,  
 with   $\alpha_1\cdot \alpha_2=-1$,  and $\alpha_1^2=\alpha_2^2=2$.

The corresponding Euler-Lagrange equations are given by
\be
\frac{1}{12}\left[ \partial_t^2 {\vec \phi} -\partial_x^2 {\vec \phi}\right] = -{\vec \nabla}_{\phi}\,V,
\lab{veceqmotgen}
\ee
where ${\vec \nabla}_{\phi}$ is the gradient in $\phi$-space. In terms of the components fields $\phi_a$ one gets 
\br
\partial_{+}\partial_{-}\phi_1 &=& 2\, \frac{\delta\, V}{\delta\,\phi_1}+ \frac{\delta\, V}{\delta\,\phi_2},
\nonumber\\
\partial_{+}\partial_{-}\phi_2 &=& 2\, \frac{\delta\, V}{\delta\,\phi_2}+ \frac{\delta\, V}{\delta\,\phi_1}.
\lab{generaleq}
\er
Here we have introduced the light-cone coordinates (with the speed of light set to unity)
\be
x_{\pm}=x\pm t,\qquad\qquad \partial_{\pm}=\frac{1}{2}\(\partial_x\pm\partial_t\),
\qquad\qquad \partial_{+}\partial_{-}=-\frac{1}{4}\(\partial_t^2-\partial_x^2\).
\lab{lightconedef}
\ee

The integrable $SU(3)$ Affine Toda model corresponds to the potential
\br
V_{\rm Toda}&=&-\frac{1}{3}\left[ e^{i\(2\phi_1-\phi_2\)}+e^{i\(2\phi_2-\phi_1\)}+e^{-i\(\phi_1+\phi_2\)}-3 \right]
\nonumber\\
&=& -\frac{1}{3}
\left[ e^{i\,{\vec \alpha}_1\cdot {\vec \phi}}+e^{i\,{\vec \alpha}_2\cdot {\vec \phi}}+e^{i\,{\vec\alpha}_0\cdot {\vec \phi}}-3\right],
\lab{todapotential}
\er
where  ${\vec \alpha}_0=-{\vec\alpha}_1-{\vec\alpha}_2$ (see \rf{vecphidef}). 

In this paper we consider deformations of the integrable Affine Toda model, such that we keep the kinetic term in \rf{lagrangian} unchanged, but take the potential to be of the form 
\be
V_{{\vec v}}=-\frac{1}{3}\left[ e^{i\,{\vec v}_1\cdot {\vec \phi}}+e^{i\,{\vec v}_2\cdot {\vec \phi}}+e^{i\,{\vec v}_0\cdot {\vec \phi}}-3\right],
\lab{deformedpot}
\ee
where ${\vec \phi}$ is still given by \rf{vecphidef}, and ${\vec v}_j$, $j=0,1,2$, are vectors in the root space of the $SU(3)$ Lie algebra, which are deformations of the roots ${\vec \alpha}_j$. The choice of the vectors ${\vec v}_j$ is restricted by some conditions which we will discuss below.

The Hamiltonian density and energy associated to \rf{lagrangian} are given respectively by
\be
{\cal H} =  \frac{1}{24}\, \left[\(\partial_{t} {\vec \phi}\)^2+\(\partial_{x} {\vec \phi}\)^2\right] + 
V, 
\qquad\qquad\qquad\qquad
 E=\int_{-\infty}^{\infty} dx\, {\cal H}.
 \lab{hamiltonian}
\ee
Since the fields are complex, so are the Hamiltonian density and energy. Therefore, such models do not 
possess vacua solutions that minimize the energy. However, in order for the energy to be conserved in time, it is necessary to require that the flows of momenta at both ends of spatial infinity are equal, {\it i.e.} that 
\be
\frac{d\,E}{d\,t}= \frac{1}{12}\,\partial_x{\vec \phi}\cdot  \partial_t{\vec \phi}\mid_{x=-\infty}^{x=\infty} = 0.
\ee
For the solutions which we consider in this paper this condition is satisfied as space and time derivatives of the fields vanish at spatial infinity. For static configurations there is a further point to take into account.  It is well known that for theories of the type we are considering the quantity 
\be
{\cal E} =  \frac{1}{24}\, \(\partial_{x} {\vec \phi}\)^2 - V
\ee
is independent of $x$ for static solutions of the equations of motion, {\it i.e.} $\frac{d\,{\cal E}}{d\,x}=0$. This corresponds to the energy of a mechanical problem of a particle moving in $\phi$-space in an inverted potential with $x$ playing the role of time. Therefore, for static solutions for which the space derivatives of the fields vanish at spatial infinity one finds that the conservation of ${\cal E}$ in $x$, implies that
\be
V\({\vec \phi}_{(+)}\)=V\({\vec \phi}_{(-)}\),
\lab{condpotgen}
\ee
where ${\vec \phi}_{(\pm)}$ are the asymptotic values of the fields at spatial infinity, {\it i.e.} $ {\vec \phi} \rightarrow {\vec \phi}_{(\pm)}$, as $x\rightarrow \pm \infty$. For the deformed potentials 
\rf{deformedpot} the condition \rf{condpotgen} becomes 
\be
\sum_{j=0}^2 e^{i\,{\vec v}_j\cdot {\vec \phi}_{(+)}}=\sum_{j=0}^2 e^{i\,{\vec v}_j\cdot {\vec \phi}_{(-)}}.
\lab{potconddeformed}
\ee
However, for the static equations \rf{veceqmotgen} to be satisfied at spatial infinity one requires that
\be
\sum_{j=0}^2 {\vec v}_j\, e^{i\,{\vec v}_j\cdot {\vec \phi}_{(\pm)}}=0.
\lab{condstaticeqmot}
\ee
This imposes conditions on vectors $\vec v_i$.

Let us restrict our interest to the cases where ${\vec v}_1$ and ${\vec v}_2$ are linearly independent and consider the dual basis ${\vec w}_a$, such that ${\vec w}_a\cdot{\vec v}_b=\delta_{ab}$, $a,b=1,2$. Then, taking the scalar product of \rf{condstaticeqmot} with ${\vec w}_a$ one finds that
\be
e^{i\,{\vec v}_1\cdot {\vec \phi}_{(\pm)}}
+{\vec w}_1\cdot{\vec v}_0\,e^{i\,{\vec v}_0\cdot {\vec \phi}_{(\pm)}}=0,
\qquad\qquad\qquad 
e^{i\,{\vec v}_2\cdot {\vec \phi}_{(\pm)}}
+{\vec w}_2\cdot{\vec v}_0\,e^{i\,{\vec v}_0\cdot {\vec \phi}_{(\pm)}}=0.
\lab{condstaticeqmot2}
\ee
Next we note that we have to discard the cases where ${\vec v}_0$ is orthogonal either to ${\vec w}_1$ or ${\vec w}_2$, since \rf{condstaticeqmot2} would imply that the imaginary part of ${\vec \phi}_{(\pm)}$ had to diverge, and so the derivatives of the fields would not vanish asymptotically at spatial infinity as we have assumed. One then concludes from \rf{condstaticeqmot2} that
\be
\frac{e^{i\,{\vec v}_1\cdot {\vec \phi}_{(\pm)}}}{{\vec w}_1\cdot{\vec v}_0}=
\frac{e^{i\,{\vec v}_2\cdot {\vec \phi}_{(\pm)}}}{{\vec w}_2\cdot{\vec v}_0}=
- e^{i\,{\vec v}_0\cdot {\vec \phi}_{(\pm)}}.
\lab{condstaticeqmot3}
\ee
Using \rf{condstaticeqmot3} one can conclude that \rf{potconddeformed} implies that 
\be
\left[1-\({\vec w}_1+{\vec w}_2\)\cdot {\vec v}_0\right]\,
e^{i\,{\vec v}_j\cdot \({\vec \phi}_{(+)}-{\vec \phi}_{(-)}\)}=
\left[1-\({\vec w}_1+{\vec w}_2\)\cdot {\vec v}_0\right], \qquad\qquad\qquad j=0,1,2.
\ee
Thus we have  two possibilities. Either
\be
\({\vec w}_1+{\vec w}_2\)\cdot {\vec v}_0=1\qquad\qquad\mbox{\rm and so}\qquad \qquad
{\vec v}_0=\beta\, {\vec v}_1+\(1-\beta\)\,{\vec v}_2 \qquad (\beta \; {\rm real})
\lab{firstcase}
\ee
or
\be
e^{i\,{\vec v}_j\cdot \({\vec \phi}_{(+)}-{\vec \phi}_{(-)}\)}=1,
\qquad\qquad\qquad\qquad j=0,1,2.
\lab{secondcase}
\ee
However, we are really interested in theories that can be  deformed away from the Affine Toda model in a continuous manner. If one takes $\({\vec w}_1+{\vec w}_2\)\cdot {\vec v}_0=1$ then  there is no way of having ${\vec v}_j$, $j=0,1,2$, as close as possible to ${\vec\alpha}_j$. So we shall discard the possibility \rf{firstcase}. The second case \rf{secondcase} implies that the difference of the asymptotic values of the fields has to live on a dual lattice, {\it i.e.} 
\be
{\vec \phi}_{(+)}-{\vec \phi}_{(-)}=2\,\pi\,\(m_1\,{\vec w}_1+m_2\,{\vec w}_2\),
\qquad\qquad\qquad \qquad m_1\, ,\, m_2\equiv \; {\rm integers}.
\lab{duallatticecond}
\ee
In addition we have to take ${\vec v}_0$ as
\be
{\vec v}_0=n_1\, {\vec v}_1+n_2\, {\vec v}_2,
\qquad\qquad\qquad \qquad n_1\, ,\, n_2\equiv \; {\rm integers}.
\lab{v0possible}
\ee

Let us restrict our attention to  deformations that preserve, as much as possible,  the symmetries of the Affine Toda model. For instance, the undeformed model \rf{todapotential} is invariant under the exchange $\phi_1\leftrightarrow \phi_2$. In addition,  for the solutions which satisfy either the condition  $\phi_2=-\phi_1^*$, or $\phi_a=-\phi_a^*$, $a=1,2$, the energy becomes real. So, in order to keep such symmetries  and the reality conditions for the energy, we consider in this paper the following deformation:  
\br
{\vec v}_1&=&\(1-\frac{\ve}{3}\)\,{\vec \alpha}_1- \frac{2}{3}\,\ve\,{\vec \alpha}_2,
\nonumber\\
{\vec v}_2&=&- \frac{2}{3}\,\ve\,{\vec \alpha}_1+ \(1-\frac{\ve}{3}\)\,{\vec \alpha}_2,
\lab{choicev}\\
{\vec v}_0&=&-\({\vec v}_1+{\vec v}_2\)=-\(1-\ve\)\({\vec \alpha}_1+{\vec \alpha}_2\)
\nonumber
\er
with $\ve$ being a real parameter. Note that this corresponds to taking $n_1=n_2=-1$ in \rf{v0possible} and so $\vec v_0$ is expressed in terms of $\vec v_i$ like $\alpha_0$ in terms of $\vec\alpha_i$.
It then follows that   
${\vec v}_1\cdot {\vec \alpha}_1={\vec v}_2\cdot {\vec \alpha}_2=2$, and 
${\vec v}_1\cdot {\vec \alpha}_2={\vec v}_2\cdot {\vec \alpha}_1=-\(1+\varepsilon\)$. In addition, one finds  that 
\be
{\vec v}_1^2 ={\vec v}_2^2 =2\(1+\frac{\varepsilon^2}{3}\),
\qquad\qquad 
\qquad\qquad 
{\vec v}_1\cdot{\vec v}_2= -\(1+2\,\varepsilon-\frac{\varepsilon^2}{3}\).
\lab{deformedroots}
\ee
With such a choice, the potential \rf{deformedpot} becomes 
\be
V_{\ve}= -\frac{1}{3}\left[e^{i\left[2\,\phi_1-\(1+\varepsilon\)\phi_2\right]}+
e^{i\left[2\,\phi_2-\(1+\varepsilon\)\phi_1\right]}+e^{-i\(1-\varepsilon\)\left[\phi_1+\phi_2\right]}-3\right]. 
\lab{deformedpotepsilon}
\ee
Note that  the vectors ${\vec v}_a$, $a=1,2$, correspond to the deformations of the simple roots ${\vec \alpha}_a$ of $SU(3)$ which modify the angle between them, and rescale their lengths equally, as shown in  \rf{deformedroots}. The dual basis associated to the choice \rf{choicev} is given by
\be
{\vec w}_1=\frac{2\,{\vec \alpha}_1+\(1+\ve\)\,{\vec \alpha}_2}{\(3+\ve\)\(1-\ve\)},
\qquad\qquad\qquad \qquad
{\vec w}_2=\frac{2\,{\vec \alpha}_2+\(1+\ve\)\,{\vec \alpha}_1}{\(3+\ve\)\(1-\ve\)}.
\lab{dualbasis}
\ee
Using \rf{vecphidef} and \rf{dualbasis} one finds that the condition \rf{duallatticecond} becomes
\be
\phi_1^{(+)}-\phi_1^{(-)}=2\,\pi\,\left[\frac{2\,m_1+\(1+\ve\)\,m_2}{\(3+\ve\)\(1-\ve\)}\right]\,,
\qquad\quad
\phi_2^{(+)}-\phi_2^{(-)}=2\,\pi\,\left[\frac{2\,m_2+\(1+\ve\)\,m_1}{\(3+\ve\)\(1-\ve\)}\right].
\ee
As we have remarked above, a given solution satisfying  the condition  $\phi_2=-\phi_1^*$,   has real energy. Therefore, for such static solutions one needs $m_1= - m_2$, and so 
\be
\phi_1^{(+)}-\phi_1^{(-)}=-\left[\phi_2^{(+)}-\phi_2^{(-)}\right]=\frac{2\,\pi\, m_1}{\(3+\ve\)}.
\lab{boundaryvaluesphi}
\ee
At the same time we observe that  a  solution satisfying the condition $\phi_a=-\phi_a^*$, $a=1,2$, also has real energy, and a static solution of this kind  can only exist when $m_1=m_2=0$.

%%%%%%%%%%%%%%%%%%%%%%%%%%%%%%%%%%%%%%%%%%%%%

\section{The quasi-zero curvature condition}
\label{sec:curvature}
\setcounter{equation}{0}

To discuss integrability of the model we introduce the Lax potentials as
\br
A_{+}&=& - \(V+v_0\)\, b_1 + i\,\left[\frac{\delta \, V}{\delta\,\phi_1} \(E_{\alpha_1}^0-E_{-\alpha_3}^1\) +
\frac{\delta \, V}{\delta\,\phi_2} \(E_{\alpha_2}^0-E_{-\alpha_3}^1\)\right],
\nonumber\\
A_{-}&=& b_{-1}-i\,\sum_{a=1}^2\partial_{-}\phi_a\,H_{\alpha_a}^0,
\lab{laxpot}
\er
with $v_0$ being a constant, and 
\be
b_1=E_{\alpha_1}^0+E_{\alpha_2}^0+E_{-\alpha_3}^1,\qquad\qquad
b_{-1}=E_{-\alpha_1}^0+E_{-\alpha_2}^0+E_{\alpha_3}^{-1}
\lab{bpm1def}
\ee
with $H_{\alpha_a}^n$, $a=1,2$, and $E_{\pm\alpha_s}^n$, $s=1,2,3$, $n\in \IZ$, being the Chevalley basis of the $SU(3)$ loop algebra described in appendix \ref{sec:su3}.

The curvature of such potentials takes the form
\br
F_{+-}&=& \partial_{+} A_{-}-\partial_{-} A_{+}+\sbr{A_{+}}{A_{-}}=-i\left[
\partial_{+}\partial_{-}\phi_1 - 2\, \frac{\delta\, V}{\delta\,\phi_1}- \frac{\delta\, V}{\delta\,\phi_2}\right] \, H_{\alpha_1}^0
\lab{quasizc}\\
 &-&i\left[
\partial_{+}\partial_{-}\phi_2 - 2\, \frac{\delta\, V}{\delta\,\phi_2}- \frac{\delta\, V}{\delta\,\phi_1}\right] \, H_{\alpha_2}^0
 -i\, \sum_{a=1}^2 X_a\, F_1^a
 \nonumber
\er
with
\be
F_1^1=E_{\alpha_1}^0+\omega\, E_{\alpha_2}^0+\omega^2\,E_{-\alpha_3}^1,\qquad\qquad\qquad
F_1^2=E_{\alpha_1}^0+\omega^2\, E_{\alpha_2}^0+\omega\,E_{-\alpha_3}^1.
\lab{fandef}
\ee
Here $\omega$ is a cubic root of unity other than unity itself, {\it i.e.} $\omega^3=1$ and $\omega\neq 1$, and so $1+\omega+\omega^2=0$. In addition, we have
\br
X_1&=& \frac{1}{3}\left[\(1-\omega\)\,\partial_{-} \phi_1\, W_1\(\omega\)- 
\omega\,\(1-\omega\)\,\partial_{-} \phi_2\, W_2\(\omega\)\right],
\nonumber\\
X_2&=& \frac{1}{3}\left[\(1-\omega^2\)\,\partial_{-} \phi_1\, W_1\(\omega^2\)+
\omega\,\(1-\omega\)\,\partial_{-} \phi_2\, W_2\(\omega^2\)\right],
\lab{x1x2def}
\er
where
\br
W_1\(\omega\)&=& 
\frac{\delta^2\,V}{\delta\phi_1^2}-\omega\,\frac{\delta^2\,V}{\delta\phi_1\,\delta\,\phi_2}
+i\,\omega^2 \frac{\delta\,V}{\delta\phi_1}-i\,\omega \frac{\delta\,V}{\delta\phi_2}
+\(1-\omega^2\)\,\(V+v_0\),
\nonumber\\
W_2\(\omega\)&=& 
\frac{\delta^2\,V}{\delta\phi_2^2}-\omega^2\,\frac{\delta^2\,V}{\delta\phi_1\,\delta\,\phi_2}
-i\,\omega^2 \frac{\delta\,V}{\delta\phi_1}+i\,\omega \frac{\delta\,V}{\delta\phi_2}
+\(1-\omega\)\,\(V+v_0\).
\lab{w1w2def}
\er
Note that, as $W_a$, $a=1,2$, are functions of $\omega$, in the calculation of $X_2$ one has to interchange $\omega\leftrightarrow\omega^2$ in the expressions for $W_a$ given above.

The coefficients of $H_{\alpha_a}^0$, $a=1,2$, in \rf{quasizc} are exactly the equations of motion \rf{generaleq} of the deformed models we are considering, and so they vanish when evaluated on the solutions of such models.  In order for the curvature $F_{+-}$ to vanish one needs the anomalies $X_a$, $a=1,2$ to vanish, and so one has to choose potentials that satisfy the four equations, 
$W_a\(\omega\)=W_a\(\omega^2\)=0$, for $a=1,2$. If one takes an ansatz of the form $V\sim \left[\exp\(i\, \gamma_a\, \phi_a\)-v_0\right]$, then these four equations become four algebraic equations for the unknowns $\gamma_1$ and $\gamma_2$. One can check that the only possible solutions are three choices: 
\be
\(\gamma_1 \, ,\, \gamma_2\)= \(2 \, ,\, -1\)\; ; \; \(-1 \, ,\, 2\)\; \; {\rm or} \; \;\(-1 \, ,\, -1\)
\ee
and so any linear combination of the form $V= q_1\, e^{i\, \(2 \phi_1-\phi_2\)}+q_2\, e^{i\, \(- \phi_1+2\,\phi_2\)}+q_0\, e^{i\, \(- \phi_1-\phi_2\)}-\mu_0$, leads to the vanishing of the anomalies, and so to an exactly integrable field theory. The Affine Toda model, corresponding to all $q_j\neq 0$, $j=0,1,2$, and the so-called Conformal Toda model corresponding to $q_0=0$, are examples of such integrable models.

\subsection{The quasi-conserved quantities}
\label{sec:quasicharges}

In order to calculate the quasi-conserved quantities for the theories \rf{lagrangian} we employ a modified version of the technique widely used in integrable field theories 
\cite{drinfeldsokolov,olive1,olive2,aratyn}. This procedure is called the abelianization procedure because it consists of gauge transforming the Lax potentials into an infinite abelian sub-algebra of the $SU(3)$ loop algebra. In our case, due to the fact that the potentials \rf{laxpot} are not really flat, we are able to gauge transform only one component of   \rf{laxpot} into the abelian sub-algebra.  The main ingredient of the technique relies upon the fact that the generator $b_{-1}$ introduced in \rf{bpm1def}, is a semi-simple element of the $SU(3)$ loop algebra ${\cal G}$. By this we mean that the kernel and image of the adjoint action of $b_{-1}$ have no intersection and ${\cal G}$ splits into the vector space sum of kernel and image, {\it i.e.}
\be
{\cal G}= {\rm Ker} + {\rm Im}\, ; \qquad \qquad \sbr{b_{-1}}{{\rm Ker}}=0 \, ;\qquad\quad 
{\rm Im}=\sbr{b_{-1}}{{\cal G}}\, ; \qquad\quad {\rm Ker} \cap {\rm Im}=0.
\lab{kernelimagedef}
\ee
The second important ingredient of the technique is an integer gradation of the $SU(3)$ loop algebra ${\cal G}$, such that 
\be
{\cal G} = \bigoplus_{n=-\infty}^{\infty} {\cal G}_n \, ; \qquad\quad 
\sbr{D}{{\cal G}_n}= n\, {\cal G}_n\, ; \qquad\quad \sbr{{\cal G}_n}{{\cal G}_m}\subset {\cal G}_{n+m}\, ;\qquad\quad  
 n\, , \, m \in \IZ.
\ee
The relevant gradation for our case is the so-called principal gradation performed by the grading operator 
\be
D = H_{\alpha_1}^0+H_{\alpha_2}^0+3\, \lambda\,\frac{d\;}{d\lambda},
\lab{gradingoperator}
\ee
where $H_{\alpha_a}^0$, $a=1,2$, are the generators of the Chevalley  basis of the Cartan sub-algebra of ${\cal G}$, and $\lambda$ is the so-called spectral parameter of the loop algebra (see appendix 
\ref{sec:su3} for details). 

The calculations become  simpler if one uses a special basis for ${\cal G}$, described in appendix \ref{sec:su3}, where the generators of the kernel are denoted as $b_{3n\pm 1}$, $n\in \IZ$, and the generators of the image as $F_n^a$, $n\in \IZ$, $a=1,2$, and they have well defined grades {\it w.r.t.} $D$, {\it i.e.}
\be
\sbr{D}{b_{3n\pm 1}}=\(3n\pm 1\)\, b_{3n\pm 1}\; ;\qquad\qquad\qquad
\sbr{D}{F_n^a}= n\, F_n^a.
\ee
In terms of such a basis the Lax potentials 
\rf{laxpot} become (see appendix \ref{sec:su3} for the definition of the new basis)
\be
A_{-} = b_{-1} - i\, \sum_{a=1}^2 \pa_{-}  \vp_a\, F_0^a \; ;\qquad\qquad\quad
A_{+} = - \(V+v_0\)\, b_1 +\frac{i}{3}\left[ \frac{\delta\, V}{\delta \,\vp_1}\, F_1^{2}+
 \frac{\delta\, V}{\delta \,\vp_2}\, F_1^{1}\right], 
\lab{gpabelian}
\ee
where we have redefined the fields as
\be
\sum_{a=1}^2\phi_a\,H_{\alpha_a}^0=\sum_{a=1}^2\vp_a\, F^a_0\qquad\quad \rightarrow \qquad \quad
\(\vp_1\,,\,\vp_2\)= \frac{1}{3}\(\phi_1+\omega^2\,\phi_2\;,\;\phi_1+\omega\,\phi_2\)
\lab{curlyphidef}
\ee

Next we perform a gauge transformation  with a group element which is an exponentiation of the positive grade elements of the image of the adjoint action of $b_{-1}$, {\it i.e.} 
\be
A_{\mu}\rightarrow a_{\mu}=g\,A_{\mu}\,g^{-1}-\partial_{\mu}g\,g^{-1}\, ; \quad\quad {\rm with} \qquad
g=\exp\(\sum_{n=1}^{\infty} {\cal F}_n\)\, ; \quad\quad {\rm and} \quad 
{\cal F}_n=\sum_{a=1}^2\zeta^{(n)}_a\, F^a_n.
\lab{gaugetransfgdef}
\ee
We first consider the $a_{-}$-component of the transformed Lax potential, and split it into the 
eigensubspaces of the grading operator \rf{gradingoperator} as 
$a_{-}=\sum_{n=-1}^{\infty}{\cal A}_{-}^{(n)} $, with $\sbr{D}{{\cal A}_{-}^{(n)}}=n\,{\cal A}_{-}^{(n)}$. We then get that 
\br
{\cal A}_{-}^{(-1)}&=& b_{-1},
\nonumber\\
{\cal A}_{-}^{(0)}&=& -\sbr{b_{-1}}{{\cal F}_1}- i\, \sum_{a=1}^2 \pa_{-}  \vp_a\, F_0^a
\lab{splitgaugetransf},\\
{\cal A}_{-}^{(1)}&=& -\sbr{b_{-1}}{{\cal F}_2}- i\, \sum_{a=1}^2\pa_{-}  \vp_a\, \sbr{{\cal F}_1}{F_0^a}+\frac{1}{2!}\sbr{{\cal F}_1}{\sbr{{\cal F}_1}{b_{-1}}} 
- \partial_{-}{\cal F}_1,
\nonumber\\
&\vdots&
\nonumber\\
{\cal A}_{-}^{(n+1)}&=& -\sbr{b_{-1}}{{\cal F}_n} + \ldots 
\nonumber
\er
One can now choose the parameters $\zeta^{(n)}_a$ in ${\cal F}_n$, order by order in the grade decomposition, to cancel the image component of $a_{-}$. Indeed, if one takes 
\be
\zeta_a^{(1)}=i\,\partial_{-}\varphi_a, \qquad\qquad\qquad\qquad  a=1,2
\lab{zeta1def}
\ee
one can check that the components of ${\cal A}_{-}^{(0)}$ in the direction of $F_0^a$ are cancelled, and so  ${\cal A}_{-}^{(0)}=0$. Note that the element ${\cal F}_n$, of grade $n$, first appears in the grade expansion in the component ${\cal A}_{-}^{(n+1)}$ of grade $n+1$. Since the image subspaces  are always two dimensional for any grade $n$, one can choose the parameters $\zeta^{(n)}_a$ in ${\cal F}_n$ recursively, to cancel the image component of ${\cal A}_{-}^{(n+1)}$. In addition, note that 
$\zeta^{(n)}_a$ is a polynomial in $x_{-}$-derivatives of the  fields $\vp_a$, and each term of such polynomials contains precisely $n$ $x_{-}$-derivatives. Note also that in such a recursive process of canceling the image components of ${\cal A}_{-}^{(n+1)}$ we do not use the equations of motion. Thus we find that the 
$a_{-}$-component of the transformed Lax potential becomes
\be
a_{-}=b_{-1}+ \sum_{M\geq 1}^{\infty} a_{-}^{(M)}\,b_{M}\,; \qquad\qquad\qquad M\equiv 3n\pm1, \qquad\qquad 
n\in \IZ.
\lab{aminusexpand}
\ee

Note that this procedure has used up all freedom of the choice of parameters $\zeta^{(n)}_a$. So what can we say about the transformed  $a_{+}$-component of the Lax potentials? Well, we can restrict our attention to fields which satisfy the equations of motion and use them, or equivalently the quasi-zero curvature condition to determine its form. The curvature $F_{+-}$, given in \rf{quasizc}, gets transformed into 
\br
\pa_{+}a_{-}-\pa_{-}a_{+}+\sbr{a_{+}}{a_{-}}= -i\,\sum_{a=1}^2 X_a\, g\,F_1^a\,g^{-1}, 
\lab{transflax}
\er
where in the last equality we have imposed the equations of motion \rf{generaleq} (see \rf{quasizc}).
Since the group element $g$ is an exponentiation of generators of strictly positive grades, it follows that $g\,F_1^a\,g^{-1}$ has also strictly positive grades only, and so we can split it into its image and kernel components as  
\be
g\,F_1^a\,g^{-1}= F_1^a
+\sum_{M\geq 2}^{\infty} \alpha^{(M,a)}\,b_{M}
+\sum_{n=2}^{\infty} \sum_{b=1}^2\beta_b^{(n,a)}\,F^b_n\, ; \qquad\qquad \qquad M\equiv 3n\pm1;\qquad n\in \IZ.
\lab{anomalyexpand}
\ee
From \rf{gpabelian} we observe that $A_{+}$ has grade one components only, and so $a_{+}$ has strictly positive grades only. Thus the split of $a_{+}$ into its image and kernel components gives us: 
\be
a_{+}= \sum_{M\geq 1}^{\infty} a_{+}^{(M)}\,b_{M}
+\sum_{n=1}^{\infty} \sum_{a=1}^2 a_{+}^{(n,a)}\,F^a_n.
\lab{aplusexpand}
\ee
Next we put \rf{aminusexpand}, \rf{aplusexpand} and \rf{anomalyexpand} into \rf{transflax}, and find that the kernel component leads to 
\br
\pa_{+}a_{-}^{(1)}-\pa_{-}a_{+}^{(1)}&=&0,
\lab{quasiconslaw}\\
\pa_{+}a_{-}^{(M)}-\pa_{-}a_{+}^{(M)}&=&-i\,\sum_{a=1}^2 X_a\,\alpha^{(M,a)}
\,; \qquad\qquad\qquad M\equiv 3n\pm1\geq 2 \qquad\qquad n\in \IZ
\nonumber
\er
and the image component of \rf{transflax} leads to 
\br
\sum_{n=1}^{\infty} \sum_{a=1}^2 a_{+}^{(n,a)}\,\sbr{b_{-1}}{F^a_n}&=&
-\sum_{n=1}^{\infty} \sum_{a=1}^2 \partial_{-}a_{+}^{(n,a)}\,F^a_n
-\sum_{n=1}^{\infty} \sum_{a=1}^2  \sum_{M\geq 1}^{\infty} a_{-}^{(M)}\,a_{+}^{(n,a)}\,\sbr{b_{M}}{F^a_n}
\nonumber\\
&+& i\,\sum_{a=1}^2 X_a\, F_1^a
+i \sum_{n=2}^{\infty} \sum_{a,b=1}^2 X_a\,\beta_b^{(n,a)}\,F^b_n.
\lab{imagecompzc}
\er
Note that the {\it r.h.s.} of \rf{imagecompzc} does not have components of zero grade but the {\it l.h.s.} does. Therefore one concludes that $ a_{+}^{(1,a)}=0$.  For exactly integrable field theories for which the anomalies $X_a$ vanish, one can conclude that the {\it r.h.s.} of \rf{imagecompzc} does not have a component of grade one, if $ a_{+}^{(1,a)}=0$. Thus the {\it l.h.s.} would not have one too, and so one must have that $ a_{+}^{(2,a)}=0$. Continuing such a process one observes that the zero curvature condition implies that the $a_{+}$-component of the Lax potential is also transformed into the abelian kernel generated by $b_M$. In addition, for integrable theories $X_a$ vanish and so one gets that the {\it r.h.s.} of \rf{quasiconslaw} also vanishes for any $M$. For non-integrable field theories where the anomalies $X_a$ do not vanish, none of this happens. However, the fact that $a_{+}$ is not transformed into the kernel does not affect \rf{quasiconslaw}, and so we can get quasi-conservation laws as we explain next. 

From \rf{lightconedef} we see that the $x$ and $t$ components of the Lax potentials are $a_x=a_{+}+a_{-}$ and $a_t=a_{+}-a_{-}$. So we introduce the charges
\be
Q^{(M)}=\int_{-\infty}^{\infty}dx\,a_{x}^{(M)}\, ; \qquad\qquad \qquad M\equiv 3n\pm 1\geq 1\,; \qquad n\in \IZ.
\lab{chargedef}
\ee
By imposing appropriate boundary conditions at spatial infinity on the $a_t$ component of the Lax potential one gets from \rf{quasiconslaw} that
\br
\frac{d\, Q^{(1)}}{d\,t}&=&0,
\lab{chargeconserv}\\
\frac{d\, Q^{(M)}}{d\,t}&=&-2\, i\,\int_{-\infty}^{\infty}dx\,\sum_{a=1}^2\alpha^{(M,a)} \,X_a,
\qquad\qquad \qquad M\equiv 3n\pm 1\geq 2\, ;  \qquad n\in \IZ.
\nonumber
\er
From \rf{gpabelian} and \rf{aplusexpand} one observes that $a_{+}^{(1)} = -\(V+v_0\)$, and so it turns out that $a_x^{(1)}$ is a linear combination of the energy and momentum densities. This explains the origin of the conservation of the charge $Q^{(1)}$, given in \rf{chargeconserv}, even for the non-integrable case.   

In our numerical simulations we have studied the behaviour of the charge $Q^{(2)}$, and so the important quantities for evaluating the anomalies are then
\be
\alpha^{(2,1)}=i\,\(\omega-\omega^2\)\,\partial_{-}\varphi_2, \qquad\qquad\hbox{and}\qquad\qquad
\alpha^{(2,2)}=-i\,\(\omega-\omega^2\)\,\partial_{-}\varphi_1.
\lab{firstanomaly}
\ee
Choosing $v_0=-1$ in \rf{w1w2def}, it follows that quantities $X_a$, $a=1,2$, given in \rf{x1x2def}, evaluated for the potential \rf{deformedpotepsilon}, become
\br
X_1^{(\ve)}&=& \frac{\ve}{3}\,\(1-\omega\)\left[ e^{i\left[2\,\phi_1-\(1+\varepsilon\)\phi_2\right]}\( 
\omega\,\partial_{-}\phi_1+\omega^2\,\(1-\frac{\ve}{3}\,\omega^2\)\partial_{-}\phi_2\)
\right. 
\nonumber\\
&+& \left. e^{i\left[2\,\phi_2-\(1+\varepsilon\)\phi_1\right]}\( 
-\omega^2\,\(1-\frac{\ve}{3}\,\omega\)\partial_{-}\phi_1-\partial_{-}\phi_2\)
\right. 
\\
&+& \left. e^{-i\(1-\varepsilon\)\left[\phi_1+\phi_2\right]}\( 
-\(1-\frac{\ve}{3}\,\(1-\omega\)\)\partial_{-}\phi_1
+\omega\, \(1-\frac{\ve}{3}\,\(1-\omega^2\)\)\partial_{-}\phi_2\)
 \right]
 \nonumber
\er
and
\br
X_2^{(\ve)}&=& \frac{\ve}{3}\,\(1-\omega\)\left[ e^{i\left[2\,\phi_1-\(1+\varepsilon\)\phi_2\right]}\( 
-\omega\,\partial_{-}\phi_1-\(1-\frac{\ve}{3}\,\omega\)\partial_{-}\phi_2\)
\right. 
\nonumber\\
&+& \left. e^{i\left[2\,\phi_2-\(1+\varepsilon\)\phi_1\right]}\( 
\(1-\frac{\ve}{3}\,\omega^2\)\partial_{-}\phi_1+\omega^2\,\partial_{-}\phi_2\)
\right. 
\\
&+& \left. e^{-i\(1-\varepsilon\)\left[\phi_1+\phi_2\right]}\( 
\omega^2\,\(1-\frac{\ve}{3}\,\(1-\omega^2\)\)\partial_{-}\phi_1
-\omega\, \(1-\frac{\ve}{3}\,\(1-\omega\)\)\partial_{-}\phi_2\)
 \right].
 \nonumber
\er
Using \rf{firstanomaly} and \rf{curlyphidef} we then find that 
\br
\sum_{a=1}^2\alpha^{(2,a)}\,X_a^{(\ve)}&=&-\, i\, \frac{\varepsilon}{9}\, \left[
\left[6\,\(\partial_{-}\phi_1\)^2-\(3-\varepsilon\)\,\(\partial_{-}\phi_2\)^2
-2\,\(3+\varepsilon\)\,\partial_{-}\phi_1\partial_{-}\phi_2\right]\, e^{i\left[2\,\phi_1-\(1+\varepsilon\)\phi_2\right]}
\right.\nonumber\\
&+&\left.
\left[\(3-\varepsilon\)\,\(\partial_{-}\phi_1\)^2-6\,\(\partial_{-}\phi_2\)^2
+2\,\(3+\varepsilon\)\,\partial_{-}\phi_1\partial_{-}\phi_2\right]\, e^{i\left[2\,\phi_2-\(1+\varepsilon\)\phi_1\right]}
\right.\nonumber\\
&+&\left.
3\,\(1-\varepsilon\)\left[\(\partial_{-}\phi_1\)^2-\(\partial_{-}\phi_2\)^2\right]\, 
e^{-i\(1-\varepsilon\)\left[\phi_1+\phi_2\right]}
\right].
\lab{alphax2}
\er
We have also investigated the quasi-conservation of the second charge which satisfies 
(see \rf{chargeconserv}) 
\be
\frac{d\, Q^{(2)}}{d\,t}=-i\,2\,\int_{-\infty}^{\infty}dx\,\alpha^{(2,a)} \,X_a^{(\ve)} \equiv \beta^{(2)}.
\ee
Thus using \rf{alphax2} we find that the total anomaly is given by  
\br
\beta^{(2)}&=&-\frac{2}{9} \, \,\varepsilon\, \int_{-\infty}^{\infty}dx\,\left[
\left[6\,\(\partial_{-}\phi_1\)^2-\(3-\varepsilon\)\,\(\partial_{-}\phi_2\)^2
-2\,\(3+\varepsilon\)\,\partial_{-}\phi_1\partial_{-}\phi_2\right]\, e^{i\left[2\,\phi_1-\(1+\varepsilon\)\phi_2\right]}
\right.\nonumber\\
&+&\left.
\left[\(3-\varepsilon\)\,\(\partial_{-}\phi_1\)^2-6\,\(\partial_{-}\phi_2\)^2
+2\,\(3+\varepsilon\)\,\partial_{-}\phi_1\partial_{-}\phi_2\right]\, e^{i\left[2\,\phi_2-\(1+\varepsilon\)\phi_1\right]}
\right.\nonumber\\
&+&\left.
3\,\(1-\varepsilon\)\left[\(\partial_{-}\phi_1\)^2-\(\partial_{-}\phi_2\)^2\right]\, 
e^{-i\(1-\varepsilon\)\left[\phi_1+\phi_2\right]}
\right].
\lab{alphax2b}
\er

%%%%%%%%%%%%%%%%%%%%%%%%%%%%%%%%%%%%%%%%%%%%%%

\section{The Lorentz  transformation and the charge conservation} 
\label{sec:lorentz}
\setcounter{equation}{0}

Consider the Lorentz transformation in $(1+1)$-dimensions (see \rf{lightconedef})
\be
\Lambda\; : \qquad x_{\pm}\rightarrow e^{\mp \lambda}\, x_{\pm}\qquad\qquad {\rm or} \qquad\qquad
x\rightarrow \frac{x-v\,t}{\sqrt{1-v^2}}, \qquad 
t\rightarrow \frac{t-v\,x}{\sqrt{1-v^2}} 
\lab{lorentztransf}
\ee
with $\lambda$ being the rapidity, related to the velocity $v$   by $v=\tanh \lambda$. Note that the Lax potentials \rf{laxpot}, or equivalently \rf{gpabelian}, do not transform as vectors under the Lorentz transformation \rf{lorentztransf}. The Lorentz group in  $(1+1)$-dimensions  is a non-compact one-parameter group, namely $SO(1,1)$. Consider also an internal one-parameter group generated by the grading operator $D$, defined in \rf{gradingoperator}, and acting on the loop algebra $SU(3)$ as an automorphism, {\it i.e.} 
\be
\Sigma\; :\;T\;\rightarrow\; \qquad \Sigma\(T\)=e^{\lambda\, D}\, T\,e^{-\lambda\, D},
\qquad\qquad\qquad 
\Sigma\(\sbr{T}{T^{\prime}}\)=\sbr{\Sigma\(T\)}{\Sigma\(T^{\prime}\)}.
\lab{internallorentz}
\ee

The structure of the Lax potentials \rf{gpabelian} is such that they transform as vectors under the diagonal subgroup, {\it i.e.} (the fields $\phi_a$, or equivalently $\vp_a$, are scalars under the Lorentz group \rf{lorentztransf})
\be
\Omega\(A_{\pm}\)= e^{\pm \lambda}\, A_{\pm},\qquad\qquad \quad 
{\rm where} 
\qquad \qquad \quad \Omega\equiv \Lambda\,\Sigma.
\lab{diagonallorntz}
\ee
In consequence, the curvature is invariant under such a diagonal subgroup, and so is the anomalous term appearing in \rf{quasizc}, {\it i.e.}
\be
\Omega\(F_{+-}\)=F_{+-},\qquad\qquad\qquad \qquad
\Omega\( \sum_{a=1}^2 X_a\, F_1^a\)= \sum_{a=1}^2 X_a\, F_1^a.
\lab{invlorentz}
\ee
Let us now analyse how the transformed Lax potentials $a_{\pm}$, transform under $\Omega$. First we consider $a_-$ and we look at the second line of  \rf{splitgaugetransf} and observe that 
\be
\Omega\(\sum_{a=1}^2 \pa_{-}  \vp_a\, F_0^a\)= e^{-\lambda}\, \sum_{a=1}^2 \pa_{-}  \vp_a\, F_0^a.
\ee

%{\bf  lines above have to be a bit clearer}

However, as ${\cal A}_{-}^{(0)}=0$, this expression has to be cancelled by the transforms of $\sbr{b_{-1}}{{\cal F}_1}$ and of  $b_{-1}$ and so we see that it must be that
\be
\Omega\({\cal F}_1\)= {\cal F}_1
\lab{f1transf}
\ee
since $\Omega\(b_{-1}\)=e^{-\lambda}\, b_{-1}$. Indeed, one observes from \rf{zeta1def} that $\Omega\(\zeta_a^{(1)}\)=e^{-\lambda}\,\zeta_a^{(1)}$, and so we find that $\Omega \(F_1^a\)=e^{\lambda}\, F_1^a$. This demonstrates the validity of \rf{f1transf}. Looking at the terms in the next lines of \rf{splitgaugetransf}   and using \rf{f1transf} we observe that, under the action of $\Omega$, the last three terms of the third line of  \rf{splitgaugetransf} get multiplied by $e^{-\lambda}$. Thus, in order for the term  $\sbr{b_{-1}}{{\cal F}_2}$ to cancel the image part of these three terms one needs that 
$\Omega\(\sbr{b_{-1}}{{\cal F}_2}\)=e^{-\lambda}\,\sbr{b_{-1}}{{\cal F}_2}$. Consequently, it must be that
\be
\Omega\({\cal F}_2\)= {\cal F}_2.
\lab{f2transf}
\ee
Continuing this process recursively, order by order in the grades, one concludes that all ${\cal F}_n$ have to be invariant under $\Omega$, and so the group element $g$ of  the gauge transformation \rf{gaugetransfgdef}, {\it i.e.}
satisfies  
\be
\Omega\(g\)=g.
\lab{ginv}
\ee 
In consequence the transformed Lax potentials $a_{\pm}$ transform as vectors under the diagonal Lorentz subgroup in the same way as $A_{\pm}$, {\it i.e.} they satisfy
\be
\Omega\(a_{\pm}\)=e^{\pm\lambda}\, a_{\pm}.
\ee

Moreover, one of the  consequences of the fact that all ${\cal F}_n$'s are invariant under $\Omega$, is that from \rf{gaugetransfgdef} we see that $\Omega\(\zeta_a^{(n)}\)=e^{-n\,\lambda}\, \zeta_a^{(n)}$. Since the parameters $\zeta_a^{(n)}$ were so chosen  that the $a_{-}$-component of the Lax operator is gauge transformed into the kernel of the adjoint action of $b_{-1}$, it follows that it depends only on $x_{-}$-derivatives of the fields, and not of their $x_{+}$-derivatives. So, from its transformation under $\Omega$, we see that each parameter  $\zeta_a^{(n)}$ of the gauge transformation has to be a polynomial in the derivatives of the fields with all of its terms containing only $n$ $x_{-}$-derivatives. Moreover,  from \rf{ginv} it then follows that $\Omega\(g\,F_1^a\,g^{-1} \)= e^{\lambda}\,g\,F_1^a\,g^{-1}$, and so each term on the {\it r.h.s.} of \rf{anomalyexpand} under the action of $\Omega$ gets multiplied by $e^{\lambda}$. Since $\Omega\(b_M\)=e^{M\,\lambda}\, b_M$, this then implies that 
\be
\Omega\( \alpha^{(M,a)}\)= e^{\(-M+1\)\lambda}\, \alpha^{(M,a)}\; ;
\qquad\qquad \qquad \qquad M\equiv 3n\pm1\geq 2;\qquad n\in \IZ.
\ee

From \rf{anomalyexpand} we then see that $\alpha^{(M,a)}$ is a function of the parameters $\zeta_a^{(n)}$, and so depends only on the $x_{-}$-derivatives of the fields. Therefore, each term in $\alpha^{(M,a)}$ has to contain exactly $(M-1)$ $x_{-}$-derivatives of the fields. Looking at \rf{firstanomaly} we note  that $\alpha^{(2,a)}$ is indeed linear in the $x_{-}$-derivative. 
Then from \rf{invlorentz} and the fact that $\Omega\(F_1^a\)=e^{\lambda}\, F_1^a$, it follows that $\Omega\(X_a\)=e^{-\lambda}\, X_a$. In consequence, we have demonstrated that the anomalies of the charges, appearing in \rf{chargeconserv}, satisfy 
\be
\Omega\(dt\, \int_{-\infty}^{\infty}dx\,\sum_{a=1}^2\alpha^{(M,a)} \,X_a \) = 
e^{-M\,\lambda}\; dt\, \int_{-\infty}^{\infty}dx\,\sum_{a=1}^2\alpha^{(M,a)} \,X_a. 
\lab{anomalieslorentz}
\ee

\begin{figure}
  \centering
  \subfigure[]{\includegraphics[trim = 0.0cm 0.0cm 0.8cm 1.8cm, width=0.40\textwidth]{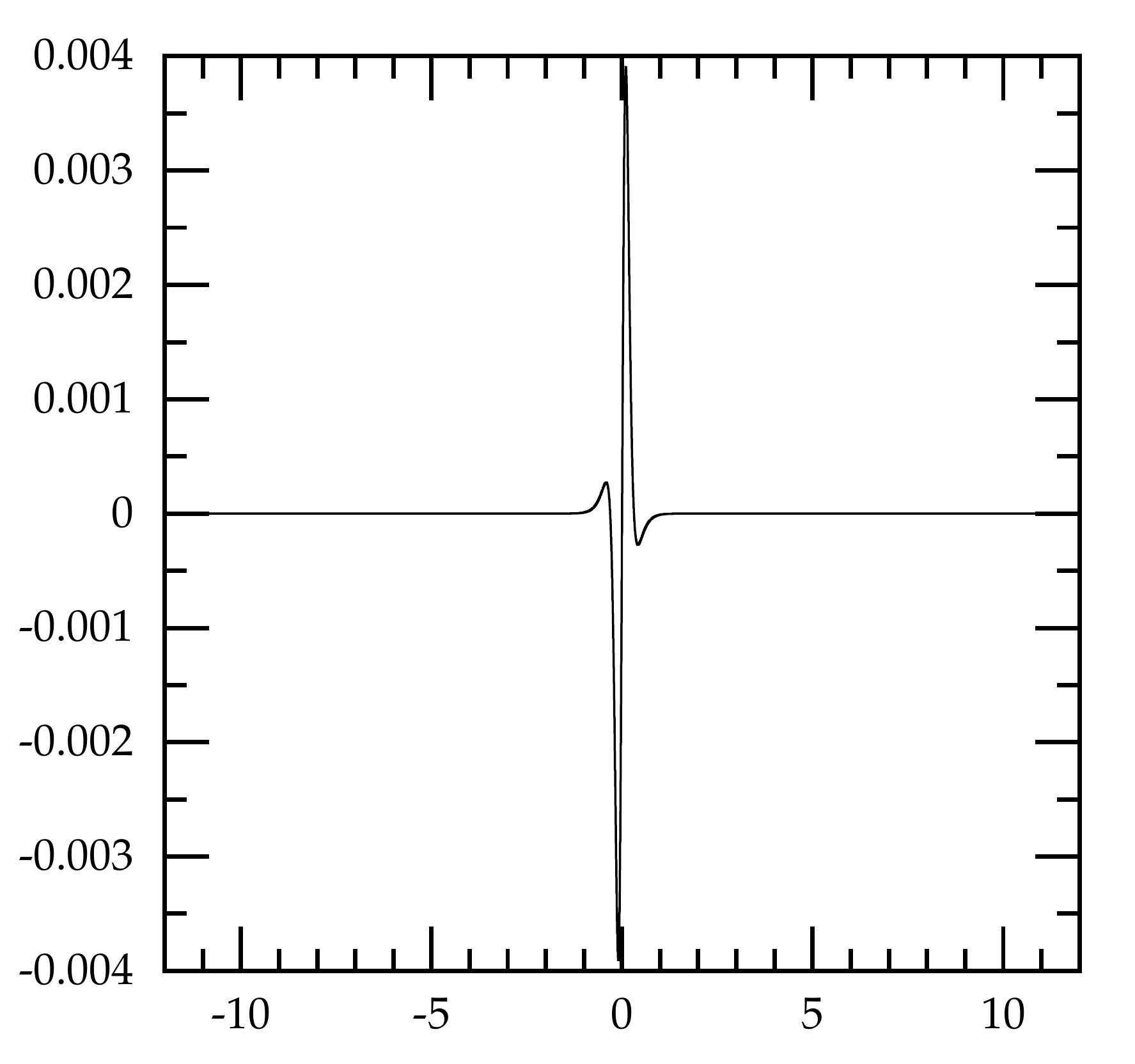}}                
  \subfigure[]{\includegraphics[trim = 0cm 0cm 1.8cm 1.8cm,  width=0.40\textwidth]{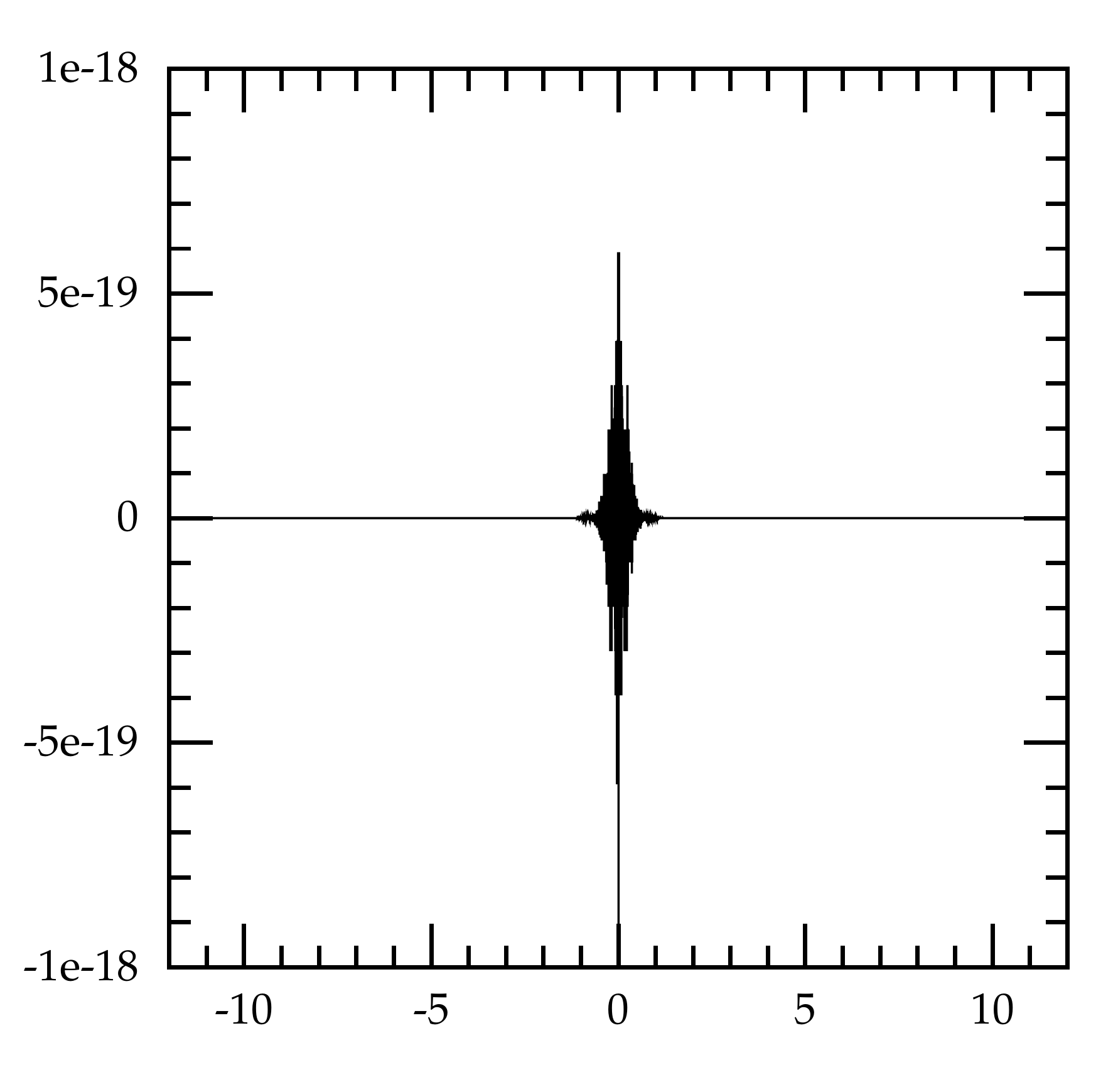}}
  \caption{The real (a) and imaginary parts (b) of the density of the anomaly $\beta^{(2)}$, given in \rf{alphax2b}, as  functions of $x$, in the rest frame of the one-soliton. $\varepsilon=0.0005$ }
  \label{fig:anomalydensity1sol}
\end{figure}

This observation proves a very important property of the charges $Q^{(M)}$. Consider a solution of the equations of motion \rf{generaleq} which is in the form of a traveling wave, {\it i.e.} $\phi_a=\phi_a\(x-v\,t\)$. By a Lorentz transformation one can go to the rest frame of such a solution where it is time-independent. Clearly, the charges $Q^{(M)}$ evaluated on such a static solution, should be time independent and so the anomalies appearing on the {\it r.h.s.} of the second equation in    \rf{chargeconserv} should vanish. But, from \rf{anomalieslorentz} it follows that, if the anomalies vanish in one reference frame, they vanish also in in any other reference frame connected by a Lorentz transformation. Thus, we conclude that all the charges $Q^{(M)}$, for any $M$ in the infinite set of them defined in \rf{chargedef}, are exactly conserved for any traveling wave solution and, in particular, they are conserved for the one-soliton type solutions. That is a highly non-trivial result since the densities of the anomalies, namely $\sum_{a=1}^2\alpha^{(M,a)} \,X_a$, do not vanish in general when evaluated on a traveling wave solution. It is their integral over the whole one-dimensional space that has to vanish.  Note also that for finite energy solutions of the equations of motion the space and time derivatives of the fields have to vanish at spatial infinity. In consequence, the $\alpha^{(M,a)}$ and $X_a$ expressions have to vanish at spatial infinity, since as we have seen above, they are polynomials in the $x_{-}$-derivatives of the fields (see \rf{x1x2def}). So, for any one-soliton solution the densities of the anomalies $\sum_{a=1}^2\alpha^{(M,a)} \,X_a$, are localized in space, and their space integral vanishes. One possible reason for the vanishing of such an integral is that the  densities of the anomalies are odd functions of $x$, in the rest frame of the traveling wave solution.  We have verified that this is exactly what happens for the one-soliton solutions of the theories \rf{lagrangian} with potentials given by \rf{deformedpotepsilon}.  In section \ref{sec:numerical} we explain how the one-solitons of such theories can be constructed numerically. One can then evaluate the anomalies on such solutions numerically. In Fig.  \ref{fig:anomalydensity1sol} we plot the real and imaginary parts of the density of the anomaly  $\beta^{(2)}$, given in \rf{alphax2b}, as  functions of $x$, in the rest frame of the one-soliton.
The value shown there is for $\varepsilon=0.0005$. Note that the complex density of the anomaly is indeed an odd function of $x$ (the imaginary part is essentially zero; its infinitesimal values are numerical artifacts).  

We have not  understood yet the phenomenon of the cancellation of the anomalies. However,  the conservation of the infinite set of charges for traveling wave solutions is clear from the  argument based on the Lorentz transformation given above. In the case of traveling wave solutions like one-solitons this argument  implies that the anomalies have to vanish irrespective of their densities being odd functions of $x$ or not. For the case of two-soliton solutions (moving with different velocities) we have found that in all examples where the anomalies cancel, there is a space-time parity transformation playing a role. It would be interesting to investigate if there is a relation between the roles of the space parity in the case of one-solitons and the space-time parity in the case of two-solitons. In the next section we discuss the role of the space-time parity in the cancellation of the anomalies.  

\subsection{The parity transformation and charge conservation}

The  properties of field configurations, specially those describing one and two soliton solutions, under space-time parity transformations do seem to play a role in the vanishing of 'total' anomalies, {\it i.e.} when the anomalies are integrated not only over space but also over time. Consider a space-time parity transformation given by
\be
P\; : \qquad\qquad \({\tilde x}\, , \, {\tilde t}\) \rightarrow \(-{\tilde x}\, , \, -{\tilde t}\) \; ; \qquad\qquad\qquad 
{\tilde x}=x-x_{\Delta}\;; \qquad \qquad {\tilde t}=t-t_{\Delta},
\lab{paritydef}
\ee
where $x_{\Delta}$ and $t_{\Delta}$ are constants depending on the parameters of the solution under consideration. Let us look at the solutions of the equations of motion such that the fields, evaluated on them, behave as follows under this parity transformation:
\be
P\(\phi_1\)=\phi_2+c_2,\qquad\qquad\qquad \qquad P\(\phi_2\)=\phi_1+c_1,
\lab{parityonfields}
\ee
where $c_1$ and $c_2$ are constants. In addition, we are interested in potentials that are invariant under the parity, {\it i.e.}
\be
P\(V\(\phi_1\, ,\,\phi_2\)\)= V\(\phi_1\, ,\,\phi_2\).
\lab{paritypotential}
\ee
Note that \rf{paritydef} and \rf{parityonfields} imply that
\be
P\(\partial_{\mu}\phi_1\)=-\partial_{\mu}\phi_2, \qquad\qquad\qquad 
P\(\partial_{\mu}\phi_2\)=-\partial_{\mu}\phi_1,
\lab{parityderivative}
\ee
where $\partial_{\mu}$ stands for the space-time derivatives, and 
 \be
P\(\delta \phi_1\)=\delta \phi_2, \qquad\qquad\qquad 
P\(\delta \phi_2\)=\delta \phi_1,
\lab{parityvariation}
\ee
where $\delta$ stands for the functional variations of the fields.

Using  \rf{paritypotential} and \rf{parityvariation} we find from \rf{w1w2def} that 
\be
P\;:\qquad \qquad W_1\(\omega\) \leftrightarrow W_2\(\omega^2\), \qquad\qquad \qquad 
 W_2\(\omega\) \leftrightarrow W_1\(\omega^2\).
 \lab{w1w2parity}
 \ee
Then,  \rf{parityderivative} and \rf{x1x2def}, give us that
\be
P\(X_1\)=-\omega^2\, X_2,\qquad\qquad\qquad\qquad P\(X_2\)=-\omega\, X_1.
\lab{x1x2parity}
\ee

Next we check how  the quantities $\alpha^{(M,a)}$ and the anomaly densities  transform under this parity transformation.  To determine this  we need to use another automorphism  of the $SU(3)$ loop algebra which involves the following order two outer automorphism of the finite simple $SU(3)$ Lie algebra ($\sigma^2=1$)
\be
\sigma\(H_{\alpha_1}\)=H_{\alpha_2}\; ;\qquad\qquad
\sigma\(E_{\pm\alpha_1}\)=-E_{\pm\alpha_2}\; ; \qquad \quad 
\sigma\(E_{\pm\alpha_3}\)=-E_{\pm\alpha_3}.
\lab{sigmafinitesu3}
\ee
One can check that \rf{sigmafinitesu3} is indeed an automorphism of the algebra $SU(3)$ given in 
\rf{su3comrel}. 
This automorphism is insensitive to the value of the $\lambda$ parameter of the loop algebra, and so we find that (see appendix \ref{sec:su3})
\br
\sigma\(b_{3n\pm 1}\)&=&-b_{3n\pm 1},
\nonumber\\
\sigma\(F^1_{3n}\)&=&\omega\, F^2_{3n},\qquad\qquad\qquad\;\;
\sigma\(F^2_{3n}\)=\omega^2\, F^1_{3n},
\nonumber\\
\sigma\(F^1_{3n+1}\)&=&-\omega\, F^2_{3n+1},\qquad\qquad
\sigma\(F^2_{3n+1}\)=-\omega^2\, F^1_{3n+1},
\nonumber\\
\sigma\(F^1_{3n-1}\)&=&-\omega\, F^2_{3n-1},\qquad\qquad
\sigma\(F^2_{3n-1}\)=-\omega^2\, F^1_{3n-1}.
\lab{loopalgebraauto}
\er
Next we consider the combined action of the space-time parity $P$ and this automorphism  $\sigma$  
\be
S\equiv P\,\sigma.
\lab{sparitydef}
\ee
From \rf{curlyphidef} and \rf{parityderivative} we see that
\be
P\(\partial_{\mu}\vp_1\)=-\omega^2\,\partial_{\mu} \vp_2\; ; \qquad\qquad\hbox{and}\qquad \qquad
P\(\partial_{\mu}\vp_2\)=-\omega\,\partial_{\mu} \vp_1.
\ee
Thus \rf{loopalgebraauto} gives us:
\be
S\(\sum_{a=1}^2\partial_{-}\vp_a\, F_0^a\)=- \sum_{a=1}^2\partial_{-}\vp_a\, F_0^a.
\ee
Then applying $(1+S)$ to both sides of  the second equation in \rf{splitgaugetransf} we  get 
\be
\(1+S\){\cal A}_{-}^{(0)}= -\sbr{b_{-1}}{\(1-S\){\cal F}_1}.
\lab{1plusa0}
\ee

Let us recall that the procedure in \rf{splitgaugetransf} involved choosing the group element $g$ and so also the ${\cal F}_n$'s in such a way that the new Lax potential $a_{-}$ was transformed into the kernel of the adjoint action of $b_{-1}$. Hence, as a result of this procedure ${\cal A}_{-}^{(0)}$ belongs to the kernel. But since $\sigma$, and  so $S$, maps kernel into kernel (see \rf{loopalgebraauto}), we note that the {\it l.h.s.} of \rf{1plusa0} belongs to the kernel. However, since the {\it r.h.s.} of \rf{1plusa0} is the commutator of $b_{-1}$ with something, it belongs to the image of the adjoint action of $b_{-1}$. Since image and kernel do not possess common elements (see \rf{kernelimagedef}), then both sides of  \rf{1plusa0} have to vanish. Also, since $\sigma$, and so $S$, maps image into image (see \rf{loopalgebraauto}), it follows that $\(1-S\){\cal F}_1$ belongs to the image, and so it cannot commute with $b_{-1}$. Thus it must be that
\be
\(1+S\){\cal A}_{-}^{(0)}=0\; ; \qquad\qquad\qquad \(1-S\){\cal F}_1=0.
\ee
Then applying $(1+S)$ to both sides of  the third equation in \rf{splitgaugetransf} we find 
\be
\(1+S\){\cal A}_{-}^{(1)}= -\sbr{b_{-1}}{\(1-S\){\cal F}_2}.
\lab{1plusa1}
\ee
Using very similar arguments to those presented above one can also conclude that
\be
\(1+S\){\cal A}_{-}^{(1)}=0\; ; \qquad\qquad\qquad \(1-S\){\cal F}_2=0.
\ee
Continuing this process recursively, order by order in the grade expansion of $a_{-}$, one concludes that all ${\cal F}_n$'s are invariant under $S$ and so that
\be
S\(g\)=g.
\lab{ginvs}
\ee 

Next, using \rf{loopalgebraauto} and \rf{ginvs} one finds that
\be
S\(g\, F_1^1\,g^{-1}\)= -\omega \, g\,F_1^2 \,g^{-1}\; ; \qquad\qquad \qquad \qquad 
S\(g\, F_1^2\,g^{-1}\)= -\omega^2 \, g\,F_1^1 \,g^{-1}.
\lab{songfgminus1}
\ee
Then from \rf{anomalyexpand}, \rf{loopalgebraauto}, \rf{sparitydef} and \rf{songfgminus1} one also finds that
\be
P\(\alpha^{(M,1)}\)= \omega \,\alpha^{(M,2)}\; ; \qquad\quad \quad  
P\(\alpha^{(M,2)}\)= \omega^2 \,\alpha^{(M,1)}\;; \qquad\quad\quad M=3n\pm 1;\qquad n\in\IZ. 
\ee
In consequence, \rf{x1x2parity} allows us to conclude that
\be
P\(\sum_{a=1}^2 \alpha^{(M,a)}\, X_a\)=-\,\sum_{a=1}^2 \alpha^{(M,a)}\, X_a.
\lab{anomalyunderpaprity}
\ee

Thus we have demonstrated that the anomaly densities are odd under our parity transformation. This implies that if we integrate them on a rectangle with centre at $\(x\, , \, t\)= \(x_{\Delta}\, , \, t_{\Delta}\)$, (see \rf{paritydef}), they  vanish, {\it i.e.}
\be
\int_{-{\tilde t}_0}^{{\tilde t}_0} dt\,\int_{-{\tilde x}_0}^{{\tilde x}_0} dx\; \sum_{a=1}^2 \alpha^{(M,a)}\, X_a =0.
\ee
Finally, taking ${\tilde x}_0\rightarrow \infty$, we find from \rf{chargeconserv} that the charges satisfy the mirror type symmetry
\be
Q^{(M)}\({\tilde t}_0\)=Q^{(M)}\(-{\tilde t}_0\)\,;
\qquad\qquad \qquad M\equiv 3n\pm 1\geq 2\, ;  \qquad n\in \IZ.
\lab{mirrorsymcharges}
\ee
So, if one considers the scattering of two one-soliton fields (which make a two-soliton solution satisfying \rf{parityonfields} and \rf{paritypotential}), the values of the infinite number of charges $Q^{(M)}$ do vary in time, but after the scattering they all return to the values they had before the scattering. Since in a scattering process what matters are the asymptotic states, we see that the properties of such scatterings resemble those of an integrable theory, and that is why we call such theories quasi-integrable.

%%%%%%%%%%%%%%%%%%%%%%%%%%%%%%%%%%%%%%%%%%%%

\section{The exact soliton solutions of the  integrable Affine Toda Models}
\label{sec:solitons}
\setcounter{equation}{0}

The exact soliton solutions for the Affine Toda theories (AT) can be constructed by a variety of methods, all of which are based in one way or another on the zero curvature condition or the Lax-Zakharov-Shabat equation \cite{lax}.   Among the several  methods that have been used to study such theories, we have the inverse scattering method \cite{ism}, B\"acklund transformations \cite{liao}, the dressing transformation method  
\cite{dress}, the solitonic specialization \cite{solitonic} of the Leznov-Saveliev solution \cite{leznovsaveliev}, the direct Hirota method \cite{hirota}, and others (see \cite{babelonbook} for a more complete account). The soliton solutions for the $SU(N)$ Affine Toda field theories were first constructed by Hollowood \cite{hollo} using the Hirota method. The generalization of the construction to AT models associated to other algebras were presented  in  \cite{clis1,clis2,zhu,mackay} using the Hirota method, and in 
\cite{olive1st,vertex,solitonic} using the Leznov-Saveliev method and the representation theory of Kac-Moody algebras based on vertex operators. 

The Hirota method is perhaps the most efficient procedure for constructing explicit analytical soliton solutions. However, it does not provide a way of finding the  so-called tau-functions which are crucial for the Hirota method. Such functions can however be easily found using  the dressing transformation method and the representation theory of Kac-Moody algebras  based on vertex operators \cite{vertex}. Therefore, the most efficient method for constructing soliton solutions is perhaps a hybrid procedure based on the dressing transformation and the Hirota methods  as explained in 
\cite{miramontes,bueno}. An additional advantage is that this procedure can be easily adapted to be carried out with the help a computer package for algebraic manipulations.  In fact, the magic of the Hirota method, which produces exact solutions by truncations of a formal series expansion, can be understood through the nilpotency of  vertex operators in highest weight representations of the Kac-Moody algebras. In such representations  the central element of these algebras cannot vanish, and so the Lax potentials, like the ones given in \rf{laxpot}, have to live in the full Kac-Moody algebra and not only in the loop algebra. This requires the extension of the AT models to the so-called Conformal Affine Toda models (CAT) by the introduction of one extra field (or two if one wants conformal symmetry). Such an extension explains  the need for one extra tau-function for the Hirota method to work, as compared to the number of fields of the AT models (see \cite{clis2} for details).   Therefore,  for an AT model associated to a Kac-Moody algebra 
${\hat{\cal G}}$, affine to a finite simple Lie algebra ${\cal G}$, of rank $r$, there are $r+1$ tau-functions $\tau_j$, $j=0,1,\ldots r$, satisfying coupled partial differential equations, the so-called Hirota's equations.  These equations are quadratic, cubic or quartic, in the tau-functions, depending on the connectivity of the Cartan matrix of ${\hat{\cal G}}$ (see \cite{clis2} for details).  Then an $N$-soliton solution is obtained through the Hirota ansatz for the tau-functions 
\be
\tau_j= \delta_j^{(0)} +\kappa\,\sum_{k=1}^N \delta_{j,(k)}^{(1)}\,e^{\Gamma\(z_k\)}+
\kappa^2\,\sum_{k,l=1}^N \delta_{j,(k,l)}^{(2)}\,e^{\Gamma\(z_k\)+\Gamma\(z_l\)}+\ldots
\qquad\qquad j=0,1,\ldots r,
\lab{multisolitontau}
\ee
where $ \delta_j^{(0)}$ are constants corresponding to the values of the tau-functions on a vacuum solution of the theory. The other constants $\delta_{j,(k)}^{(1)}$, $\delta_{j,(k,l)}^{(2)}$, {\it etc} are obtained, recursively, from the  expansion the Hirota equation in powers of $\kappa$. In the expression above the $\Gamma$ function stands for
\be
\Gamma\(z_k\)= m_k\(z_k\, x_{+}+\frac{x_{-}}{z_k}\)+\xi_k= 
2\, m_k\, \eta_k\,\frac{\(x- v_k\,t\)}{\sqrt {1-v_k^2}}+\xi_k,
\lab{gammadef}
\ee
where  $z_k=\eta_k\,e^{-\alpha_k}$ and $v_k=\tanh\alpha_k$, with $\alpha_k$ real and $\eta_k=\pm 1$.
So, $v_k$ is the velocity and $\alpha_k$ is the rapidity of the soliton $k$. The parameters $\xi_k$ fix the positions of the solitons at $t=0$, but in some cases they can even be taken to be  complex. The square of the parameter $m_k$, and the first order vectors $\delta^{(1)}$'s,  are determined from the first order (in $\kappa$) Hirota's equations, which lead to the eigenvalue problem \cite{clis2}
\be
L_{ij}\delta_{j,(k)}^{(1)}= m_k^2\, \delta_{i,(k)}^{(1)}.
\lab{eigenvaluehirota}
\ee
Here $L_{ij}=l_i^{\psi}\,K_{ij}$,  $K_{ij}$ are elements of the extended Cartan matrix of the affine Kac-Moody algebra ${\hat{\cal G}}$, and $l_i^{\psi}$ are positive integers appearing in the expansion of the highest co-root $\psi/\psi^2$, in terms of the simple co-roots $\alpha_a/\alpha_a^2$, of ${\cal G}$, {\it i.e.} $\psi/\psi^2=\sum_{a=1}^r l_a^{\psi}\, \alpha_a/\alpha_a^2$, and $l_0^{\psi}=1$. Moreover, the parameters $m_k$ 
 label, together with the topological charges,  the species of the soliton solutions, and they also  fix the masses of the one-soliton solutions. Note that the Hirota method fixes the moduli of $m_k$, through \rf{eigenvaluehirota}, but not their sign. In fact, the sign of $\Gamma\(z_k\)$ can be changed by flipping either the sign of $z_k$ or of $m_k$, and this changes the sign of the topological charge of the solitons. So, such a flip of the signs turns a soliton of a given species into an anti-soliton of the same species and vice-versa.  The higher order  vectors $\delta^{(n)}$'s are determined, recursively, through the expansion of the Hirota  equations in powers of $\kappa$ \cite{clis2,zhu}.

The solitons have in general short range non-trivial interactions, but there is an interesting  situation, first observed in \cite{clis2}, where the existence multi-soliton solutions, which are at rest with respect to each other was first pointed out and which, consequently, do not have static interactions. Such solutions are more easily constructed by considering the Hirota ansatz for one-soliton solution given by 
\be
\tau_j= \delta_j^{(0)} +\kappa\, \delta_{j}^{(1)}\,e^{\Gamma\(z\)}+
\kappa^2\,\delta_{j}^{(2)}\,e^{2\, \Gamma\(z\)}+\ldots 
\qquad\qquad\quad j=0,1,\ldots r
\lab{onesoltauansatz}
\ee
with $\delta_j^{(0)}$ as before,  $\delta_{j}^{(1)}$ being determined by \rf{eigenvaluehirota}, and $\Gamma\(z\)$ being given by \rf{gammadef}.

 The phenomenon of the existence of static multi-soliton configurations occurs whenever a given eigenvalue of the matrix $L_{ij}$ is degenerate. In general such degeneracy is related to a symmetry of the Dynkin diagram of ${\cal G}$, but it can also be an accidental degeneracy. If a given  eigenvalue of  $L_{ij}$ is degenerate, the vector $\delta_{j}^{(1)}$, associated to that solution,  can be taken as a generic linear combination of the degenerate eigenvectors. This situation introduces new parameters into the solutions which can make the Hirota expansion truncate at higher orders. If one takes all but one such parameters to be zero one gets a one-soliton solution. However, by taking them different from zero one gets solutions which can be interpreted as multi-soliton solutions in which solitons are at rest with respect to each other. So, there are no static interactions among them which would have set them to move. There can be, however, interactions depending on their relative velocities.  The number of solitons in a given static multi-soliton solution is equal to the degree of the degeneracy of the corresponding eigenvalue $m_k^2$ (see \rf{eigenvaluehirota}). The details of such construction can be found in \cite{clis2}, and  the results can be summarized as follows: associated to the symmetries of the Dynkin diagrams one has   static two-soliton solutions for the AT models associated to the algebras $SU(N)$, $SO(2N)$ ($N$ a positive integer) and $E_6$, and static three-soliton solution  for the $SO(8)$ AT model. Associated to accidental degeneracies one has static two-soliton solutions in the AT models associated to the algebras $SO(6N+2)$ and $SO(6N+1)$ ($N$ a positive integer).

The list however does not end there. The higher order vectors $\delta^{(n)}$'s are determined by  algebraic equations of the form \cite{clis2}
\be
 \( L_{ij}-n^2\,\lambda\, \delta_{ij}\)\delta_j^{(n)}= V_i^{(n-1)},
 \ee
 where $\lambda$ is an eigenvalue of $L_{ij}$, and $ V_j^{(n-1)}$ is a vector made out of the vectors $\delta^{(m)}$'s with $m<n$. Therefore, if the matrix $L_{ij}$ has two eigenvectors $\lambda$ and $\lambda^{\prime}$, such that $\lambda^{\prime}-n^2\, \lambda=0$, then one can add to $\delta_j^{(n)}$ a term proportional to the eigenvector associated to  $\lambda^{\prime}$. This brings an extra parameter into the solution which makes the Hirota expansion truncate at higher orders, and so gives the solution the character of a static multi-soliton configuration. The cases where such a behaviour had occured, were first discussed in \cite{clis2} through  a theorem which involves  Galois theory in its proof, and they corresponded to the algebras $SU(6\,N)$ and $Sp(3\, N)$ ($N$ a positive integer). Therefore the  AT models associated to the algebras  $Sp(3\, N)$ present static two-soliton solutions, and those associated to $SU(6\,N)$ can be described as representing static three-soliton solutions, since two of the solitons come from the degeneracy of any $SU(N)$ associated to the symmetry of its Dynkin diagram. 
 
Finally we would like to point out that static two-soliton solutions can be constructed out of solitons and anti-solitons of the same species.  As we have mentioned above solitons and  anti-solitons of the same species are associated to the same eigenvalue $m_k^2$ of $L_{ij}$, since they correspond to opposite choices of the signs of $m_k$ (not determined by \rf{eigenvaluehirota}). Therefore one can have in \rf{multisolitontau} the same eigenvector $\delta^{(1)}$ associated to two exponentials of $\Gamma$'s with opposite signs, {\it i.e.} the Hirota tau-functions are given by:
\be
\tau_j= \delta_j^{(0)} +\kappa\, \delta_{j}^{(1)}\(e^{\Gamma\(z\)}+ e^{-\Gamma\(z\)}\) +
\kappa^2\,\delta_{j}^{(2)}+\ldots
\qquad\qquad\quad j=0,1,\ldots r.
\lab{multisolitontausamespecies}
\ee
Since the velocity is solely determined by $z$, there is a rest frame where such a solution can be made static. 
 
 The phenomenon of static multi-soliton solutions which was first observed in \cite{clis2}, has been also explored further in  some papers, in particular in those dealing with the construction of multi-soliton solutions of the AT models \cite{zhu,mcghee}.  More recently, the behaviour of the energy density of such static multi-soliton solutions has been studied in the case of $SU(N)$ AT models  by one of us \cite {klimas}.

 \subsection{The solitons of the $SU(3)$ Affine Toda Model}
 \label{subsec:solutions}

Here we discuss the exact soliton solutions of the integrable $SU(3)$ affine Toda model, which corresponds to the theory \rf{lagrangian} with potential being given by \rf{todapotential}. According to \rf{generaleq} the Euler-Lagrange equations for such a theory are given by
\br
\partial_{+}\partial_{-}\phi_1 &=& -i\left[ e^{i\(2\phi_1-\phi_2\)}-e^{-i\(\phi_1+\phi_2\)}\right],
\nonumber\\
\partial_{+}\partial_{-}\phi_2 &=& -i\left[ e^{i\(2\phi_2-\phi_1\)}-e^{-i\(\phi_1+\phi_2\)}\right].
\lab{todaeq}
\er

 For the case of the $SU(3)$ affine Toda model  the Hirota tau-functions $\tau_j$, $j=0,1,2$, are defined by the following field transformation
\be
\phi_a=i\ln\frac{\tau_a}{\tau_0}\,,\qquad\qquad\qquad a=1,2.
\lab{taudef}
\ee
When one substitutes \rf{taudef} into \rf{todaeq} one gets two equations for three tau-functions. However, as mentioned above one needs the conformal affine extension of the model to get the Hirota's equation for the tau-functions and so these tau-functions must satisfy: 
\be
\tau_j\partial_{+}\partial_{-}\tau_j-\partial_{+}\tau_j\,\partial_{-}\tau_j=\tau_j^2-\tau_{j-1}\,\tau_{j+1},\qquad\quad\qquad j=0,1,2; \qquad\qquad \tau_{j+3}=\tau_j.
\lab{hirotaeqs}
\ee
One can easily check that any solution of  \rf{hirotaeqs},  by substitution into \rf{taudef}, leads to a solution of  \rf{todaeq}.

For the case of $SU(3)$ we have that the positive integers $l_i^{\psi}$ introduced below \rf{eigenvaluehirota} are all equal to unity. Therefore, the matrix $L_{ij}$ is the same as the extended Cartan matrix of $SU(3)$ and is given by
\br
L=\(\begin{array}{ccc}
2&-1&-1\\
-1&2&-1\\
-1&-1&2
\end{array}\).
\lab{su3matrixl}
\er
Its eigenvalues are $0$ and $3$, with $3$ being doubly degenerate. The zero eigenvalue  leads to solutions traveling with the speed of light and do not correspond to solitons. We then have two species of one-solitons associated to the degenerate eigenvalue $m_k^2=3$, and they can lead to static two-soliton solutions as explained above (see \cite{clis2}).  Therefore, from \rf{gammadef} we have
\be
\Gamma\(z_k\)= \sqrt{3}\(z_k\, x_{+}+\frac{x_{-}}{z_k}\)+\xi_k= 
2\, \sqrt{3}\, \eta_k\,\frac{\(x- v_k\,t- x_0^{(k)}\)}{\sqrt {1-v_k^2}},
\lab{gammadefsu3}
\ee
where we have introduced $x_0^{(k)}$ as 
$\xi_k=-\,2\,\sqrt{3}\,\eta_k\,\frac{\,x_0^{(k)}}{\sqrt {1-v_k^2}}$. 
Note that $\tau_j=1$, $j=0,1,2$, solves the Hirota equation \rf{hirotaeqs} and corresponds, in fact, to a vacuum solution. Therefore, using the Hirota ansatz \rf{onesoltauansatz} with $\delta_j^{(0)}=1$ one obtains two one-soliton solutions (of two  different species). The one-soliton solution of the  species-1 is given by:
\br
\( \begin{array}{c}
\tau_0\\
\tau_1\\
\tau_2
\end{array}\) = 
\( \begin{array}{c}
1\\
1\\
1
\end{array}\) +
\( \begin{array}{c}
1\\
\omega\\
\omega^2
\end{array}\) \, e^{\Gamma\(z\)},
\lab{onesol1}
\er 
and the  one-soliton solution of the species-2 is
\br
\( \begin{array}{c}
\tau_0\\
\tau_1\\
\tau_2
\end{array}\) = 
\( \begin{array}{c}
1\\
1\\
1
\end{array}\) +
\( \begin{array}{c}
1\\
\omega^2\\
\omega
\end{array}\) \, e^{\Gamma\(z\)}
\lab{onesol2}
\er
with $\Gamma\(z\)$ given by \rf{gammadefsu3}, and where $\omega$ is a cubic root of unity, different from unity itself. So we take
\be
\omega = e^{i\,2\,\pi/3},\qquad\qquad \qquad \qquad 1+\omega+\omega^2=0.
\ee
From \rf {todapotential} and \rf{hamiltonian} we find that the Hamiltonian for the $SU(3)$ AT model is given by
\be
{\cal H}_{{\rm Toda}} =  \frac{1}{24}\, \left[\(\partial_{t} {\vec \phi}\)^2+\(\partial_{x} {\vec \phi}\)^2\right] -\frac{1}{3}
\left[ e^{i\,{\vec \alpha}_1\cdot {\vec \phi}}+e^{i\,{\vec \alpha}_2\cdot {\vec \phi}}+e^{i\,{\vec\alpha}_0\cdot {\vec \phi}}-3\right],
\lab{su3hamiltonian}
\ee
where ${\vec \phi}$ is defined in \rf{vecphidef}. 
Therefore, the discrete transformations:
\be
{\vec \phi} \rightarrow  {\vec \phi} +2\pi\, {\vec \mu} 
\ee
are symmetries of the Hamiltonian, if ${\vec \mu}\cdot {\vec\alpha} \in \IZ$ for any root ${\vec\alpha} $ of $SU(3)$.  The vectors $\vec\mu$ are called co-weights of the algebra, and they form the co-weight lattice. Such a lattice describes the degenerate vacua of the theory and gives rise to topological solitons. Indeed, the topological current is defined as
\be
{\vec j}_{\mu}=-\frac{1}{2\,\pi}\, \varepsilon_{\mu\nu}\,\partial^{\nu} {\vec \phi}
\ee
and
\be
{\vec Q}_{\rm top.}=\int_{-\infty}^{\infty} dx\, {\vec j}_0=\frac{1}{2\,\pi}\left[{\vec \phi}\(\infty\)-{\vec \phi}\(-\infty\)\right].
\ee
One can check that the topological charges of the species-1 and species-2 one-solitons, given by \rf{taudef} and
 \rf{onesol1} or \rf{onesol2}  are given, respectively,  by
\be
{\vec Q}^{(1)}_{\rm top.}=-\eta\, \frac{1}{3}\({\vec \alpha_1}+2\,{\vec \alpha_2}\)
=-\eta\, {\vec \lambda_2}
\ee
and
\be
{\vec Q}^{(2)}_{\rm top.}=-\eta\, \frac{1}{3}\(2\,{\vec \alpha_1}+{\vec \alpha_2}\)
=-\eta\, {\vec \lambda_1},
\ee
where $\eta=\pm 1$ is the sign introduced in \rf{gammadefsu3}.  Moreover,  $\lambda_a$, $a=1,2$ are the fundamental weights of $SU(3)$, and we have normalized the roots as $\alpha_a^2=2$. Note that the one-soliton solutions  \rf{onesol1} and \rf{onesol2} are such that
\be
\tau_1^*=\tau_2,\qquad\qquad \tau_0^*=\tau_0 
\lab{realitycondonesol}.
\ee
Therefore from \rf{taudef} and \rf{vecphidef} we see that 
\be
\phi_1^*=-\phi_2\qquad\qquad \mbox{\rm and so} \qquad\qquad 
{\vec \phi}^* = -\({\vec \alpha}_1\, \phi_2+ {\vec \alpha}_2\, \phi_1\).
\lab{realitycondonesolphi}
\ee
Thus the complex conjugation of ${\vec \phi}$ amounts to a sign flip and the interchange ${\vec \alpha}_1\leftrightarrow {\vec \alpha}_2$. In consequence, the Hamiltonian \rf{su3hamiltonian} is real when evaluated on the one-soliton solutions \rf{onesol1} or \rf{onesol2}.

Using the Hirota ansatz \rf{multisolitontau} one can construct also two-soliton solutions for the $SU(3)$ AT model by solving the Hirota equations \rf{hirotaeqs} recursively as explained above. By combining the two species of one-solitons one gets three types of two-soliton solutions. The species-11  two-soliton solution is given by
\br
\( \begin{array}{c}
\tau_0\\
\tau_1\\
\tau_2
\end{array}\) = 
\( \begin{array}{c}
1\\
1\\
1
\end{array}\) +
\( \begin{array}{c}
1\\
\omega\\
\omega^2
\end{array}\)\,e^{\Gamma\(z_1\)}+
\( \begin{array}{c}
1\\
\omega\\
\omega^2
\end{array}\) \, e^{\Gamma\(z_2\)}+
\( \begin{array}{c}
1\\
\omega^2\\
\omega
\end{array}\) \,e^{\Gamma\(z_1\)+\Gamma\(z_2\)+\Delta_{11}}.
\lab{twosolsu311}
\er
The species-22  two-soliton solution is
\br
\( \begin{array}{c}
\tau_0\\
\tau_1\\
\tau_2
\end{array}\) = 
\( \begin{array}{c}
1\\
1\\
1
\end{array}\) +
 \( \begin{array}{c}
1\\
\omega^2\\
\omega
\end{array}\)\,e^{\Gamma\(z_1\)}+
\( \begin{array}{c}
1\\
\omega^2\\
\omega
\end{array}\) \, e^{\Gamma\(z_2\)}+
\( \begin{array}{c}
1\\
\omega\\
\omega^2
\end{array}\) \,e^{\Gamma\(z_1\)+\Gamma\(z_2\)+\Delta_{11}}.
\lab{twosolsu322}
\er
 The species-12  two-soliton solution is given by:
\br
\( \begin{array}{c}
\tau_0\\
\tau_1\\
\tau_2
\end{array}\) = 
\( \begin{array}{c}
1\\
1\\
1
\end{array}\) +
 \( \begin{array}{c}
1\\
\omega\\
\omega^2
\end{array}\)\,e^{\Gamma\(z_1\)}+
\( \begin{array}{c}
1\\
\omega^2\\
\omega
\end{array}\) \, e^{\Gamma\(z_2\)}+
\( \begin{array}{c}
1\\
1\\
1
\end{array}\) \,e^{\Gamma\(z_1\)+\Gamma\(z_2\)+\Delta_{12}},
\lab{twosolsu312}
\er
where  $\Gamma\(z_k\)$ is given in \rf{gammadefsu3}, and the quantities $\Delta_{11}$ and $\Delta_{12}$ are given by 
\br
 e^{\Delta_{11}}=\left\{
 \begin{array}{ll}
\frac{4\,\sinh^2\(\frac{\alpha_2-\alpha_1}{2}\)}{4\,\cosh^2\(\frac{\alpha_2-\alpha_1}{2}\)-1}\quad &{\rm if} \qquad\eta_1\,\eta_2=1\\ 
&\\
\frac{4\,\cosh^2\(\frac{\alpha_2-\alpha_1}{2}\)}{4\,\sinh^2\(\frac{\alpha_2-\alpha_1}{2}\)+1}
&{\rm if} \quad\eta_1\,\eta_2=-1
\end{array}
\right.
\lab{expdelta11}
\er
and
\be
 e^{\Delta_{12}}=\left\{
 \begin{array}{ll}
\frac{4\,\sinh^2\(\frac{\alpha_2-\alpha_1}{2}\)+1}{4\,\cosh^2\(\frac{\alpha_2-\alpha_1}{2}\)}
&{\rm if} \qquad\eta_1\,\eta_2=1\\
&\\
\frac{4\,\cosh^2\(\frac{\alpha_2-\alpha_1}{2}\)-1}{4\,\sinh^2\(\frac{\alpha_2-\alpha_1}{2}\)}
&{\rm if} \qquad\eta_1\,\eta_2=-1
\end{array}
\right..
\lab{expdelta12}
\ee
In these expressions $\alpha_a$, $a=1,2$, are the rapidities introduced in \rf{gammadef}, and related to the velocities by $v_a=\tanh\alpha_a$. Note that the two-soliton solutions \rf{twosolsu311}, \rf{twosolsu322} and \rf{twosolsu312} satisfy the conditions \rf{realitycondonesol} and \rf{realitycondonesolphi}, and so the Hamiltonian  \rf{su3hamiltonian} is real when evaluated on them. 

As explained in \cite{clis2} and mentioned above, whenever the matrix $L_{ij}$ has degenerate eigenvalues  one can construct static multi-soliton solutions. The eigenvalue $3$ of the matrix  
\rf{su3matrixl} is doubly degenerate and so we can obtain a static two-soliton solution. Such a solution is obtained using the Hirota one-soliton ansatz \rf{onesoltauansatz} and it is given by 
\br
\( \begin{array}{c}
\tau_0\\
\tau_1\\
\tau_2
\end{array}\) = 
\( \begin{array}{c}
1\\
1\\
1
\end{array}\) +\left[
 \( \begin{array}{c}
1\\
\omega\\
\omega^2
\end{array}\)\,y_1+
\( \begin{array}{c}
1\\
\omega^2\\
\omega
\end{array}\) \, y_2\right]e^{\Gamma\(z\)}+
\( \begin{array}{c}
1\\
1\\
1
\end{array}\) \,\frac{y_1\,y_2}{4}\,e^{2\,\Gamma\(z\)},
\lab{statictwosolsu312}
\er
where $y_1$ and $y_2$ are the free parameters used in the expression of $\delta_j^{(1)}$ which is a linear combination  of the degenerate eigenvectors of \rf{su3matrixl}. Similarly, this solution could have been  obtained from the two-soliton solution \rf{twosolsu312} by setting $v_1=v_2$ (or equivalently 
$\alpha_1=\alpha_2$) and $\eta_1=\eta_2$. Note that the parameters $y_a$, $a=1,2$, can be absorbed into the exponential as $y_a e^{\Gamma\(z\)}= e^{\Gamma\(z\)+x_0^{(a)}}$, and so they are related to the positions of each one-soliton forming the static two-soliton solution.   In fact \rf{statictwosolsu312} is  a particular case of the static two-soliton solution for $SU(N)$ AT models given in eq.  (4.13) of \cite{clis2}.

As we have explained in \rf{multisolitontausamespecies} one can easily obtain static two-soliton solutions by combining soliton and anti-soliton of the same species. For the species-1 solitons we get the solution 
\br
\( \begin{array}{c}
\tau_0\\
\tau_1\\
\tau_2
\end{array}\) &=& 
\( \begin{array}{c}
1\\
1\\
1
\end{array}\) +
 \( \begin{array}{c}
1\\
\omega\\
\omega^2
\end{array}\)\,\(a_1\, e^{2\,\sqrt{3}\,\frac{\(x-v\,t\)}{\sqrt{1-v^2}}}
+a_2\, e^{-2\,\sqrt{3}\,\frac{\(x-v\,t\)}{\sqrt{1-v^2}}}\)+
\( \begin{array}{c}
1\\
\omega^2\\
\omega
\end{array}\) \,4\, a_1\,a_2 
\nonumber\\\
\lab{statictwosolsu311}
\er
 and for the species-2 one gets the solution 
\br
\( \begin{array}{c}
\tau_0\\
\tau_1\\
\tau_2
\end{array}\) &=& 
\( \begin{array}{c}
1\\
1\\
1
\end{array}\) +
 \( \begin{array}{c}
1\\
\omega^2\\
\omega
\end{array}\)\,\(a_1\, e^{2\,\sqrt{3}\,\frac{\(x-v\,t\)}{\sqrt{1-v^2}}}
+a_2\, e^{-2\,\sqrt{3}\,\frac{\(x-v\,t\)}{\sqrt{1-v^2}}}\)+
\( \begin{array}{c}
1\\
\omega\\
\omega^2
\end{array}\) \,4\, a_1\,a_2.
\nonumber\\\
\lab{statictwosolsu322}
\er
The  solutions \rf{statictwosolsu311} and \rf{statictwosolsu322} can be obtained from the two-soliton solutions \rf{twosolsu311} and \rf{twosolsu322} respectively, by setting $v_1=v_2=v$, $\eta_1=-\eta_2=1$, and absorbing the parameters $\xi_a$, $a=1,2$ (see \rf{gammadef}) into the definition of $a_a$, $a=1,2$.

\subsection{The parity properties.}

In our discussions of quasi-integrability in \cite{recent,vrecent} we have tried to relate it to the parity properties of the field
configurations. So let us briefly discuss here such properties of our two-soliton configurations even though our un-deformed model is fully integrable. We will later use these results when we consider the deformed models.

To consider the parity properties we define the following quantities:
\be
X_{+}\equiv \frac{1}{2}\left[\Gamma\(z_1\)+\Gamma\(z_2\)+\Delta\right],
\qquad\quad 
X_{-}\equiv \frac{1}{2}\left[\Gamma\(z_1\)-\Gamma\(z_2\)\right],
\qquad\quad \Delta\equiv \Delta_{11} \; {\rm or}\; \Delta_{12}
\ee
with $\Gamma\(z_k\)$ defined in \rf{gammadef} and $\Delta_{11}$ and $\Delta_{12}$ defined in \rf{expdelta11} and \rf{expdelta12}, respectively.  We then  consider the following parity transformation
\be
P\; :\qquad\qquad \(X_{+}\, , \, X_{-}\)\rightarrow \(-X_{+}\, , \, -X_{-}\).
\lab{paritydeftwosoliton}
\ee
The  two-soliton solution \rf{twosolsu311} can be rewritten as 
\br
\( \begin{array}{c}
\tau_0\\
\tau_1\\
\tau_2
\end{array}\) = e^{X_{+}}\left[
\( \begin{array}{c}
1\\
1\\
1
\end{array}\)\,e^{-X_{+}}
+
\( \begin{array}{c}
1\\
\omega^2\\
\omega
\end{array}\) \,e^{X_{+}}
+
e^{-\Delta_{11}/2}
\( \begin{array}{c}
1\\
\omega\\
\omega^2
\end{array}\)\,\(e^{X_{-}}+e^{-X_{-}}\) 
\right]. 
\er
Thus, under our parity transformation, we have
\be
P\; :\qquad\qquad \frac{\tau_1}{\tau_0}\rightarrow \omega^2 \,\frac{\tau_2}{\tau_0},\qquad\qquad
\frac{\tau_2}{\tau_0}\rightarrow \omega \,\frac{\tau_1}{\tau_0},
\ee
which implies that
\be
P\; :\qquad\qquad \phi_1\rightarrow \phi_2 - \frac{4\,\pi}{3},
\qquad\qquad \phi_2\rightarrow \phi_1 - \frac{2\,\pi}{3}.
\lab{paritytwosol11}
\ee

The  two-soliton solution \rf{twosolsu322} can be rewritten as
\br
\( \begin{array}{c}
\tau_0\\
\tau_1\\
\tau_2
\end{array}\) = e^{X_{+}}\left[
\( \begin{array}{c}
1\\
1\\
1
\end{array}\) e^{-X_{+}} +
\( \begin{array}{c}
1\\
\omega\\
\omega^2
\end{array}\) \,e^{X_{+}}
+
e^{-\Delta_{11}/2} \( \begin{array}{c}
1\\
\omega^2\\
\omega
\end{array}\)\,\(e^{X_{-}}+e^{-X_{-}}\)
\right].
\er
In this case we see that under our parity transformation we have
\be
P\; :\qquad\qquad \frac{\tau_1}{\tau_0}\rightarrow \omega \,\frac{\tau_2}{\tau_0},\qquad\qquad
\frac{\tau_2}{\tau_0}\rightarrow \omega^2 \,\frac{\tau_1}{\tau_0},
\ee
which implies that
\be
P\; :\qquad\qquad \phi_1\rightarrow \phi_2 - \frac{2\,\pi}{3},
\qquad\qquad \phi_2\rightarrow \phi_1 - \frac{4\,\pi}{3}.
\lab{paritytwosol22}
\ee

The most interesting, `mixed one', two-soliton solution  \rf{twosolsu312} can be rewritten as
\br
\( \begin{array}{c}
\tau_0\\
\tau_1\\
\tau_2
\end{array}\) = e^{X_{+}}\left[
\( \begin{array}{c}
1\\
1\\
1
\end{array}\)\(e^{X_{+}}+e^{-X_{+}}\) 
+e^{-\Delta_{12}/2}\left[
 \( \begin{array}{c}
1\\
\omega\\
\omega^2
\end{array}\)\,e^{X_{-}}+
\( \begin{array}{c}
1\\
\omega^2\\
\omega
\end{array}\) \, e^{-X_{-}}\right]
\right].
\er
In this case, we have very interesting transformations properties of the fields under our parity operation as we have
\be
P\; :\qquad\qquad \frac{\tau_1}{\tau_0}\rightarrow \frac{\tau_2}{\tau_0},\qquad\qquad
\frac{\tau_2}{\tau_0}\rightarrow \frac{\tau_1}{\tau_0},
\ee
which implies that
\be
P\; :\qquad\qquad \phi_1\rightarrow \phi_2, 
\qquad\qquad \phi_2\rightarrow \phi_1. 
\lab{paritytwosol12}
\ee
The two-solitons   \rf{twosolsu311},  \rf{twosolsu322} and  \rf{twosolsu312} are solutions of the $SU(3)$ Affine Toda model which is an integrable field theory possessing an infinite number of conserved quantites. However, it is worth noting that these solutions satisfy the property \rf{parityonfields} (see \rf{paritytwosol11}, \rf{paritytwosol22} and \rf{paritytwosol12}), and that the Toda potential 
\rf{todapotential} satisfies \rf{paritypotential}. Therefore, the properties of the $SU(3)$ Affine Toda model support our criteria for quasi-integrability.  We will show in our numerical simulations that such quasi-integrability properties are preserved by some special deformations of the $SU(3)$ Affine Toda model.

%%%%%%%%%%%%%%%%%%%%%%%%%%%%%%%%%%%%%%%%%%%%%

\section{Numerical support}
\label{sec:numericalsupport} 
\setcounter{equation}{0}
\subsection{General comments}
In this and next section we present and discuss the numerical support for our results of the previous sections.

First we concentrate our attention on the undeformed models, {\it i.e.} the integrable $SU(3)$ AT model,  and then we discuss the results for the deformed model defined by the equations \rf{generaleq} corresponding to  the potential \rf{deformedpotepsilon}.
For the numerical work  and to study the time evolutions we had to solve the equations of motion  which are given by 
\br
\partial_{-}\partial_{+}\phi_1&=&-\frac{i}{3}\,\left[\(3-\varepsilon\)\, e^{i\left[2\,\phi_1-\(1+\varepsilon\)\phi_2\right]} - 2\, \varepsilon\, e^{i\left[2\,\phi_2-\(1+\varepsilon\)\phi_1\right]} - 3\(1-\varepsilon\)\, e^{-i\(1-\varepsilon\)\left[\phi_1+\phi_2\right]}\right],
\lab{eqdef}\\
\partial_{-}\partial_{+}\phi_2&=&-\frac{i}{3}\,\left[ - 2\, \varepsilon\,e^{i\left[2\,\phi_1-\(1+\varepsilon\)\phi_2\right]} + \(3-\varepsilon\)\,e^{i\left[2\,\phi_2-\(1+\varepsilon\)\phi_1\right]} - 3\(1-\varepsilon\)\, e^{-i\(1-\varepsilon\)\left[\phi_1+\phi_2\right]}\right].
\nonumber
\er
Note that if we put $\varepsilon=0$ we recover the equations of the undeformed model {\it i.e.} equations \rf{todaeq}. As these equations involve second order time derivatives of fields $\phi_i$ we treat them as a Cauchy problem and so to find their solutions we need initial values of the fields $\phi_i$ and the appropriate 
boundary conditions that the fields have to satisfy.

Of course, for $\varepsilon=0$ we have the analytical forms of the full solutions (described in Section \ref{subsec:solutions}) and so we can test our numerical methods and procedures by comparing the numerically determined solutions to the analytical ones.

\subsection{Numerical procedures}
\label{subsec:numericalprocedures}

Our numerical simulations were performed using the 4th order Runge-Kutta method of simulating time evolution.
As in \cite{recent} we experimented with various grid sizes and numbers of points and most of our simulations were performed on lattices of 40001 lattice
points with lattice spacing of 0.0006  (so they covered the region of (-12.0,\,12.0)). The time step $dt$ was 0.0002. 
At the edges of the grid ({\it i.e.} for $11.90<\vert x\vert <12.00$) we absorbed the waves reaching this region (by decreasing progressively the time change of the magnitude of the fields there). 

To perform planned numerical simulations we needed initial field configurations but unfortunately, as mentioned above,  we did not have their analytical form except for $\varepsilon=0$ ({\it i.e.} in the undeformed case). So we determined them numerically. Thus we did not have their exact form but our initial numerically determined configurations, we believe, were sufficiently close to the exact configurations so that we could trust all our results.

The procedure we adopted to determine these intial configurations was similar to the one used in \cite{LuizandVinicius}.  First we constructed approximate static one soliton field configurations. To do this we used static \rf{onesol1} configurations which we multiplied by a factor $\mu=\frac{3}{3+\varepsilon}$ (see \rf{boundaryvaluesphi}) so that they satisfied the new boundary conditions. Then, using an incredibly small time step 
($dt=1.0*10^{-7}$) we evolved these configurations using the diffusive 
equations, which were like the proper equations of motion in which the second order time derivatives were replaced by the first order ones. 
This was achieved by using the equations given by \rf{eqdef} in which $\partial_+\partial_-$ was replaced by $\frac{1}{4}(\partial_x^2-\partial_{\tau})$ where $\tau$ is an auxiliary  diffusive `time'.
This replacement had the effect of making the configuration move towards the one that solved the static equations of motion. We evolved such configurations until their energy did not change much (in practice this was the accuracy to within 
0.01\% and the fields were essentially $\tau$ independent). We then used such almost exact one soliton configurations to construct two soliton fields (static and non-static 
configurations) by exploring their symmetries and sewing the fields together 
at $x=0$ ({\it i.e.} by putting each soliton at $\pm x_0$). 
For the non-static fields we used Lorentz symmetry of the model to determine the time dependence of the one soliton fields  by calculating $\partial_t\phi_i$ from the value of the
$\partial_x\phi_i$ of the static fields. 

To be absolutely certain that this was a good procedure we compared this way of obtaining the initial conditions of the moving solitons to their exact expressions for the un-deformed 
model. When we evolved configurations from the initial conditions derived both ways - we could see no difference in the properties of fields at later times.

Then with the initial conditions so obtained we performed many simulations for various values of $\varepsilon$. In these simulations we absorbed the energy at the boundaries.
In consequence, the total energy was not conserved but the only energy which was absorbed was the energy of the radiation waves which reached the boundaries.
Hence the total remaining energy was  effectively the energy of the field configurations which we wanted to study. In fact, in most of the simulations the energy loss was extremely small showing that our model was really almost integrable; {\it i.e.} that the ideas of quasi-integrability are quite sound.

\section{Numerical results}
\label{sec:numerical} 
\setcounter{equation}{0}
\subsection{Undeformed model}
First we present our results for the un-deformed model 
{\it i.e.} for the model with $\varepsilon=0.$

Our first set of plots shows one soliton configurations.
In Fig. 2 we present the plots of $\phi_1$. The two plots show the real and imaginary parts of $\phi_1$. The plots of $\phi_2$ are very similar except that its phase rotates differently.
This similarity comes from the symmetry of the field configurations mentioned earlier.
\begin{figure}
  \centering
  \subfigure[]{\includegraphics[trim = 0cm 0cm 1.8cm 1.8cm, width=0.45\textwidth]{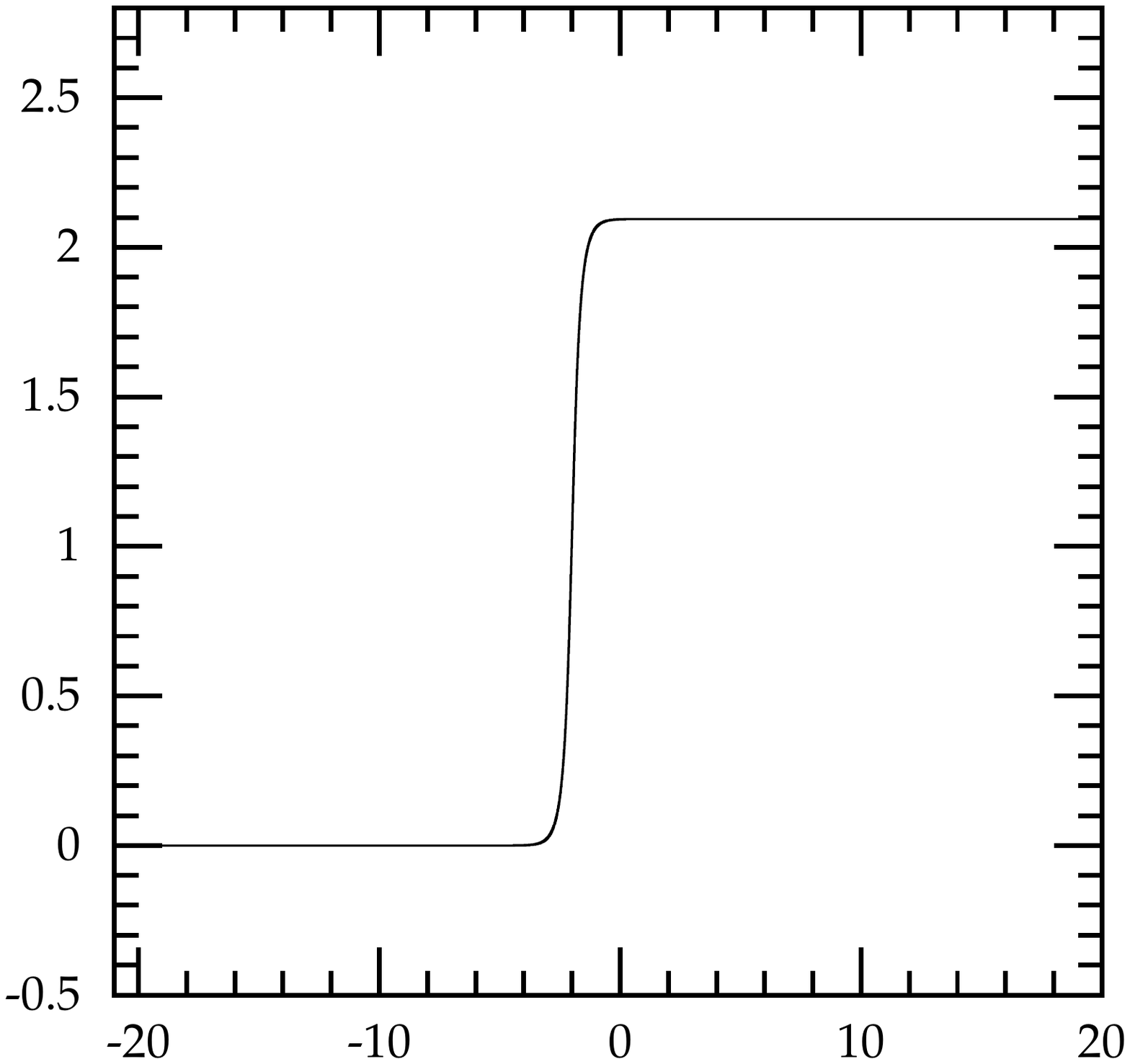}}                
  \subfigure[]{\includegraphics[trim = 0cm 0cm 1.8cm 1.8cm,  width=0.45\textwidth]{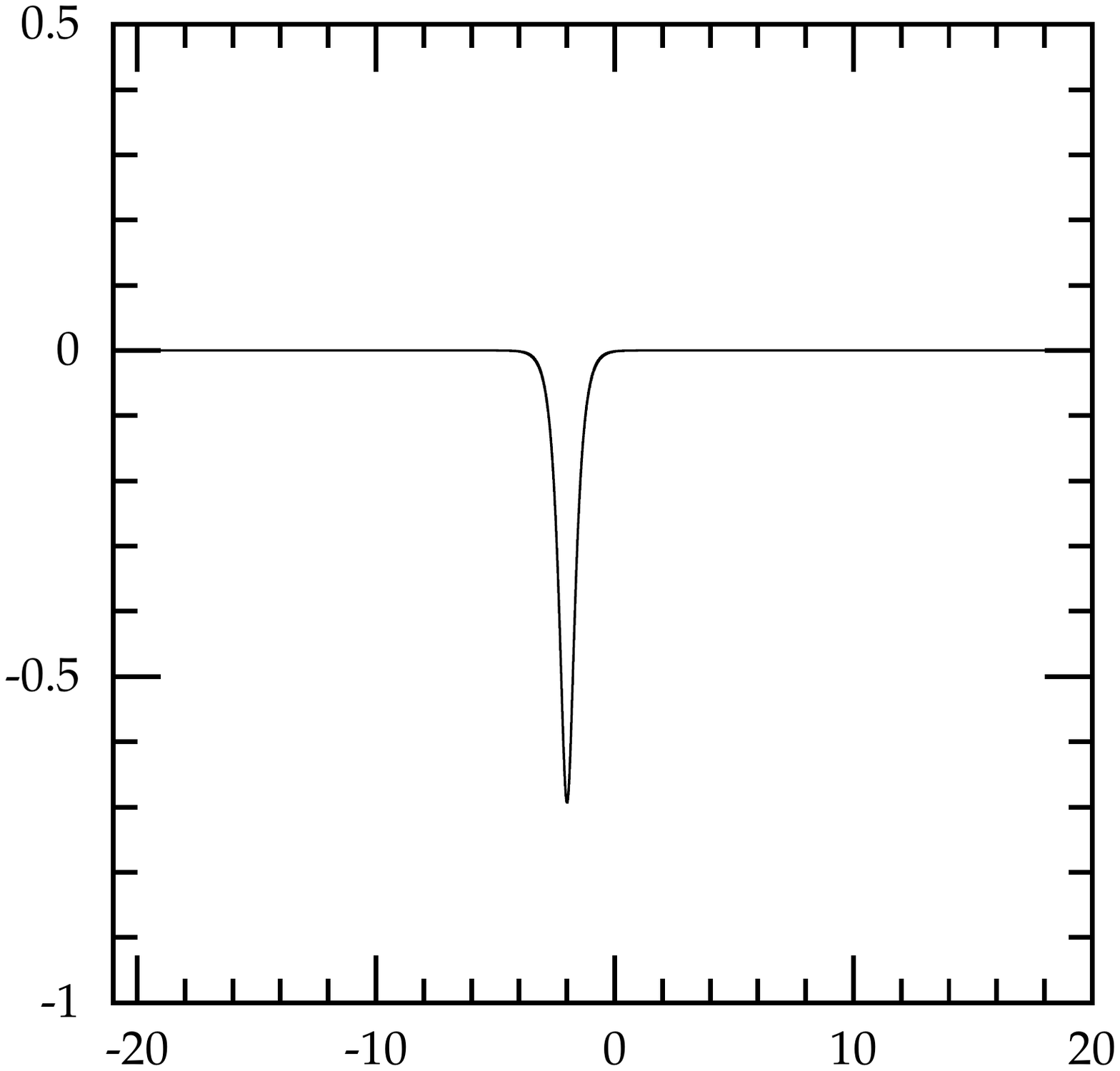}}
  \caption{The real (a) and imaginary parts (b) of the field $\phi_1$ for a typical one soliton solution.}
  \label{fig:1sol}
\end{figure}
 Note that the plots of the real parts of $\phi_i$ look very similar to those for the Sine-Gordon solitons.

 As we said earlier the model possesses also two different classes of two soliton solutions. They are shown in Fig. 3.
The plot in Fig. 3a shows the real part of $\phi_1$ of the first class (`the mixed' one), while Fig. 3b shows the configuration
of the second class (`of the two of the same' one).  Because of the symmetry $\phi_1=-(\phi_2)^{\star}$ we see that in both cases 
 $Re(\phi_2)=-Re(\phi_1)$ and the imaginary parts are the same.

\begin{figure}
  \centering
  \subfigure[]{\includegraphics[trim = 0cm 0cm 1.8cm 1.8cm, width=0.45\textwidth]{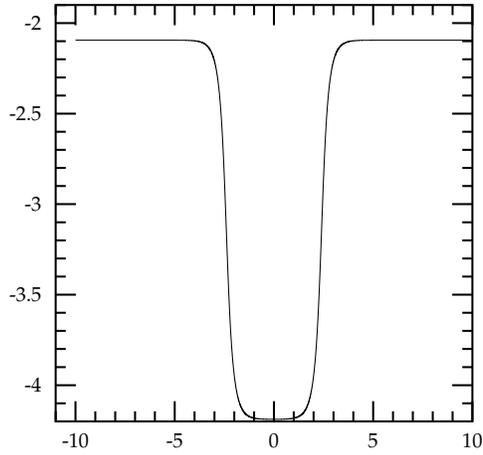}}                
  \subfigure[]{\includegraphics[trim = 0cm 0cm 1.8cm 1.8cm,  width=0.45\textwidth]{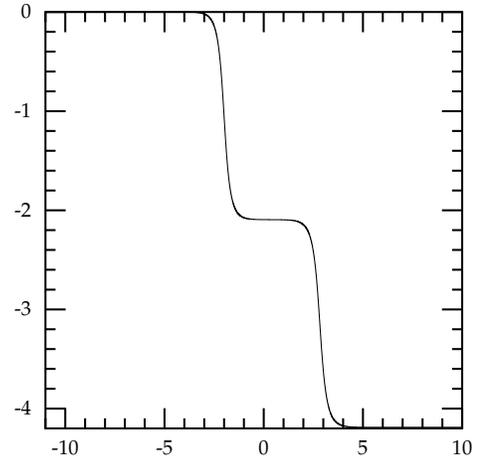}}
  \subfigure[]{\includegraphics[trim = 0cm 0cm 1.8cm 1.8cm,  width=0.45\textwidth]{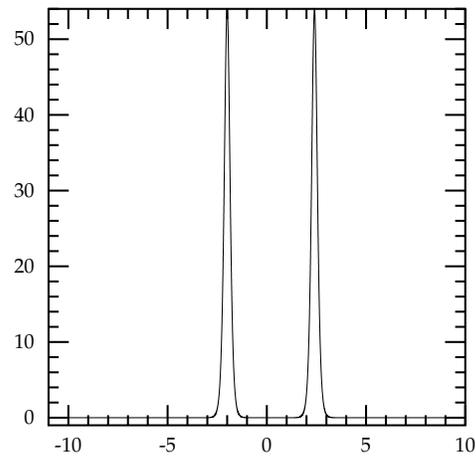}}
  \caption{The real parts of fields $\phi_1$ for the two classes of two soliton solutions; a) the solitons of the 'mixed class', b) the solitons of the 'two of the same', and c) the energy density of these solutions. }
  \label{fig:2sol}
\end{figure}

\subsection{General Comments}

Our method of generating the initial conditions by reflecting one soliton fields (when solitons ended up being far apart) gave essentially the same results as the method of taking them from the exact solutions. The results of the simulations were essentially the same; moreover, they very closely followed the analytic expressions. Hence, the method was very reliable, at least, for $\varepsilon=0$ and we hope it was also reliable for $\varepsilon\ne0$ where we do not have any analytical solutions to compare our results to. 

It is interesting to observe that the energy of all our  solutions was real. This, as stated before, can be checked for the exact solutions (when this reality is guaranteed by symmetries) but this was also true in all our simulations, which somehow preserved these symmetries. In fact, this reality was also true for the energy density.

Our simulations have also established the stability of the two soliton systems. Of course, numerical simulations introduce some small perturbations but these perturbations did not lead to any instabilities. In some ways, these numerical errors were extremely small and random and so canceled each other on average. In part, this was probably due to extra symmetries which the simulations preserved.

\subsection{Deformed model}

 As we have said earlier various deformations are possible and could be considered. 
 However, we have looked mainly at the deformation given by the potential  \rf{deformedpotepsilon}, where $\varepsilon$ took both positive or negative values.  This deformation preserves many symmetries of the original Toda system and so is very likely to lead to quasi-integrability. Indeed, like the Toda potential \rf{todapotential},  the potential \rf{deformedpotepsilon} is invariant under the interchange $\phi_1\leftrightarrow \phi_2$.  Moreover, if $\phi_1^*=-\phi_2$ the energies of the underformed and deformed models are real. 

As explained in Section \ref{subsec:numericalprocedures}, the deformed one-soliton solution was obtained through a diffusive relaxation method using the exact one-soliton solution \rf{onesol1} of the integrable $SU(3)$ AT model, as a seed. Note that if one had used as a seed, the exact one-soliton solution \rf{onesol2} of the other species, the result would have been the same as taking the previous result and interchanging $\phi_1\leftrightarrow \phi_2$.  In addition, due to the boundary condition \rf{boundaryvaluesphi}, if one has the configuration of $\phi_1$ for a deformed one-soliton, one can obtain the configuration for $\phi_2$ just by flipping the sign of the  $\phi_1$-configuration.  Therefore, the deformed two-soliton solutions associated to the exact two-soliton solutions of species-11 and species-22, given in \rf{twosolsu311} and \rf{twosolsu322} respectively, are related by the interchange $\phi_1\leftrightarrow \phi_2$, and so the numerical simulations are essentially the same. Therefore, we treat them as just one case which we refer to as {\em two of the same type}. On the other hand, the deformed 
two-soliton solution associated to \rf{twosolsu312} we call a {\em mixed case}. 

\subsubsection{Results - static cases - the `mixed case'}
Here we discuss our results corresponding to the case of two solitons of the mixed case ({\it ie} those described by $\phi_1$ and $\phi_2$  whose real parts are shown in Fig. 3a.
First, we have looked at the static case. When the solitons were too far away from each other they did not interact and they did not move. In Fig. 4 we produce plots of energy densities obtained for $\varepsilon=0.01$ at two values of time ($t=0$ and $t=1000.0$) The solitons were initially placed at $\pm 6$ and it is clear that at $t=1000$ they are still there thus we see that the solitons were initially too far apart to move.
\begin{figure}
  \centering
  \subfigure[]{\includegraphics[trim = 0cm 0cm 1.8cm 1.8cm, width=0.45\textwidth]{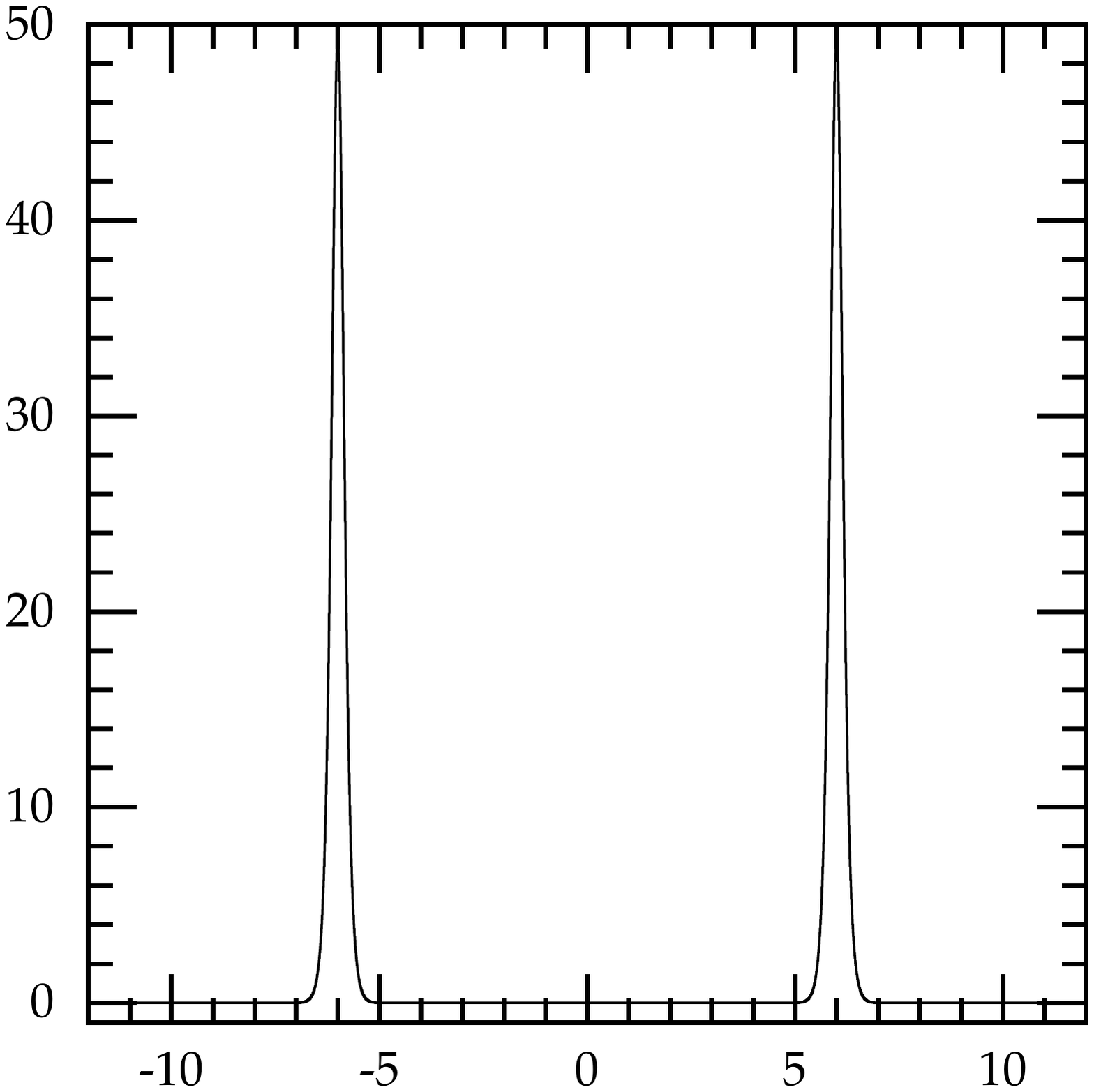}}                
  \subfigure[]{\includegraphics[trim = 0cm 0cm 1.8cm 1.8cm,  width=0.45\textwidth]{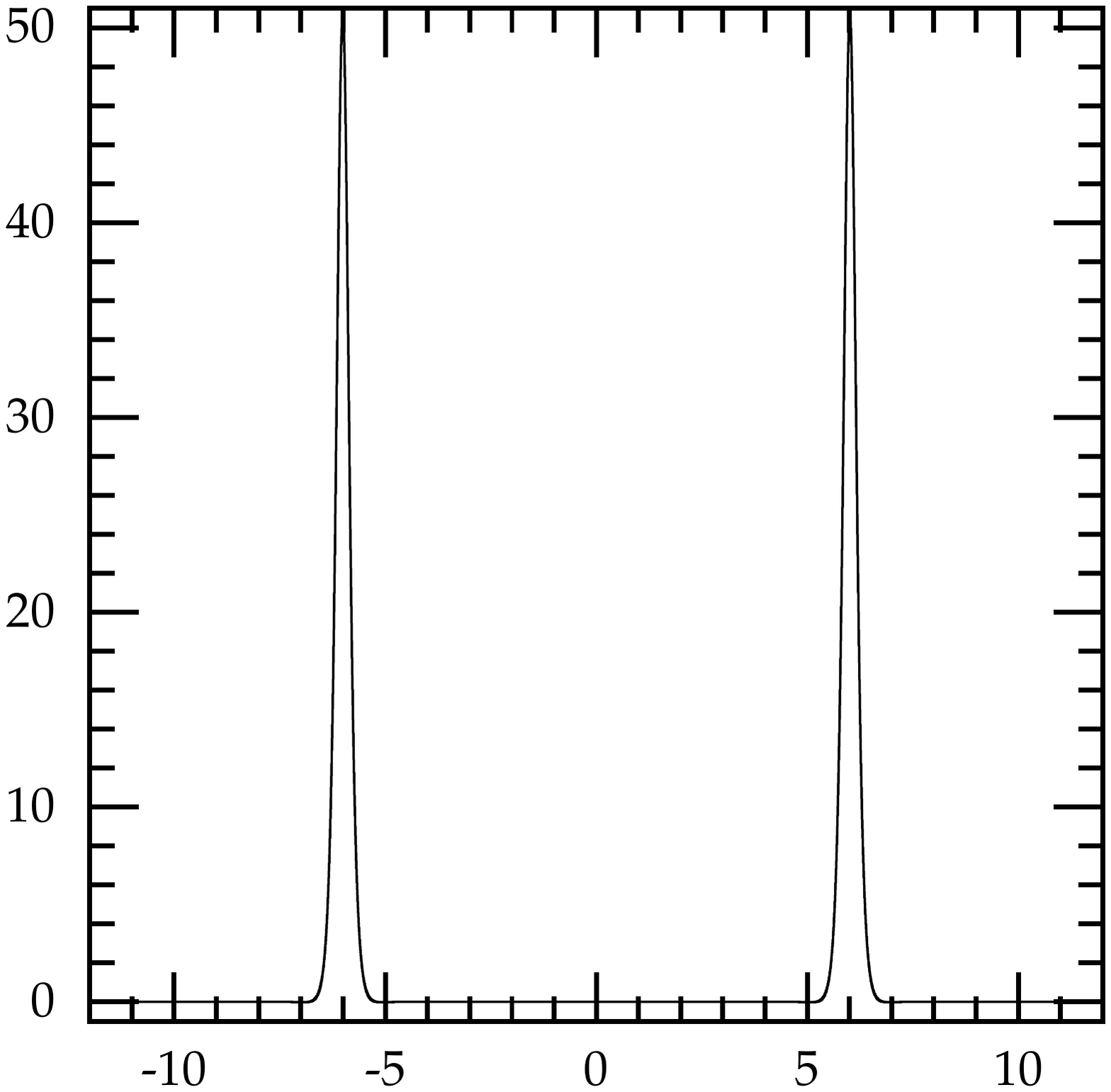}}
  \caption{Energy densities of a simulation for $\varepsilon=0.01$ (a) at $t=0$ (b) $t=1000.0$.} 
  \label{fig:3}
\end{figure}

So we started the simulations with the solitons initially placed closer together. One soliton was placed at $x=-1.5$ and the other one at $x=1.5$.  Our results can be summarised as follows. All the plots give the trajectory of one soliton (the one placed initially at $x=-1.5$, the other one followed a similar trajectory - reflected in $x=0$):

\begin{itemize}
\item The two solitons for $\varepsilon=0$ appear to be stable
and they do not move significantly.
 \item For $\varepsilon>0$ we observe repulsion.
\item For $\varepsilon<0$ we observe attraction followed by repulsion resulting in interesting oscillations.
\end{itemize}

In Fig. 5 we present a plot of `the motion' of our $x<0$  soliton for $\varepsilon=0.0$. We note essentially no motion, as to be expected from the analytical results. The small `motion' corresponds to the movement by only two lattice steps in $t=3000$ units of time and it is very likely a numerical artifact (we did not take the exact analytical solution but a field obtained by `sewing up' two one soliton expressions). 
\begin{figure}
  \centering
  \subfigure[]{\includegraphics[trim = 0cm 0cm 1.8cm 1.8cm, width=0.45\textwidth]{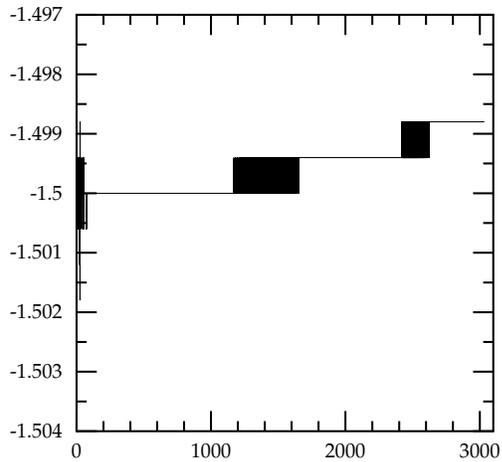}}                
   \caption{Trajectory of the soliton placed initially at $x=-1.5$ seen for $\varepsilon=0.0$.}
  \label{fig:4}
\end{figure}
In Fig. 6 we present trajectories of solitons for three simulations with negative $\varepsilon$ and in Fig. 7 two simulations for $\varepsilon>0$. 
\begin{figure}
  \centering
  \subfigure[]{\includegraphics[trim = 0cm 0cm 1.8cm 1.8cm, width=0.45\textwidth]{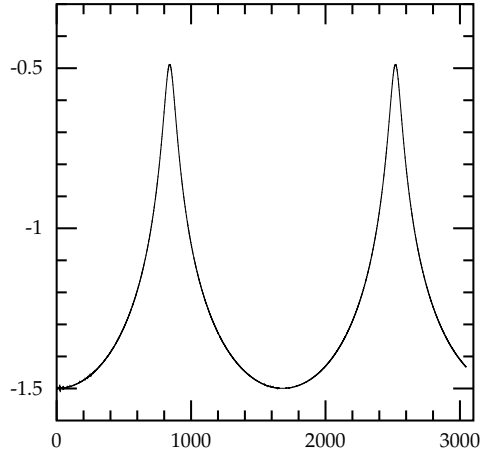}}                
  \subfigure[]{\includegraphics[trim = 0cm 0cm 1.8cm 1.8cm,  width=0.45\textwidth]{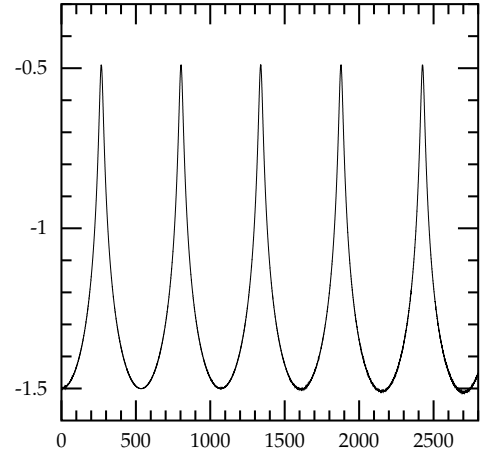}}
  \subfigure[]{\includegraphics[trim = 0cm 0cm 1.8cm 1.8cm,  width=0.45\textwidth]{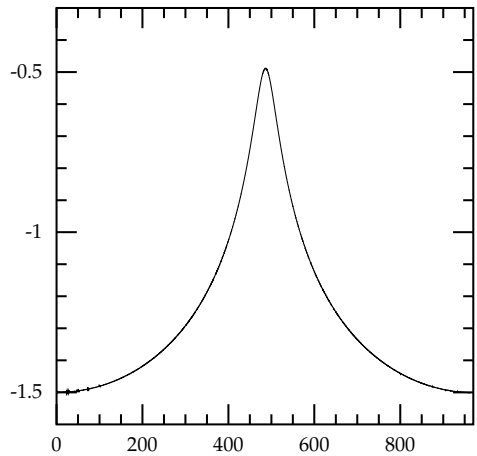}}
  \caption{Trajectiories of $x<0$ soliton  seen in  simulations for (a) $\varepsilon=-0.001$   (b) $\varepsilon=-0.01$ 
  and (c) $\varepsilon=-0.003$.} \label{fig:5}
\end{figure}
\begin{figure}
  \centering
  \subfigure[]{\includegraphics[trim = 0cm 0cm 1.8cm 1.8cm, width=0.45\textwidth]{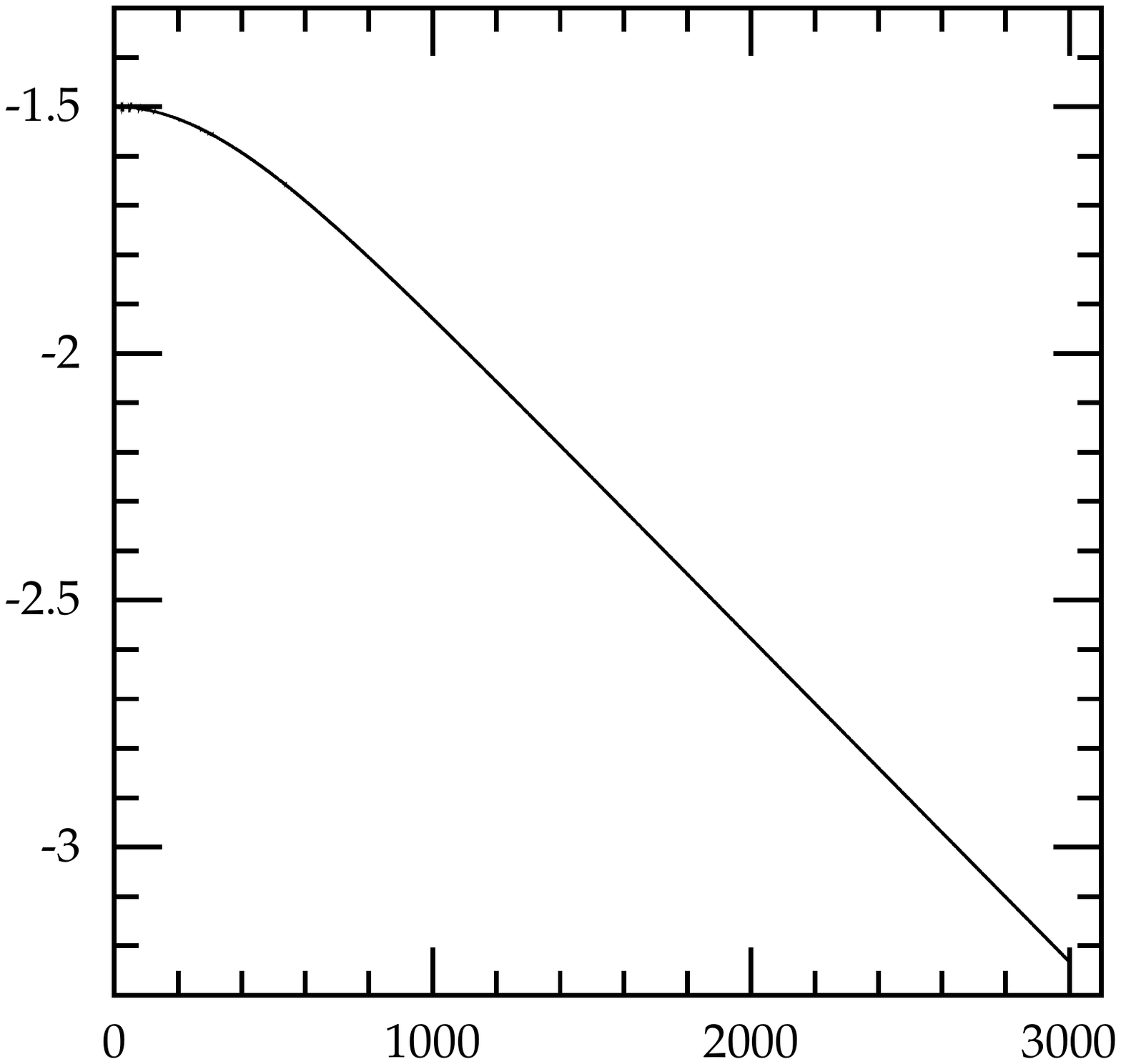}}                
  \subfigure[]{\includegraphics[trim = 0cm 0cm 1.8cm 1.8cm,  width=0.45\textwidth]{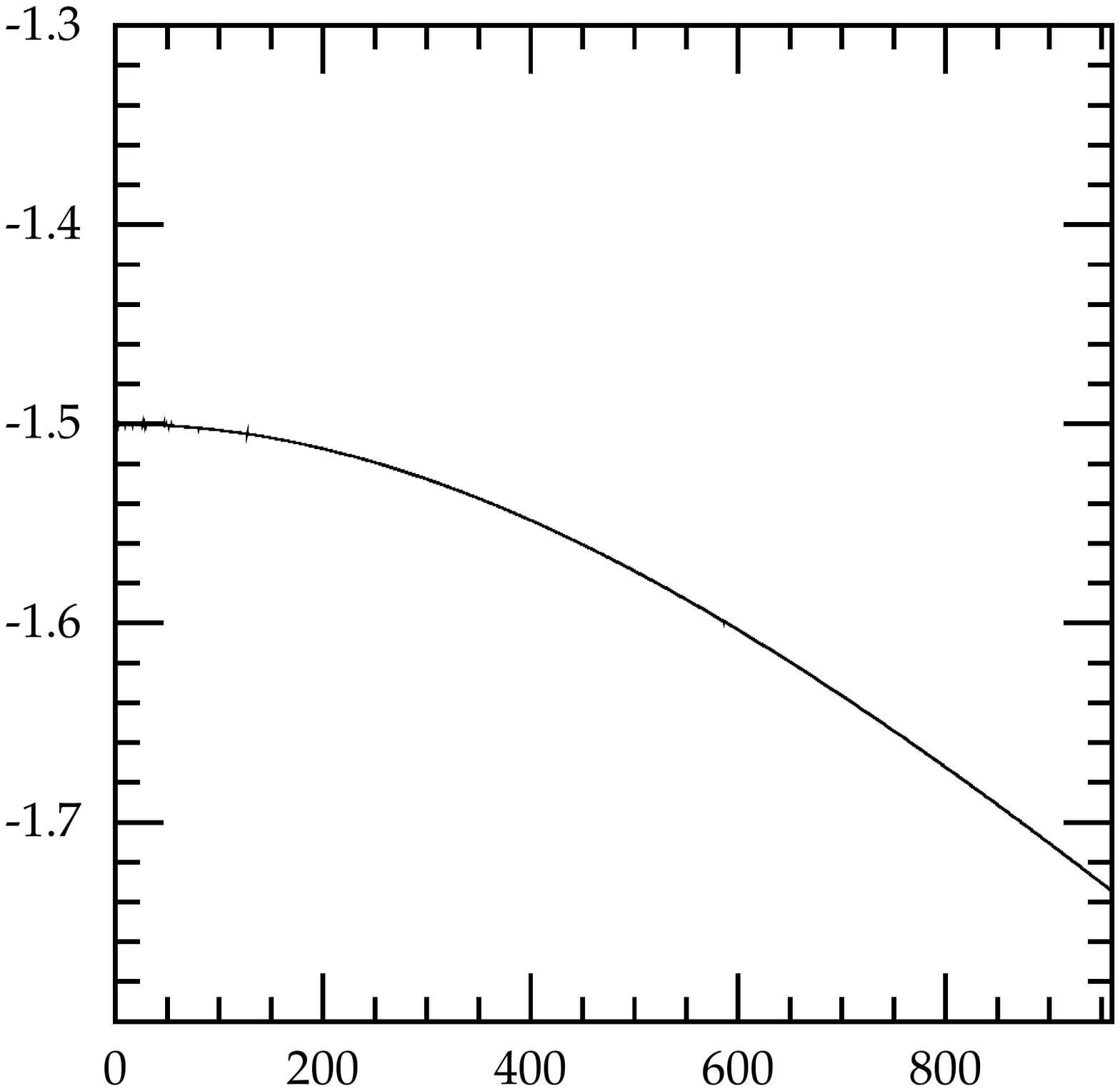}}
  \caption{Trajectiories of $x<0$ soliton  seen in  simulations for (a) $\varepsilon=0.001$ and  (b) $\varepsilon=0.005$.} 
  \label{fig:6}
\end{figure}
All these figures clearly support our claims made above. Note that for negative values of $\varepsilon$ the frequency 
of oscillations increases with the increase of $\vert\varepsilon\vert$, and in fact, as can be seen from Fig. 6b
the oscillations gradually generate a small (numerical) instability which later destabilises the process. Moreover, in all oscillations the 
solitons come close together and then bounce back.  Looking at the plots of the energy density of the solitons we find that in all these simulations the solitons 
never come closer than $r_{min}\sim 0.5+0.5=1.0$, so it would appear that they never come on top of each other (before they bounce back). This is 
further supported by the fact that the fields $\phi_1$ and $\phi_2$ look the same at all times ({\it i.e.} during the oscillations). We have tried to see
what happens when we start with the fields initially further apart or for more negative values of $\varepsilon$. In all the cases looked by us the solitons 
moved down to about the same minimal distance between them and then bounced back; the only difference was the period of  oscillations which 
increased with the decrease of the magnitude of $\varepsilon$ and/or the increase of the initial separation between the solitons.

Can they ever come on top of each other ({\it i.e.} can $r_{min}$ get smaller or even become zero)? This is difficult to assess for static solitons as we would have to start with solitons much closer together but this would introduce small perturbations due to our procedure of 'sewing'  two solitons 
together. The only way to study this would involve starting with solitons moving towards each other. This will be discussed in the next subsection.

Before we do this let us say a few words about the anomalies.
Of course, the $\varepsilon=0$ case has no anomalies so here we present the anomalies, {\it i.e.} expressions only for $\beta^{(2)}$ \rf{alphax2b} for $\varepsilon\ne0$.
In Fig. 8 we present the plots of the anomalies seen in two simulations for $\varepsilon<0$.
\begin{figure}
  \centering
  \subfigure[]{\includegraphics[trim = 0cm 0cm 1.8cm 1.8cm, width=0.45\textwidth]{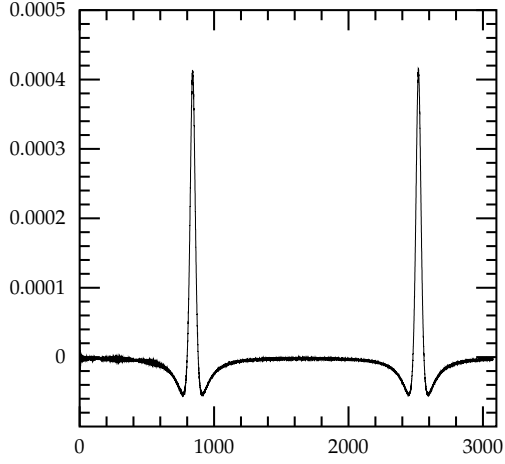}}                
  \subfigure[]{\includegraphics[trim = 0cm 0cm 1.8cm 1.8cm,  width=0.45\textwidth]{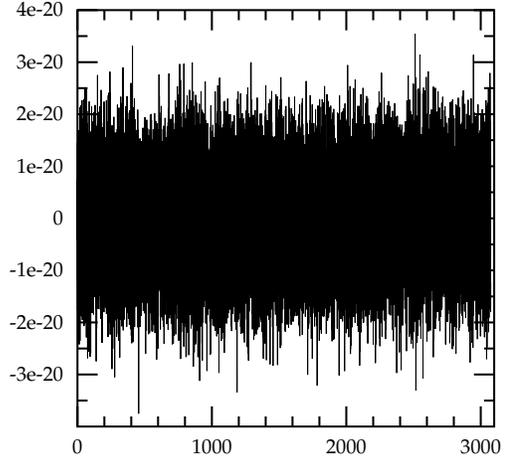}}
  \subfigure[]{\includegraphics[trim = 0cm 0cm 1.8cm 1.8cm,  width=0.45\textwidth]{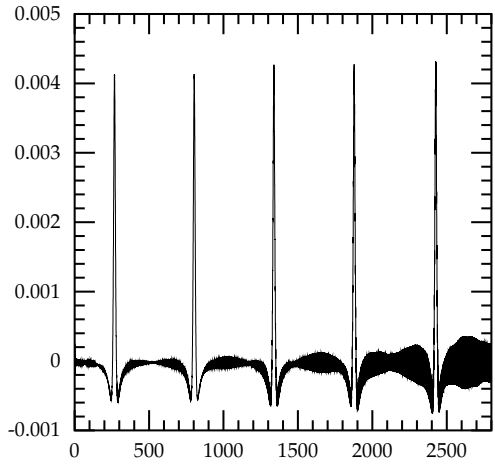}}
 \subfigure[]{\includegraphics[trim = 0cm 0cm 1.8cm 1.8cm,  width=0.45\textwidth]{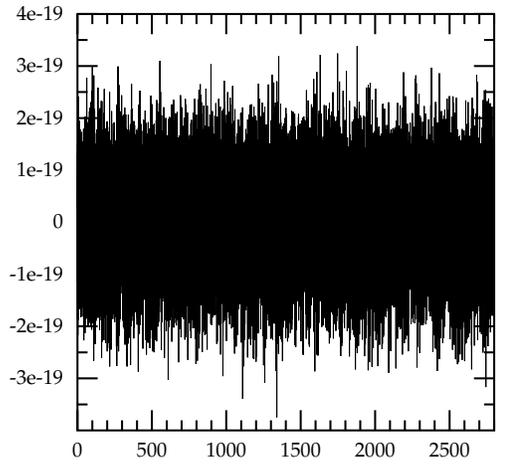}}
  \caption{Real (a), (c)  and Imaginary parts (b), (d) of anomalies  seen in  simulations for $\varepsilon=-0.001$  and  $\varepsilon=-0.01$. 
  } \label{fig:7}
\end{figure}
We clearly see that the imaginary parts of anomalies are negligible and that the real parts vary (and change when the solitons are close together) but then return to their original values. This is very much in agreement what we would expect based on the ideas of quasi-integrability.
In Fig. 9 we present similar plots of the anomaly seen in simulation for $\varepsilon=0.001$ (its trajectory is shown in Fig. 7a).
\begin{figure}
  \centering
  \subfigure[]{\includegraphics[trim = 0cm 0cm 1.8cm 1.8cm, width=0.45\textwidth]{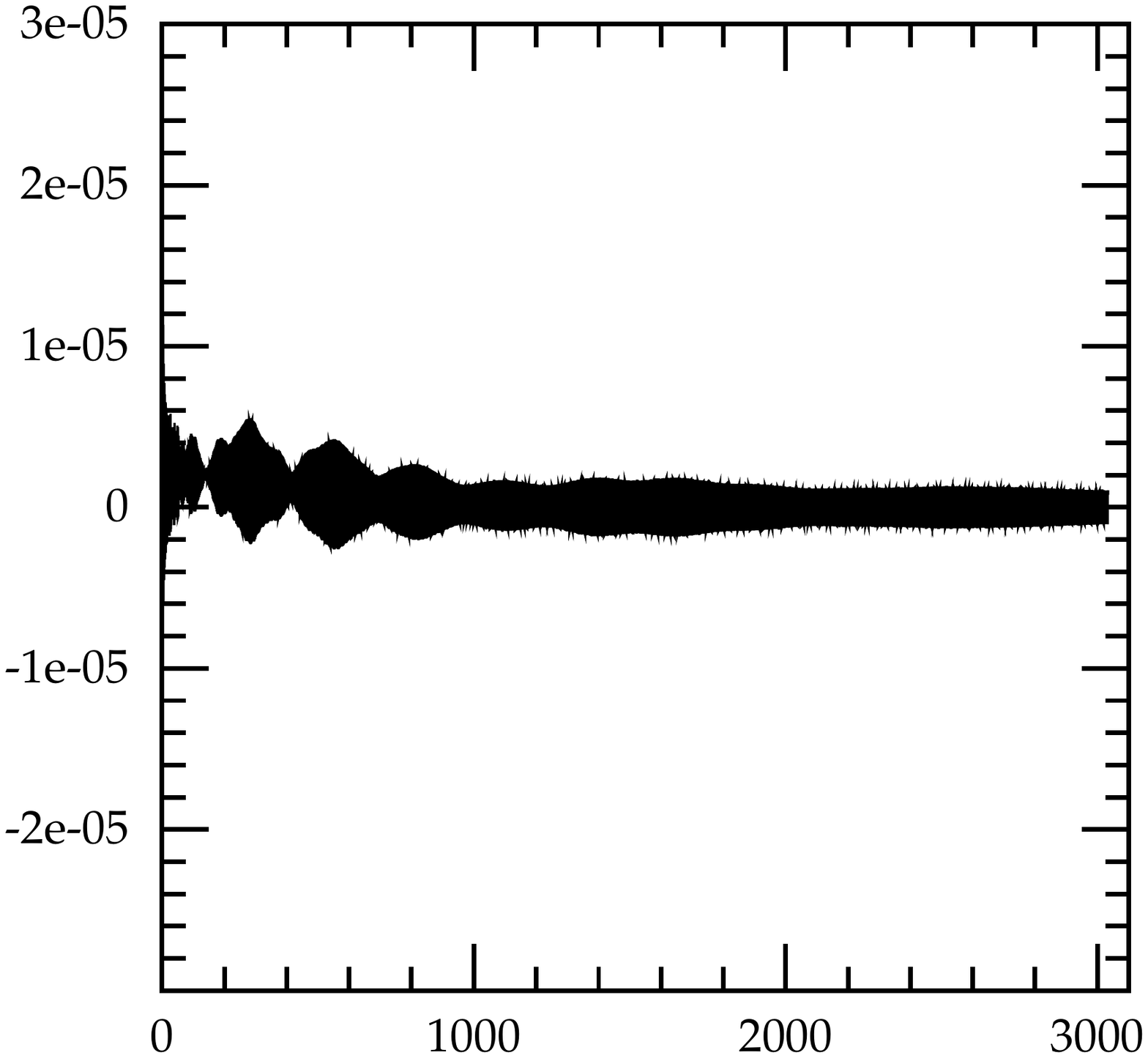}}                
  \subfigure[]{\includegraphics[trim = 0cm 0cm 1.8cm 1.8cm,  width=0.45\textwidth]{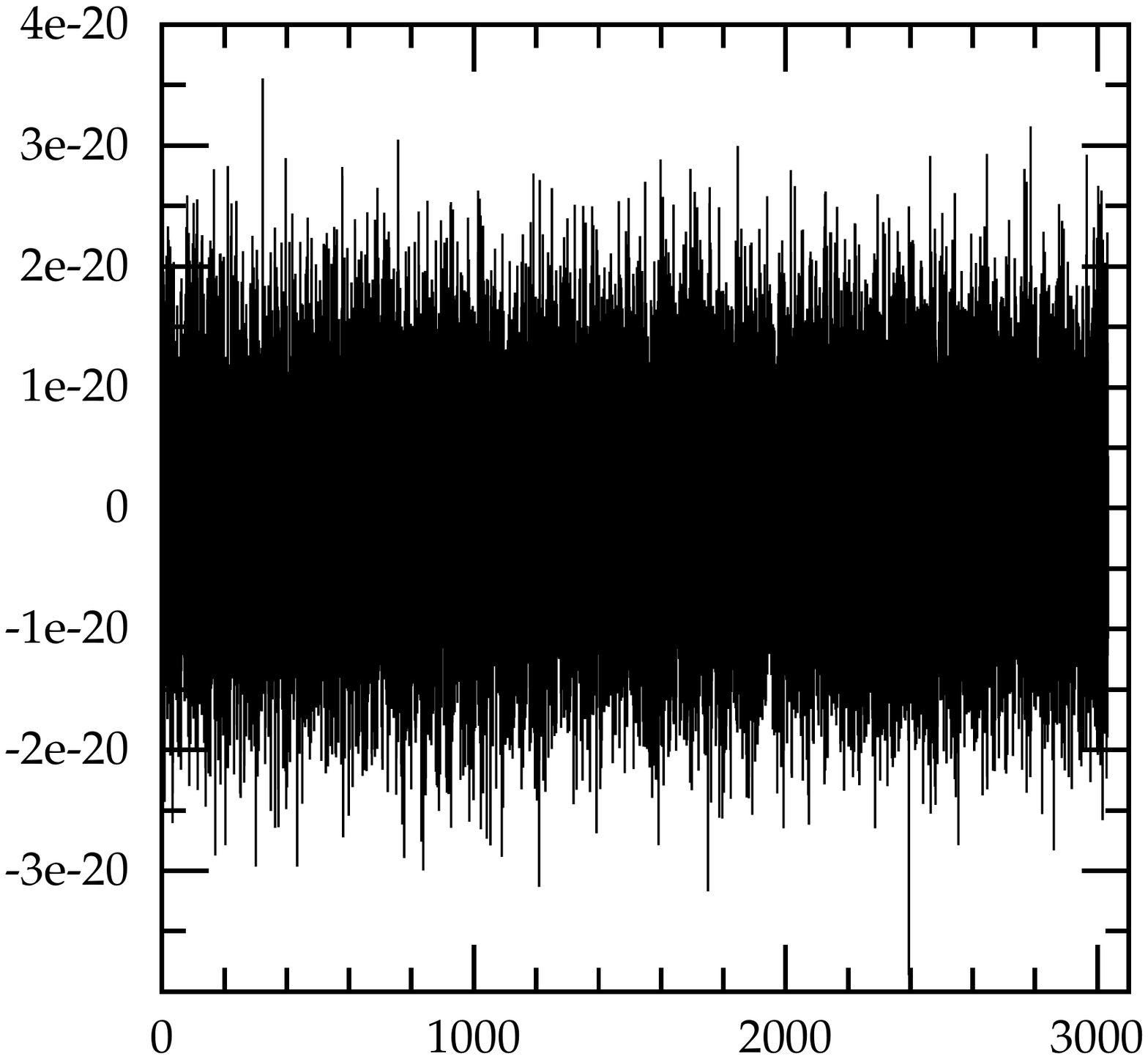}}
  \caption{Real (a) and Imaginary part (b) of the anomaly  seen in the  simulation for $\varepsilon=0.001$. 
  } \label{fig:8}
\end{figure} 
Clearly, the anomaly is again essentially real and its (real) value is very small indeed (smaller by more than two 
orders of magnitude from its value for negative values of $\varepsilon$ - this is of course, associated with the fact that solitons repel and never get very close to each other). In fact, the anomaly oscillates a little and then decreases further as the solitons move further away from each other.

\subsection{Non-static cases} 

We have also performed many interesting simulations for various values of $\varepsilon$, velocity and initial positions of solitons. Here we  discuss  the two-soliton fields of the  mixed case, and in the next section the other case. 

When we sent the solitons towards each other two things could happen - solitons could reflect with or without a `flip'.
Here, by a `flip' we denote the situation in which the two fields $\phi_1$ and $\phi_2$ swapped their shapes after the scattering. This `swapping' refers only to their real parts as the imaginary parts stay the same.
\begin{figure}
  \centering
  \subfigure[]{\includegraphics[trim = 0cm 0cm 1.8cm 1.8cm, width=0.45\textwidth]{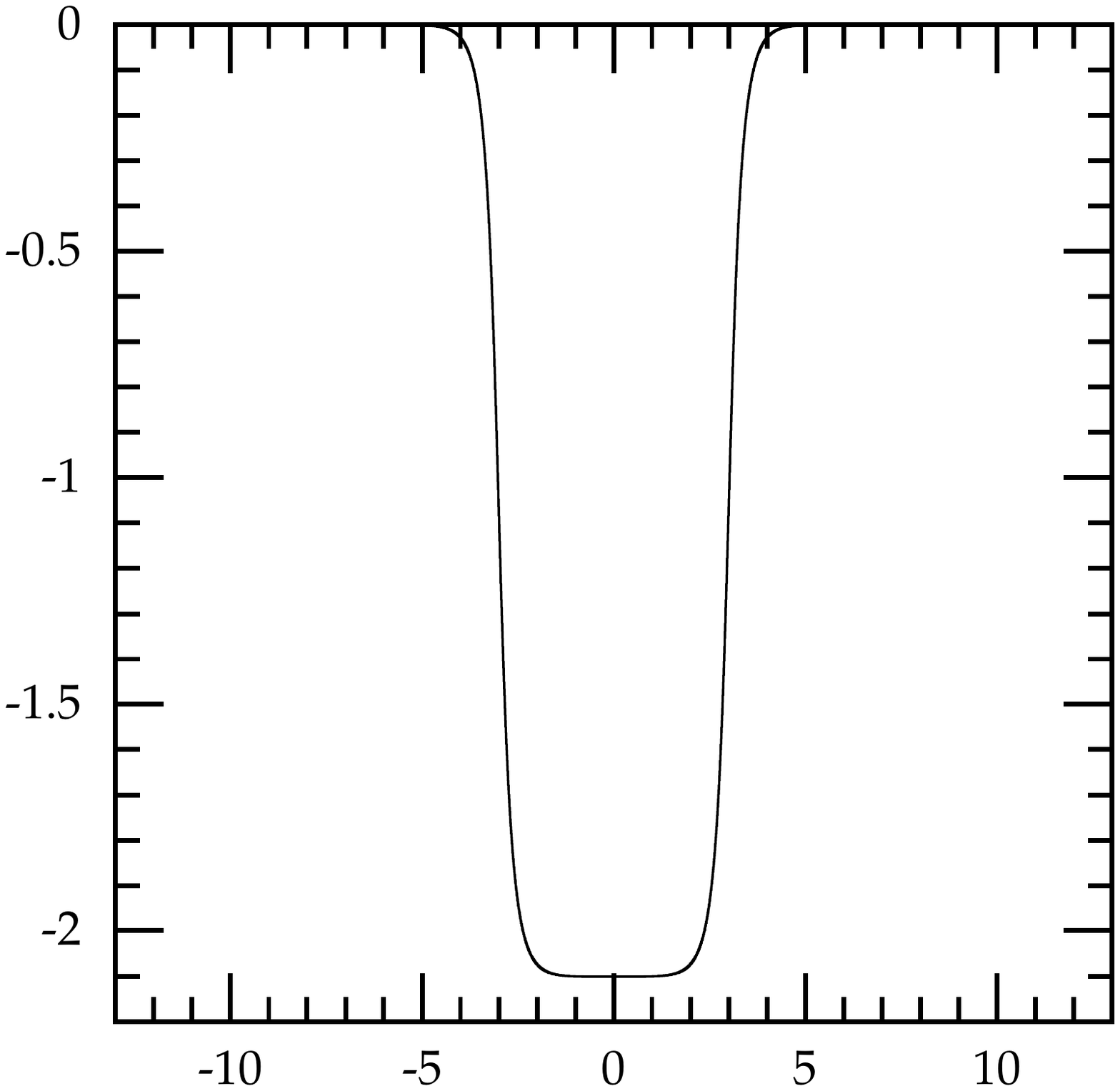}}                
  \subfigure[]{\includegraphics[trim = 0cm 0cm 1.8cm 1.8cm,  width=0.45\textwidth]{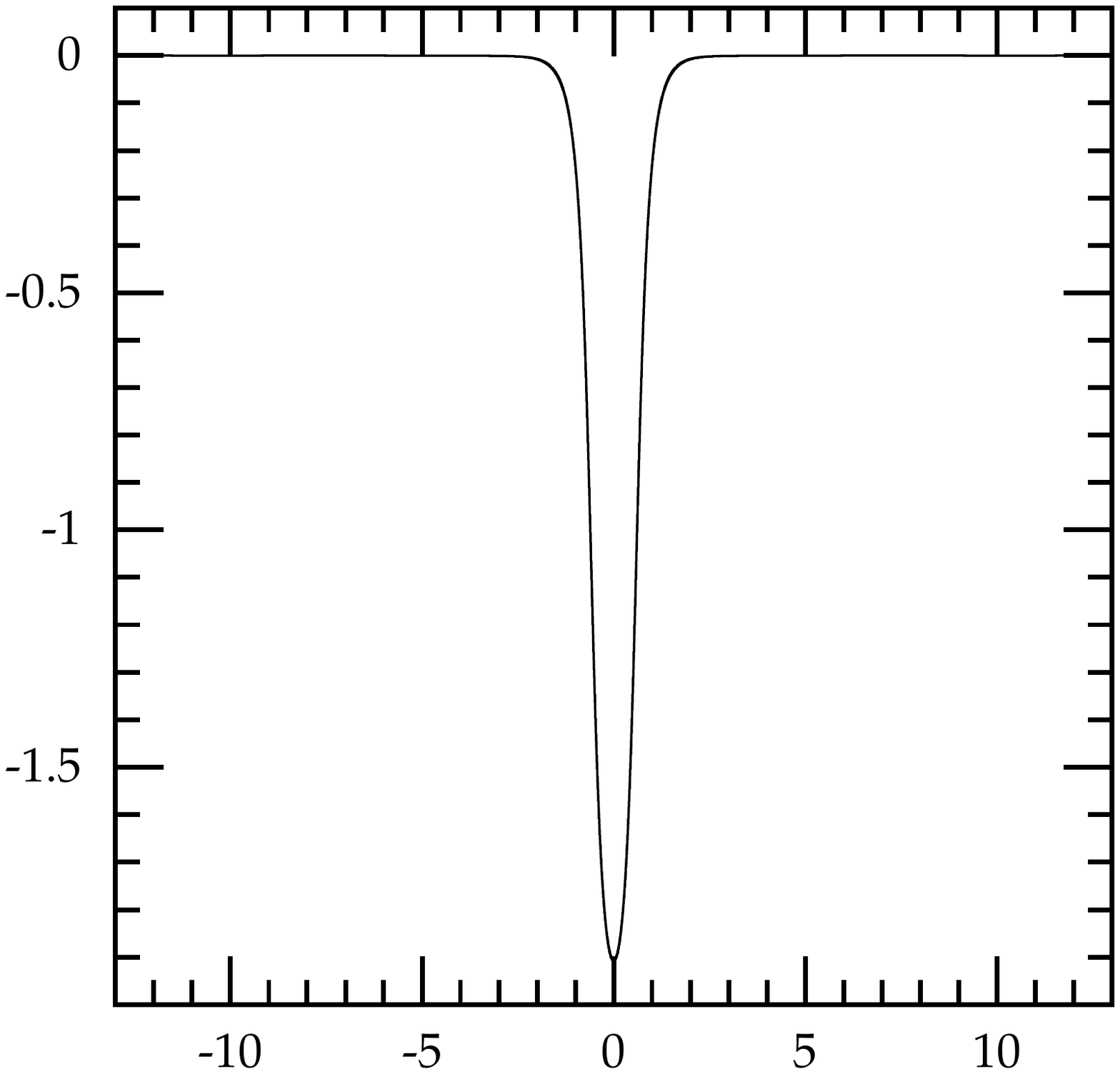}}
 \subfigure[]{\includegraphics[trim = 0cm 0cm 1.8cm 1.8cm,  width=0.45\textwidth]{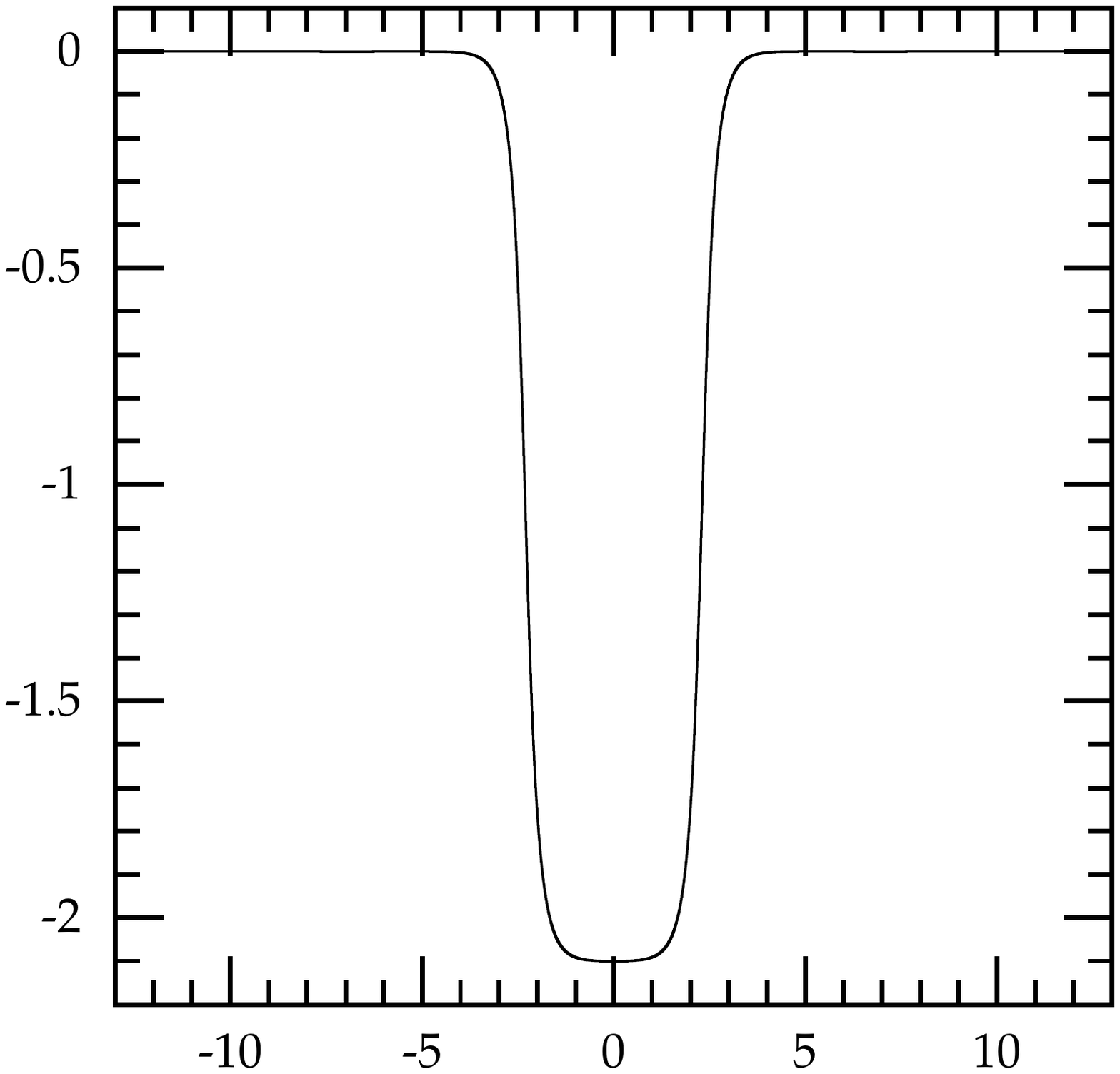}}
  \caption{Real parts of $\phi_1$ as three values of $t$ seen in a simulation started with $v=0.06$ for 
$\varepsilon=-0.01$. (a), (b), (c) correspond to, respectively, $t=$0, 40 and 80.} \label{fig:9}
\end{figure} 
In Fig. 10 we present the plots of the real parts of fields when we had a reflection, 
and in Fig. 11 the similar plots for the case when the fields performed the `flip'.
\begin{figure}
  \centering
  \subfigure[]{\includegraphics[trim = 0cm 0cm 1.8cm 1.8cm, width=0.45\textwidth]{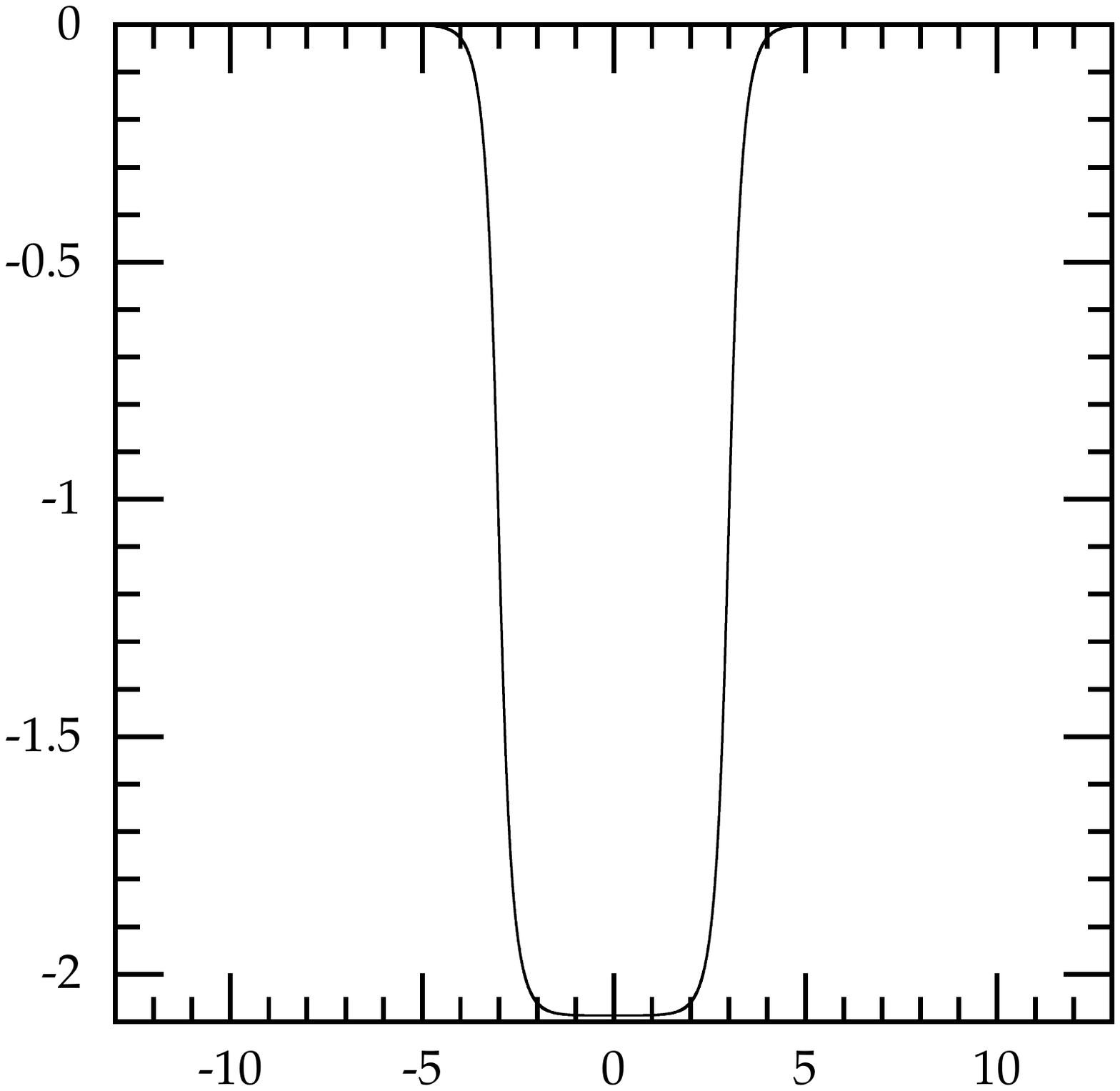}}                
  \subfigure[]{\includegraphics[trim = 0cm 0cm 1.8cm 1.8cm,  width=0.45\textwidth]{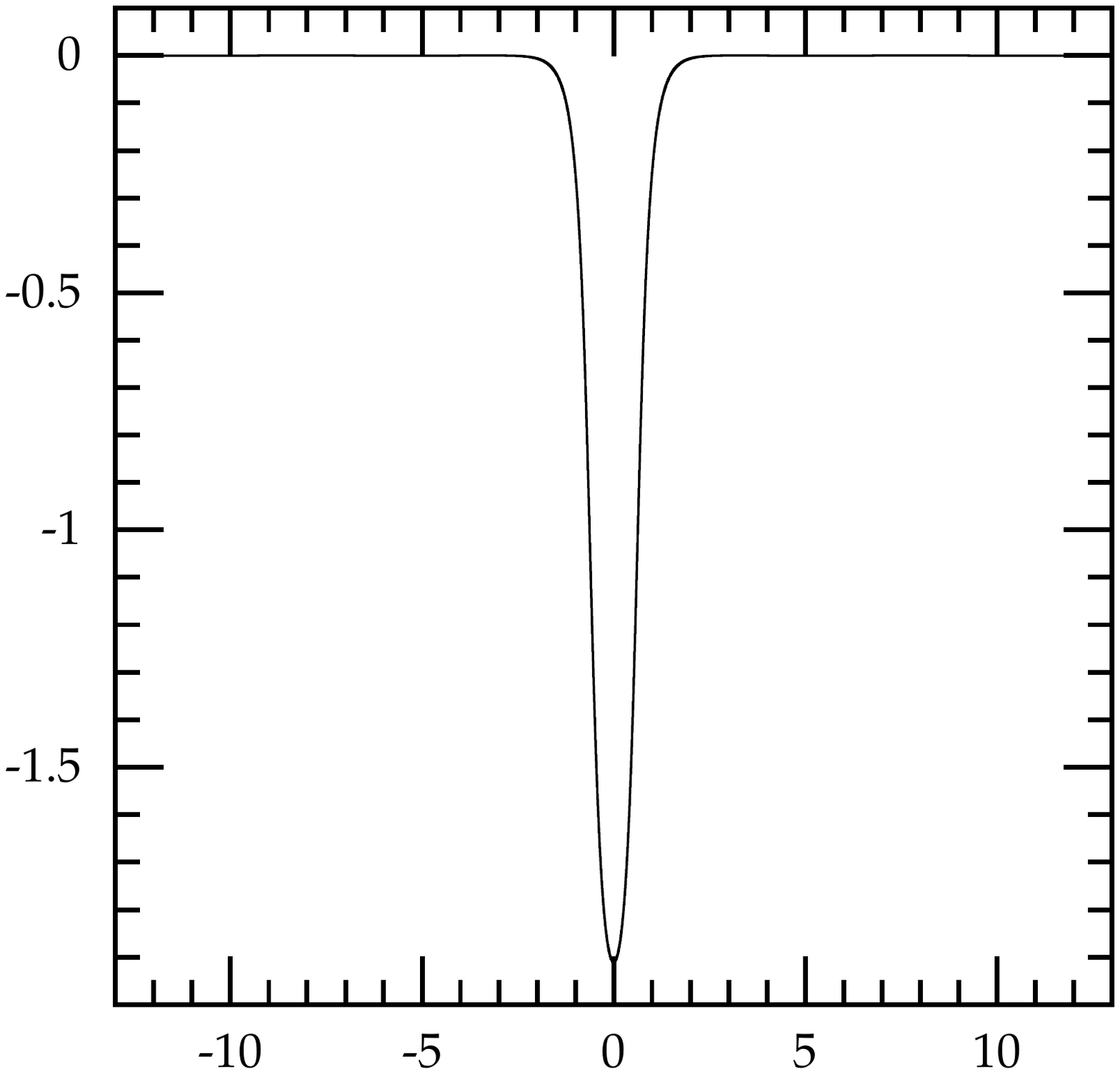}}
 \subfigure[]{\includegraphics[trim = 0cm 0cm 1.8cm 1.8cm,  width=0.45\textwidth]{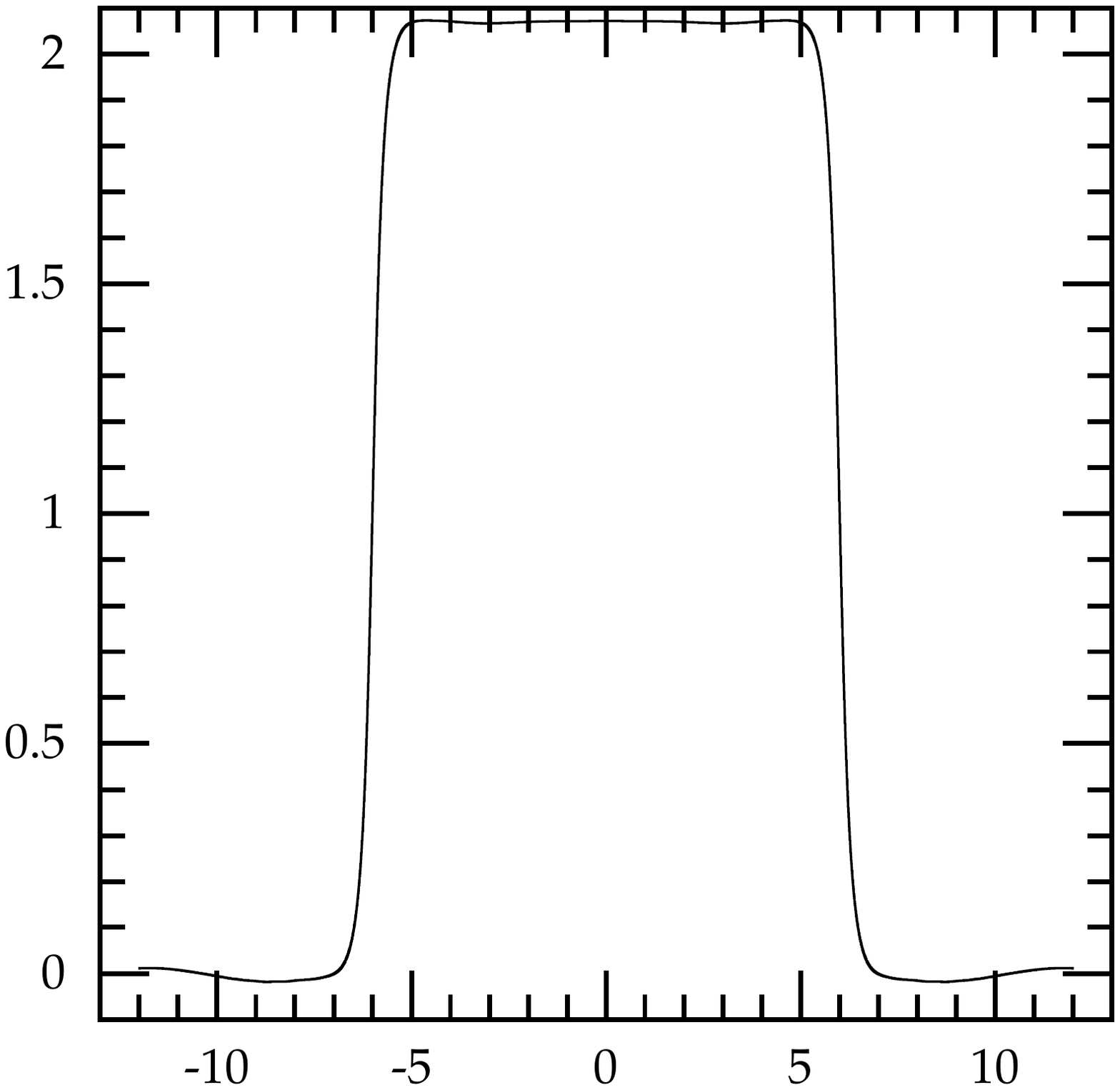}}
  \caption{Real parts of $\phi_1$ as three values of $t$ seen in a simulation started with $v=0.06$ for 
$\varepsilon=0.01$. (a), (b), (c) correspond to, respectively, $t=$0, 40 and 80.} \label{fig:10}
\end{figure} 
 We can try to relate this `flipping' to the issue of the solitons coming on top of each other or not, which we alluded to in the previous subsection.
 In fact, all the case of the `flipping' corresponded to the cases when the solitons got on top of each other.
 We have verified this in all the cases. We observed this by looking at the trajectories of solitons and comparing the plots of each field  
 and the energy densities at all relevant values of time.
 We have performed our simulations for many cases and in Fig. 12a we present the plots of the trajectory of one soliton seen in the simulation of $\varepsilon=-0.001$
 started with solitons initially at $\pm 6.00$ and moving with   velocity $v=0.1$ towards each other. We note that the trajectory reaches $x=0$ when the solitons are on top of each other,
 at which time the energy density is very localised (and in fact possesses small negative contributions) and then the field
 configuration of the solitons `flips' (basically the fields $\phi_1$ and $\phi_2$ get swapped).  From then onwards the trajectories become a bit
 irregular and a bit steeper.

 \begin{figure}
  \centering
  \subfigure[]{\includegraphics[trim = 0cm 0cm 1.8cm 1.8cm, width=0.45\textwidth]{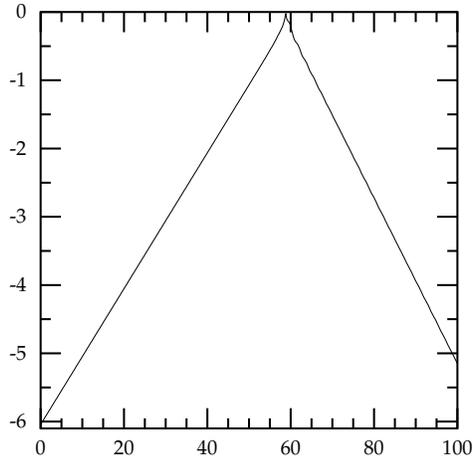}}                
  \subfigure[]{\includegraphics[trim = 0cm 0cm 1.8cm 1.8cm,  width=0.45\textwidth]{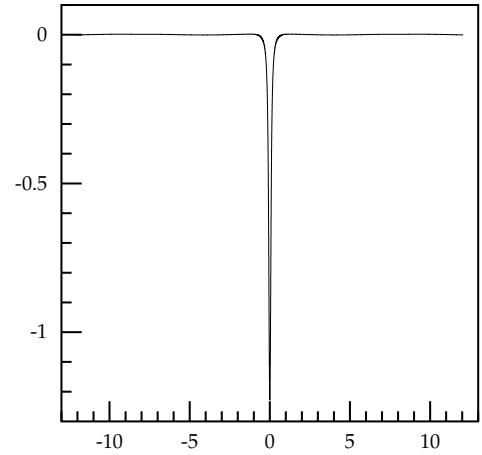}}
 \subfigure[]{\includegraphics[trim = 0cm 0cm 1.8cm 1.8cm,  width=0.45\textwidth]{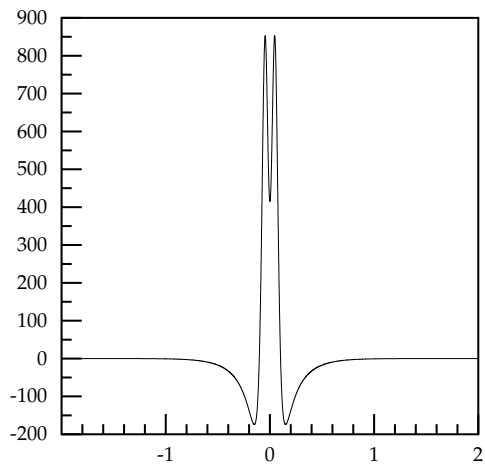}}
  \caption{The case of $\varepsilon=-0.001$. Initial velocity =0.1. (a) The trajectory of one soliton, (b) the real part of the field $\phi_1$ at $t=59.00$
(c) The energy density of the total configuration at $t=59$. The other soliton behaves in a symmetrically opposite way relative to $x=0$.} \label{fig:11}
\end{figure}

 What about the anomalies? Our simulations showed that they were always very small and were essentially real. In Fig. 13 and 14 we present plots of the anomalies  for the two simulations shown in Fig. 10 and 11.
\begin{figure}
  \centering
  \subfigure[]{\includegraphics[trim = 0cm 0cm 1.8cm 1.8cm, width=0.45\textwidth]{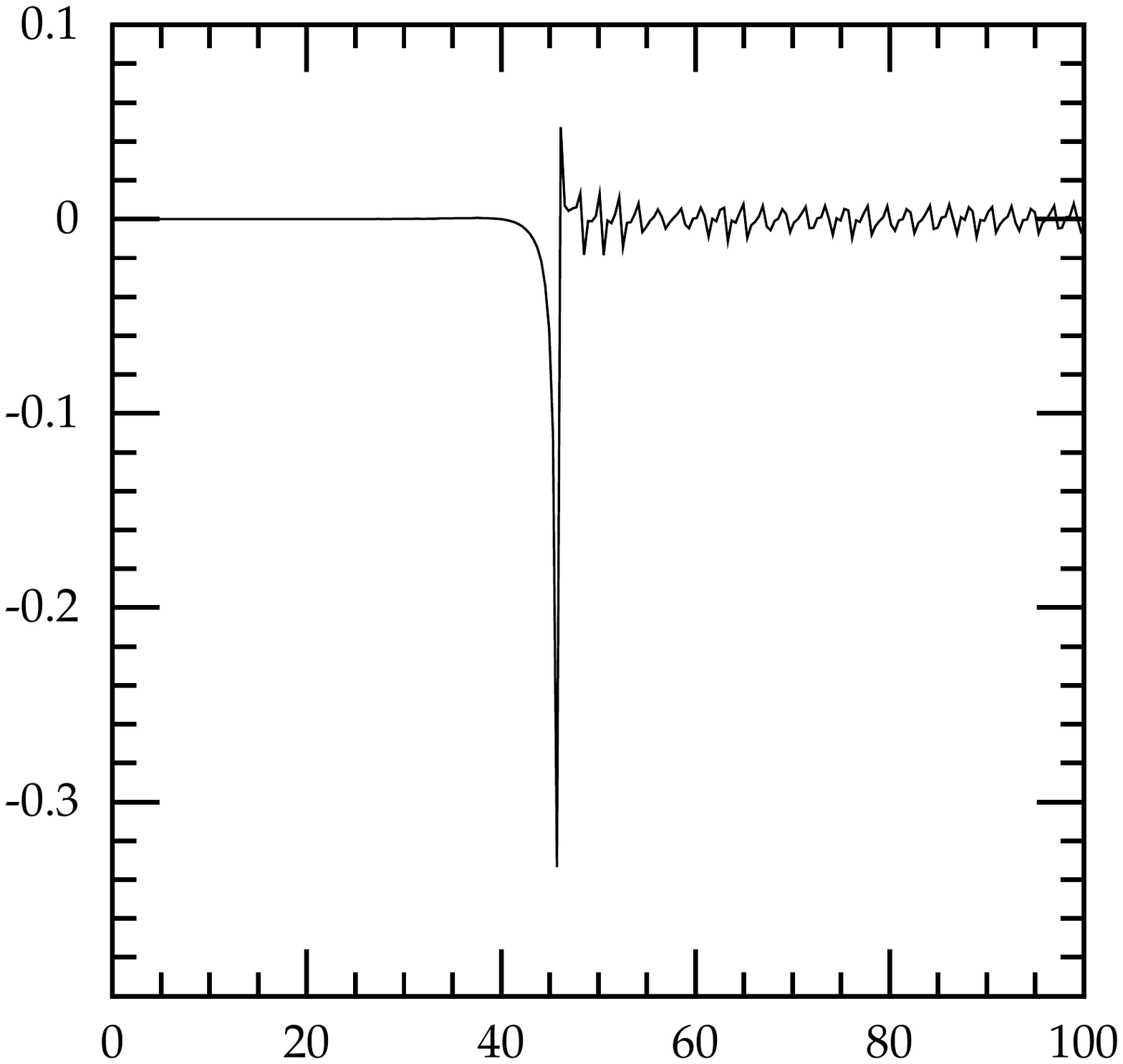}}                
  \subfigure[]{\includegraphics[trim = 0cm 0cm 1.8cm 1.8cm,  width=0.45\textwidth]{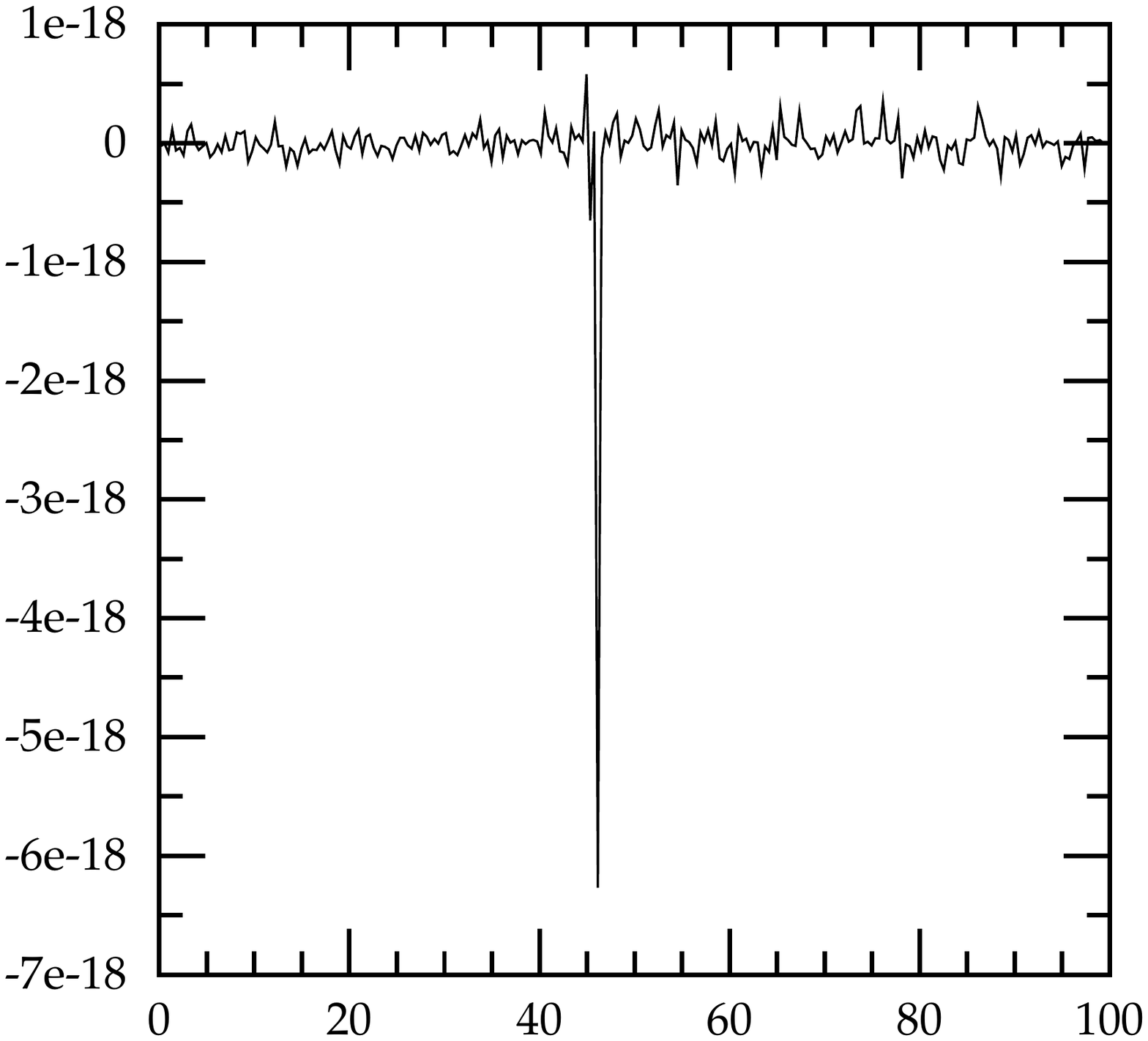}}
  \caption{Real (a) and imaginary (b) parts of the anomaly seen in the simulation for $\varepsilon=0.01$. 
  } \label{fig:12}
\end{figure} 
\begin{figure}
  \centering
  \subfigure[]{\includegraphics[trim = 0cm 0cm 1.8cm 1.8cm, width=0.45\textwidth]{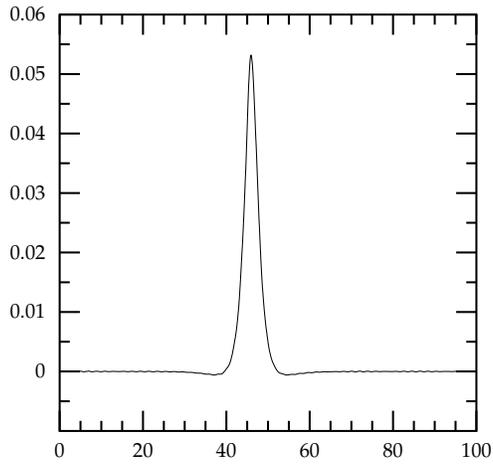}}                
  \subfigure[]{\includegraphics[trim = 0cm 0cm 1.8cm 1.8cm,  width=0.45\textwidth]{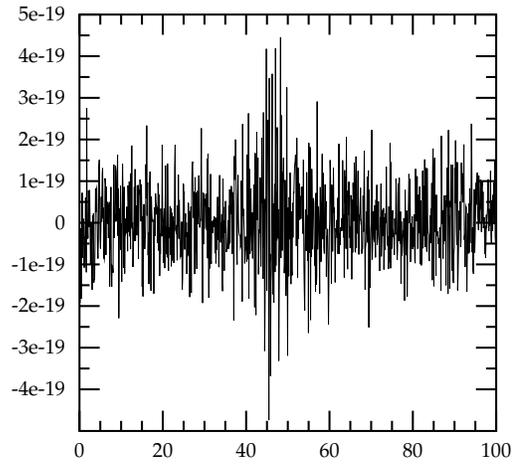}}
  \caption{Real (a) and imaginary (b) parts of the anomaly seen in the simulation for $\varepsilon=-0.01$. 
  } \label{fig:13}
\end{figure} 
We see that in the `unflipped' case the anomaly does not change as much as in the `flipped' one and so this case is more reliable.

\subsection{The other class of 2 solitonic solutions}

Next we present the results for the solitons from the other class, {\it i.e.} the one corresponding to Fig. 3b (two of the same type).
In this case we always have a repulsion so below we present  the results of only  a few simulations.

\subsubsection{Static case}

We have performed several simulations (for several values of $\varepsilon$). The results are very similar so here we present 3 plots of the position of one soliton, initially placed at $x=-1.5$ (with the other soliton placed ast $1.5$), for 3 values of $\varepsilon$. The results are shown in Fig. 15. We note that the repulsion increases with $\varepsilon$.

\begin{figure}
  \centering
  \subfigure[]{\includegraphics[trim = 0cm 0cm 1.8cm 1.8cm, width=0.45\textwidth]{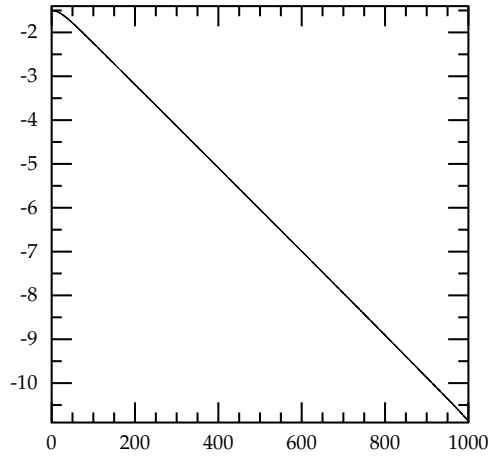}}                
  \subfigure[]{\includegraphics[trim = 0cm 0cm 1.8cm 1.8cm,  width=0.45\textwidth]{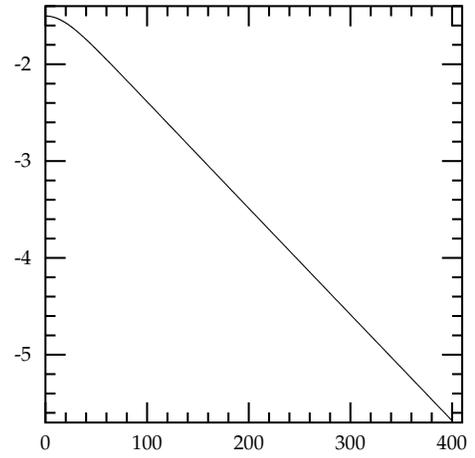}}
 \subfigure[]{\includegraphics[trim = 0cm 0cm 1.8cm 1.8cm,  width=0.45\textwidth]{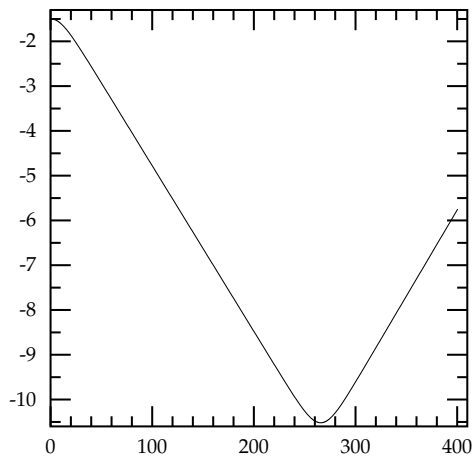}}
  \caption{Trajectories of one soliton (started at rest) for (a) $\varepsilon=-0.1$, (b) $\varepsilon=0.1$ and 
(c) $\varepsilon=0.5$. The other soliton behaves in a symmetrically opposite way relative to $x=0$.} \label{fig:14}
\end{figure}

\subsubsection{Solitons sent towards each other}

We have also performed the simulations of solitons sent towards each other with various values of velocity.
In each case we observed the repulsion (although with the increased velocity the solitons managed to get closer to each other).
In Fig. 16 we present the plots of the trajectories of solitons (sent with velocity $v=0.1$ towards each other) seen in simulations performed for several values of $\varepsilon$. As before, we plot the trajectory of one of the solitons and the other 
one moves symmetrically around $x=0$.
\begin{figure}
  \centering
  \subfigure[]{\includegraphics[trim = 0cm 0cm 1.4cm 1.8cm, width=0.35\textwidth]{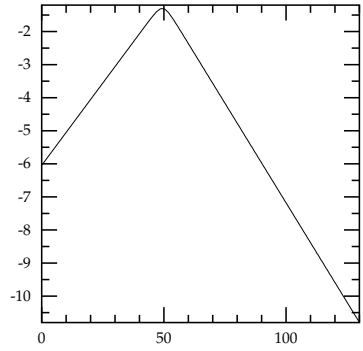}}                
  \subfigure[]{\includegraphics[trim = 0cm 0cm 1.4cm 1.8cm,  width=0.35\textwidth]{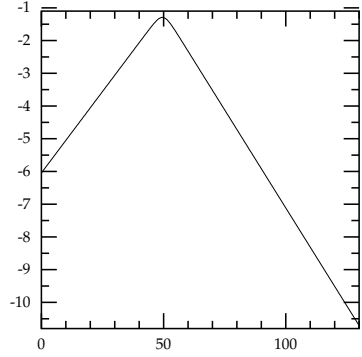}}
\subfigure[]{\includegraphics[trim = 0cm 0cm 1.4cm 1.8cm,  width=0.35\textwidth]{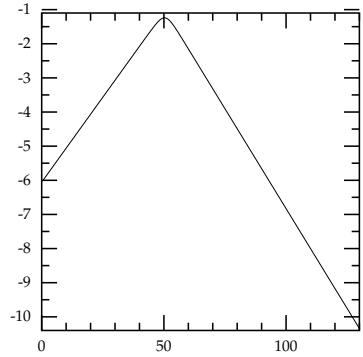}}
\subfigure[]{\includegraphics[trim = 0cm 0cm 1.4cm 1.8cm,  width=0.35\textwidth]{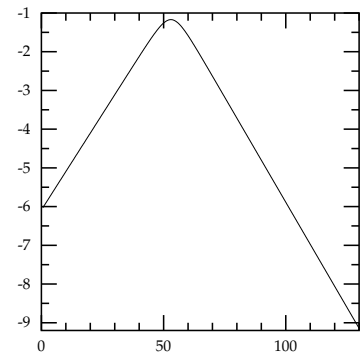}}
  \caption{Trajectories of one soliton for (a) $\varepsilon=-0.01$, (b) $\varepsilon=0.0$, (c) $\varepsilon=0.1$  and 
(d) $\varepsilon=0.5$.
  } \label{fig:15}
\end{figure} 
We do not see much difference in behaviour  between all 4 plots.
In the last figure we present the plots of the anomalies seen in the simulations described in the previous figure
(as all of them are very similar we plot the anomalies for only $\varepsilon=-0.01$ and  $\varepsilon=0.5$).
\begin{figure}
  \centering
  \subfigure[]{\includegraphics[trim = 0cm 0cm 1.4cm 1.8cm, width=0.35\textwidth]{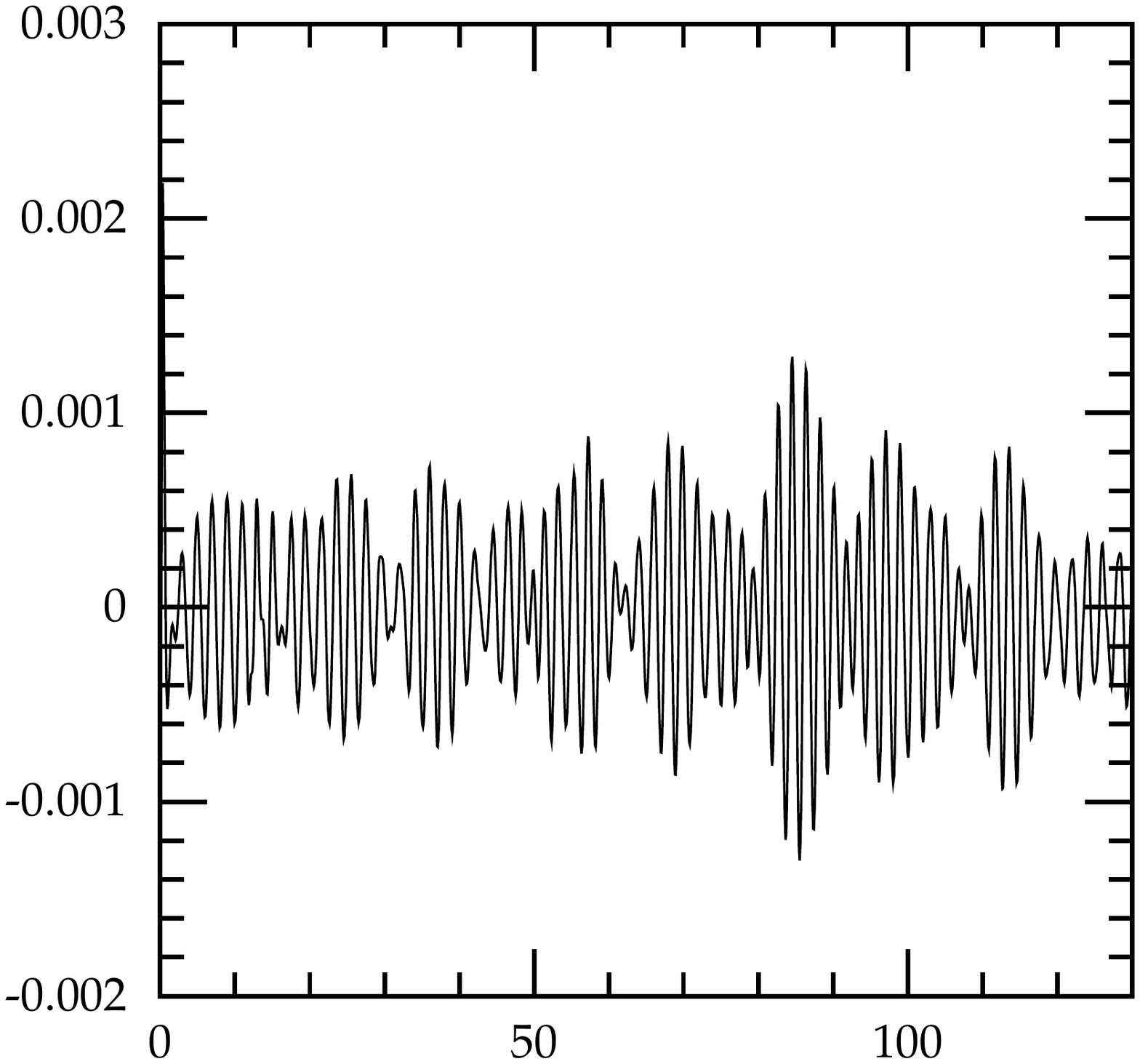}}                
  \subfigure[]{\includegraphics[trim = 0cm 0cm 1.4cm 1.8cm,  width=0.35\textwidth]{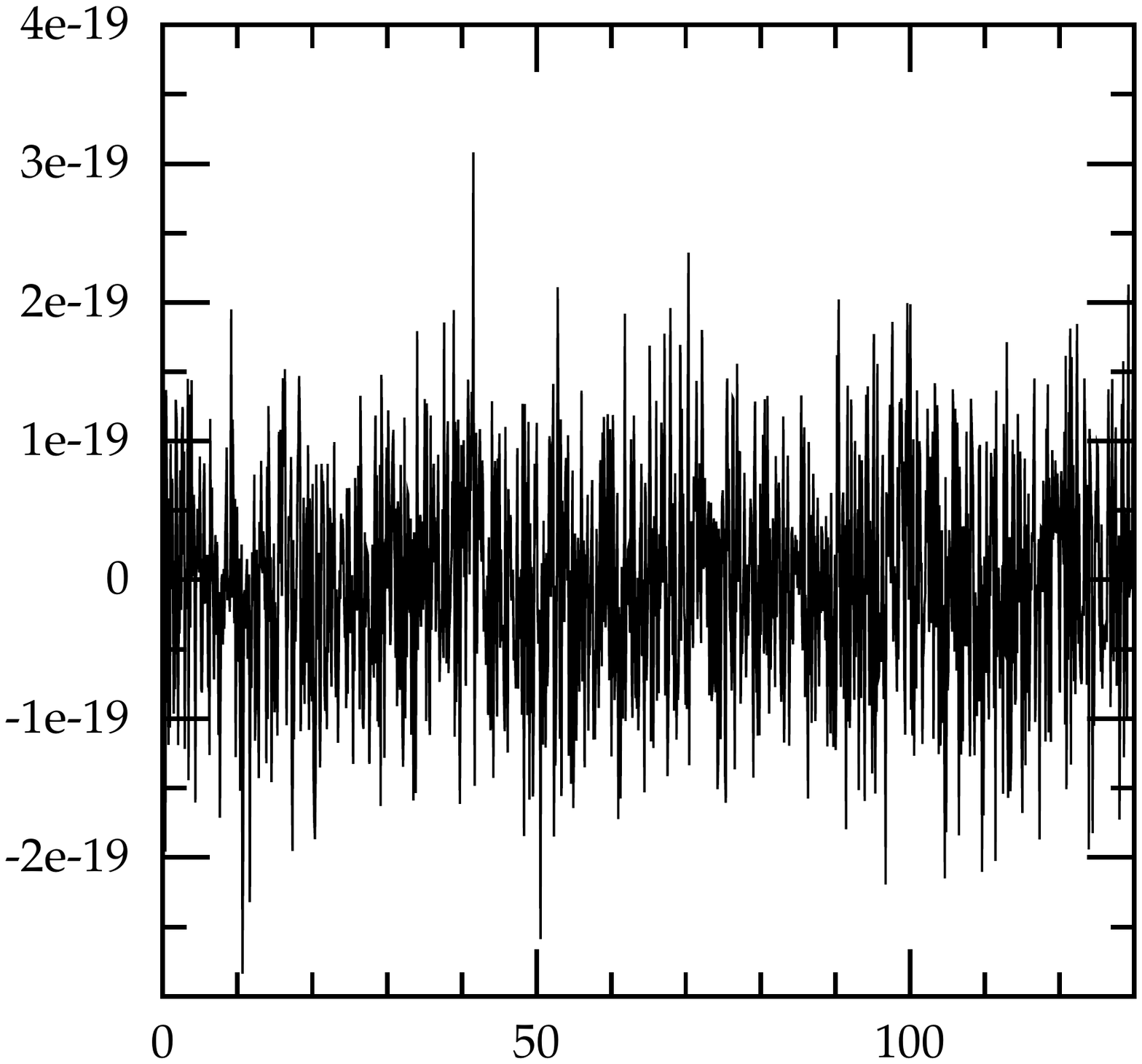}}
\subfigure[]{\includegraphics[trim = 0cm 0cm 1.4cm 1.8cm,  width=0.35\textwidth]{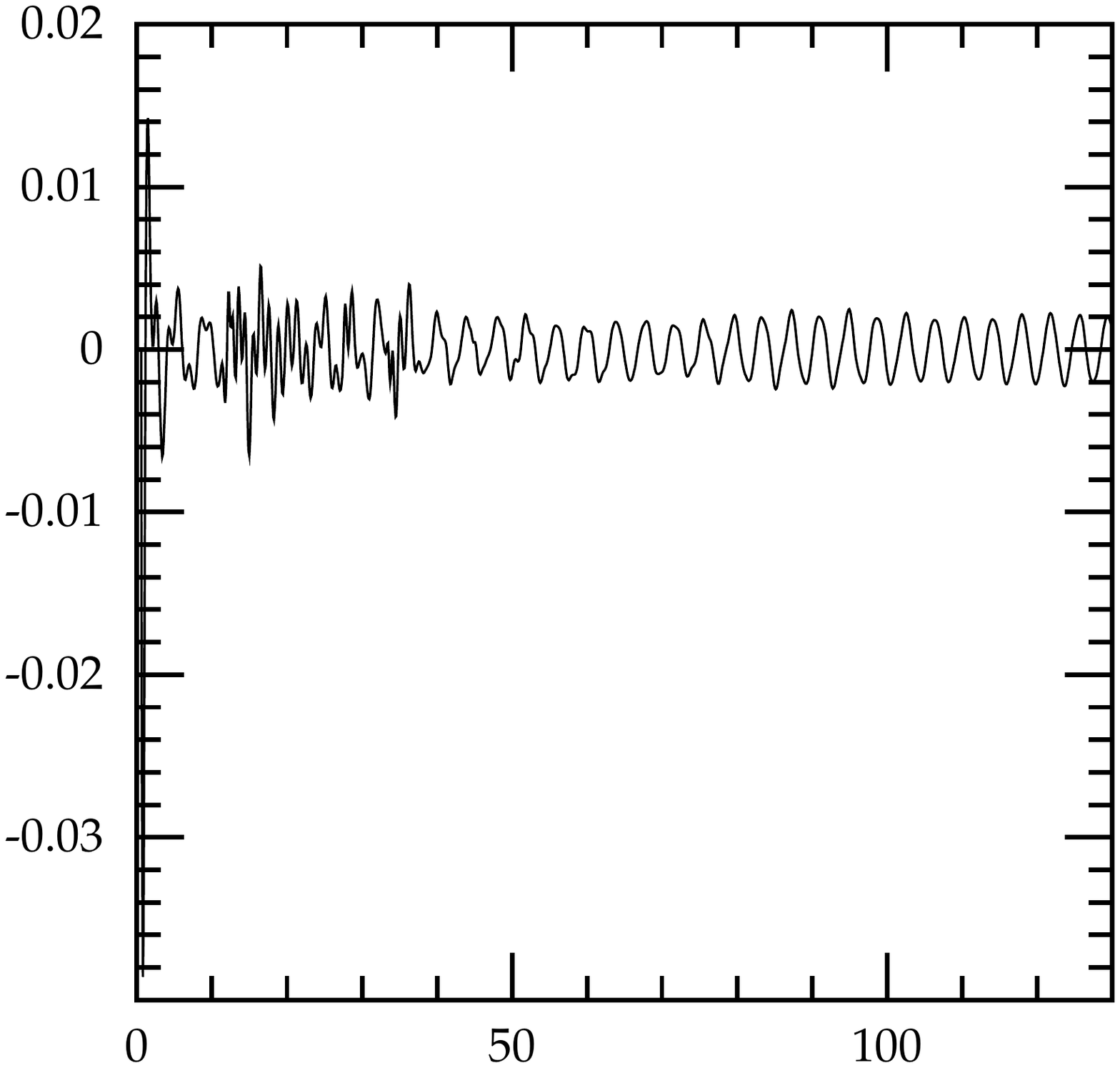}}
\subfigure[]{\includegraphics[trim = 0cm 0cm 1.4cm 1.8cm,  width=0.35\textwidth]{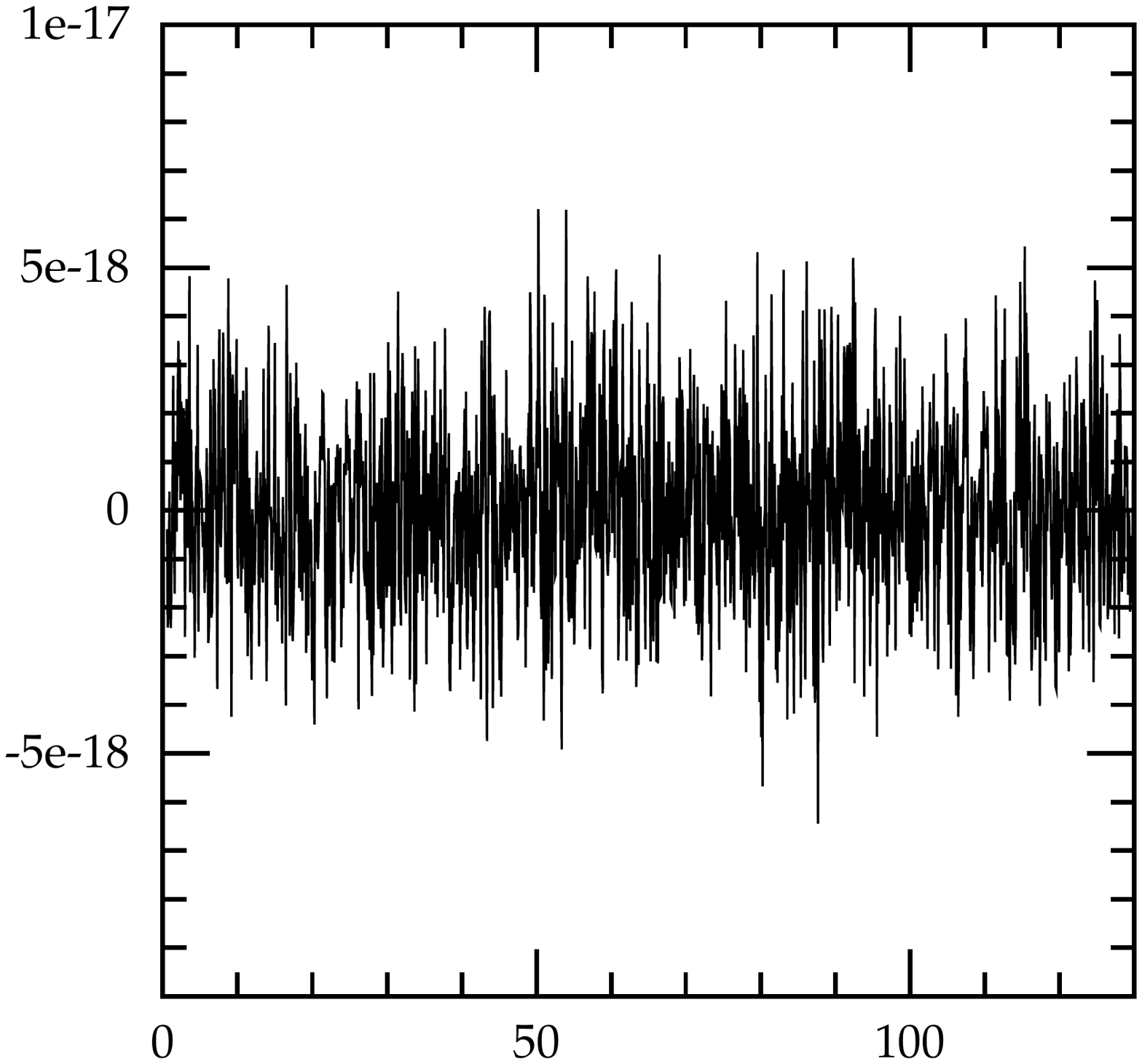}}
  \caption{Anomalies seen during the scatterings of solitons for (a), (b) real and imaginary parts for $\varepsilon=-0.01$ and 
(c) and
(d) the same for for $\varepsilon=0.5$.
  } \label{fig:16}
\end{figure} 
Again, like in the first case we see that the anomalies are essentially real (the imaginary parts are negligible) and the 
anomalies are very small (even smaller in this case). Of course, this is due to the fact that the solitons never get very close 
to each other.

\subsubsection{Further comments}

So far, in all above calculations, we have constructed the approximate (initial) two-soliton configurations by `gluing' two one-soliton ones. However, as we have two fields $\phi_1$ and $\phi_2$ we have more possibilities for performing such a construction.

For one soliton in the undeformed Toda model the fields $\phi_1$ and $\phi_2$ are related to each other by the symmetry mentioned in section 2. For two solitons we can construct the initial  $\phi_i$ fields by 'gluing' two one-soliton $\phi_1$ fields   into a two-soliton $\phi_1$ field and doing the same for $\phi_2$ fields or by taking the second one-soliton field by replacing $\phi_1$ and $\phi_2$. Both resemble the undeformed exact two-soliton fields and so at first sight both procedures can be expected to give essentially the same configurations which would then be expected to evolve in the same way (whether the initial configurations started them at rest or at a velocity towards each other).

In fact, in the discussion in the previous section the initial fields were constructed using the first approach (two $\phi_1$'s being used to construct a new $\phi_1$ and similarly for $\phi_2$). 

We have performed simulations using the second method of construction  (using both $\phi_1$ 
and $\phi_2$ fields to construct each of two soliton $\phi_i$ fields) and the results were always the same. So our expectations were correct.

%%%%%%%%%%%%%%%%%%%%%%%%%%%%%%%%%%%%%%%%%%%%%

\section{Conclusions}

In this paper we have discussed, in some detail, the results of our studies of the $SU(3)$ Toda model in (1+1) dimensions and some of its deformations.
First we looked at the undeformed model and studied some of its finite energy solutions. There were several of them, they all had real energy
and all these solutions were stable. This was checked by performing numerical simulations and comparing the results of these simulations with explicit 
analytical solutions (numerical simulations introduce small perturbations and so could be used to study their stabilities).

In our studies we looked at one and two soliton configurations. 
Amongst the solutions of the model there was one in which solitons remained at rest ({\it i.e.} the attractive and repulsive forces between them cancelled).
This cancellation of forces is very reminiscent of what is seen in systems of monopoles in (2+1) dimensions and suggests the existence of a BPS 
condition which, so far, we have not yet been able to find.

We have also perturbed the models by introducing a small perturbation. The perturbation we have considered corresponded to the change
of the angle between the root vectors of the root lattice. This changed the form of the potential $V(\phi_1,\phi_2)$ and it also changed the values of the vacua of the model. The perturbation made the model non-integrable and so we used it to see how its results fitted with our ideas on quasi-integrability.
Of course to do this we needed our perturbations to be small. In our work we have looked only at perturbations described by $\varepsilon$ and we varied $\varepsilon$ between -0.1 and +0.5.

We have performed many such simulations concentrating our attention on studying the scattering behaviour of two solitons. However to do this we needed one or two soliton field configurations which we did not have. So, first of all, we determined numerically one soliton configurations. This was done, as described in sections 6 and 7, by taking one soliton configurations of the unperturbed ({\it i.e.} $\varepsilon=0$) model and then perturbing them, so that the fields satisfied new boundary conditions, and then evolving them 
via a diffusive equation. Having determined the solutions of this equation (for various values of $\varepsilon$) we then constructed initial configurations
for our simulations by `tying' two one soliton configurations and, when we wanted to have moving solutions, boosting the solitons towards each other.
Such a procedure was successfully used in, say, \cite{LuizandVinicius} and we have tested it on the undeformed model ({\it i.e.} with $\varepsilon=0$). In the $\varepsilon\ne 0$ case the results of numerical evolutions of such static one soliton solutions were extremely close to the analytical solutions of the undeformed model (they were almost indistinguishable).
Hence, at least for small $\varepsilon$, we are confident of our results.

Then we have performed many simulations with the solitons initially at rest. First we looked at the case describing two solitons of the mixed case.
In this case we have found that for $\varepsilon>0$ the solitons repel while for 
$\varepsilon<0$ they attract and, of course, as knew originally, when $\varepsilon=0$, the forces cancel and the configuration is static.
The attractive case was found to be more interesting, as after the initial attraction, when the solitons got very close together, they started to repel and so the system oscillated.
During the oscillations the field configurations always looked the same. The energy was well conserved and the anomalies were very small.

Next we looked at the similar initial configurations but, this time, with the solitons initially moving towards each other with small velocities.
For very small velocities nothing was very different; at larger velocities the solitons could come `on top of each other'. In such cases, afterwards,
the fields $\phi_1$ and $\phi_2$ `swapped' their form, and afterwards, the solitons moved away from each other (we had a genuine `passing through each other').
For this to be the case we needed two fields, as then the rising field of one soliton in $\phi_1$ ended up in field $\phi_2$ and vice-versa.
This was observed in all cases for all values of $\varepsilon$ (for sufficiently large velocities).

The results of our simulations bring out also an additional difference between $\varepsilon=0$ and $\varepsilon\ne0$.
In the $\varepsilon=0$ case the solitons after their scattering are the same as before the scattering. In the $\varepsilon\ne0$ the solitons come out of the interaction 
region a little altered (in fact they oscillate and they move faster). This can be seen from Fig. 12a where the soliton after the scattering moves faster.
This suggests to us that the $\varepsilon\ne0$ models may have additional moving two-soliton solutions, but whether or not this is really the case,
would require further studies.

We have also looked at the solitons of the second class and in all their cases the solitons always repelled. In all the scatterings, that we have looked at (even for solitons sent towards each other with some velocity), the solitons always repelled at some short distances. And this was true for all values of $\varepsilon$ and, by this behaviour, the  scattering recalled very closely
the scattering of solitons in the Sine-Gordon model (unmodified or modified\cite{us}).
The anomaly also changed little. Thus, we note that our results, in addition to making some interesting observations about the properties of solitons of the unmodified
$SU(3)$ Toda model, also provide further support for the concept of quasi-integrability (as all the anomaly effects in the modified models were always very small).
Moreover, our results have also indicated that the static solutions of the unmodified model changed as one introduced our perturbations. 
For positive values of  $\varepsilon$ the solitons repelled and for negative values of  $\varepsilon$ they got modified to interesting oscillating fields.

\vspace{2cm}

\noindent{\bf Acknowledgements:} LAF and WJZ are very grateful for the hospitality at, respectively,  the Department of Mathematical Sciences of Durham University (UK) and the Instituto de F\'\i sica de S\~ao Carlos  (Brazil), where part of this work was developed, and to the Royal Society (UK) and FAPESP (Brazil) for financial support. LAF is partially supported by a CNPq (Brazil) research grant. PK is very grateful for hospitality at the Department of Mathematical Sciences of Durham University and the the Instituto de F\'\i sica de S\~ao Carlos.

\newpage

 %%%%%%%%%%%%%%%%%%%%%%%%%%%%%%%%%%%%%%%%%%
 %%%%%%%%%%%%%%%%%%%%%%%%%%%%%%%%%%%%%%%%%%
\appendix

\section{The $SU(3)$ loop algebra}
\label{sec:su3}

The six roots of the finite simple Lie algebra $SU(3)$ are given by $\pm{\vec \alpha}_1$, $\pm{\vec \alpha}_2$, and $\pm{\vec \alpha}_3=\pm\({\vec \alpha}_1+{\vec \alpha}_2\)$, satisfying ${\vec \alpha}_1\cdot {\vec \alpha}_2=-1$, and where we use the normalization ${\vec \alpha}_a^2=2$,  $a=1,2$, and so ${\vec \alpha}_3^2=2$. The $8$ generators of the algebra in the Chevalley basis are  the Cartan subalgebra generators $H_{\alpha_a}$, $a=1,2$, and the step operators $E_{\pm\alpha_s}$, $s=1,2,3$, which satisfy the commutation relations:
\br
\sbr{H_{\alpha_a}}{H_{\alpha_b}}&=&0 \qquad a,b=1,2;
\lab{su3comrel}\\
\sbr{H_{\alpha_1}}{E_{\pm\alpha_1}}&=&\pm 2\, E_{\pm\alpha_1};\qquad
\sbr{H_{\alpha_1}}{E_{\pm\alpha_2}}=\mp \, E_{\pm\alpha_2};
\qquad
\sbr{H_{\alpha_1}}{E_{\pm\alpha_3}}=\pm \, E_{\pm\alpha_3};
\nonumber\\
\sbr{H_{\alpha_2}}{E_{\pm\alpha_1}}&=&\mp\, E_{\pm\alpha_1};\qquad
\sbr{H_{\alpha_2}}{E_{\pm\alpha_2}}=\pm 2 \, E_{\pm\alpha_2};
\qquad
\sbr{H_{\alpha_2}}{E_{\pm\alpha_3}}=\pm \, E_{\pm\alpha_3};
\nonumber\\
\sbr{E_{\alpha_1}}{E_{-\alpha_1}}&=& H_{\alpha_1};\qquad
\sbr{E_{\alpha_2}}{E_{-\alpha_2}}= H_{\alpha_2};\qquad
\sbr{E_{\alpha_3}}{E_{-\alpha_3}}= H_{\alpha_1}+ H_{\alpha_2};
\nonumber\\
\sbr{E_{\alpha_1}}{E_{\alpha_2}}&=&E_{\alpha_3};\qquad
\sbr{E_{\alpha_1}}{E_{-\alpha_3}}=-E_{-\alpha_2};\qquad
\sbr{E_{\alpha_2}}{E_{-\alpha_3}}=E_{-\alpha_1};\qquad
\nonumber\\
\sbr{E_{-\alpha_1}}{E_{-\alpha_2}}&=&-E_{-\alpha_3};\qquad
\sbr{E_{-\alpha_1}}{E_{\alpha_3}}=E_{\alpha_2};\qquad
\sbr{E_{-\alpha_2}}{E_{\alpha_3}}=-E_{\alpha_1};\qquad
\nonumber
\er
with all the remaining commutators vanishing. In the triplet representation of $SU(3)$ the matrices satisfying \rf{su3comrel} are given by
\br
H_{\alpha_1}&=&\(\begin{array}{ccc}
1&0&0\\
0&-1&0\\
0&0&0
\end{array}\); \qquad\quad 
H_{\alpha_2}=\(\begin{array}{ccc}
0&0&0\\
0&1&0\\
0&0&-1
\end{array}\);
\lab{tripletmatrices}\\
E_{\alpha_1}&=&\(\begin{array}{ccc}
0&1&0\\
0&0&0\\
0&0&0
\end{array}\);\qquad\qquad
E_{\alpha_2}=\(\begin{array}{ccc}
0&0&0\\
0&0&1\\
0&0&0
\end{array}\);\qquad\qquad
E_{\alpha_3}=\(\begin{array}{ccc}
0&0&1\\
0&0&0\\
0&0&0
\end{array}\)
\nonumber
\er
and $E_{-\alpha_s}=E_{\alpha_s}^{\dagger}$. 
The generators of the infinite dimensional loop algebra associated to $SU(3)$ are obtained by multiplying the $SU(3)$ generators by powers of a complex parameter $\lambda$ as
\be
H_{\alpha_a}^n\equiv \lambda^n\, H_{\alpha_a};\qquad a=1,2,\qquad \quad \qquad\qquad 
E_{\pm\alpha_s}^n\equiv \lambda^n\,  E_{\pm\alpha_s};\qquad s=1,2,3
\ee
with $n$ being an integer. The commutation relations for the loop algebra are obtained from  \rf{su3comrel} by using the fact that the effect of $\lambda$ is just multiplicative, {\it i.e.} if $\sbr{T}{{\bar T}}={\tilde T}$, then 
$\sbr{T^m}{{\bar T}^n}={\tilde T}^{m+n}$, with $T$, ${\bar T}$ and ${\tilde T}$ being elements of the finite simple $SU(3)$ algebra. 

The relevant basis  appearing in the definition of the Lax potentials (see \rf{bpm1def} and \rf{fandef}) and also in the construction of the quasi-conserved charges in section \ref{sec:quasicharges} are given 
by (using the triplet matrix representation of $SU(3)$) 

\br
b_{3n+1}&=& E_{\alpha_1}^n+E_{\alpha_2}^n+E_{-\alpha_3}^{n+1}=\lambda^n\, \(
\begin{array}{ccc}
0&1&0\\
0&0&1\\
\lambda&0&0
\end{array}\);
\nonumber\\
b_{3n-1}&=& E_{-\alpha_1}^n+E_{-\alpha_2}^n+E_{\alpha_3}^{n-1}=\lambda^n\, \(
\begin{array}{ccc}
0&0&\frac{1}{\lambda}\\
1&0&0\\
0&1&0
\end{array}\);
\nonumber\\
F^1_{3n+1}&=& E_{\alpha_1}^n+\omega\,E_{\alpha_2}^n+\omega^2\,E_{-\alpha_3}^{n+1}=\lambda^n\, \(
\begin{array}{ccc}
0&1&0\\
0&0&\omega\\
\lambda\,\omega^2&0&0
\end{array}\);
\lab{matrixnewbasis}\\
F^1_{3n}&=& \(1-\omega^2\)\,H_{\alpha_1}^n+\(\omega-\omega^2\)\,H_{\alpha_2}^n=\lambda^n\, \(
\begin{array}{ccc}
1-\omega^2&0&0\\
0&-1+\omega&0\\
0&0&-\omega+\omega^2
\end{array}\);
\nonumber\\
F^1_{3n-1}&=& E_{-\alpha_1}^n+\omega\,E_{-\alpha_2}^n+\omega^2\,E_{\alpha_3}^{n-1}=\lambda^n\, \(
\begin{array}{ccc}
0&0&\frac{\omega^2}{\lambda}\\
1&0&0\\
0&\omega&0
\end{array}\);
\nonumber\\
F^2_{3n+1}&=& E_{\alpha_1}^n+\omega^2\,E_{\alpha_2}^n+\omega\,E_{-\alpha_3}^{n+1}=\lambda^n\, \(
\begin{array}{ccc}
0&1&0\\
0&0&\omega^2\\
\lambda\,\omega&0&0
\end{array}\);
\nonumber\\
F^2_{3n}&=& \(1-\omega\)\,H_{\alpha_1}^n+\(\omega^2-\omega\)\,H_{\alpha_2}^n=\lambda^n\, \(
\begin{array}{ccc}
1-\omega&0&0\\
0&-1+\omega^2&0\\
0&0&-\omega^2+\omega
\end{array}\);
\nonumber\\
F^2_{3n-1}&=& E_{-\alpha_1}^n+\omega^2\,E_{-\alpha_2}^n+\omega\,E_{\alpha_3}^{n-1}=\lambda^n\, \(
\begin{array}{ccc}
0&0&\frac{\omega}{\lambda}\\
1&0&0\\
0&\omega^2&0
\end{array}\);
\nonumber
\er
where $\omega$ is a cubic root of unity different from unity itself, {\it i.e.}
\be
\omega^3=1;\qquad\qquad\qquad 1+\omega+\omega^2=0;\qquad\qquad\qquad  \omega \neq 1.
\ee
The commutation relations of the loop algebra in such a basis can be easily obtained from their matrix construction given in \rf{matrixnewbasis}.

%%%%%%%%%%%%%%%%%%%%%%%%%%%%%%%%%%%%%%%%%%%%
%%%%%%%%%%%%%%%%%%%%%%%%%%%%%%%%%%%%%%%%%%%%

\end{document}